\documentstyle[12pt,epsf]{article}

\newcommand{\newsection}{    
\setcounter{equation}{0}
\section}

\renewcommand{\thefootnote}{\fnsymbol{footnote}}
\def\appendix#1{
  \addtocounter{section}{1}
  \setcounter{equation}{0}
  \renewcommand{\thesection}{\Alph{section}}
  \section*{Appendix \thesection\protect\indent \parbox[t]{11.715cm} {#1} }
  \addcontentsline{toc}{section}{Appendix \thesection\ \ \ #1}
  }

\newcommand{\non}{\nonumber \\*}

\newcommand{\eq}[1]{Eq.~(\ref{#1})}
\def \ov {\over }
\def\bea{\begin{eqnarray}}
\def\eea{\end{eqnarray}}

\def\LB{\left(}
\def\RB{\right)}
\def\be{\begin{equation}}
\def\ee{\end{equation}}
\def\ba{\begin{eqnarray}}
\def\ea{\end{eqnarray}}
\def\la{\label}
\def \bi{\bibitem}
\def \Tr {{\rm Tr}~}
\def\O{{\cal O}}
\def\C{{\cal C}}

\def\a{\alpha}
\def\b{\beta}
\def\A{{\cal A}}
\def\B{{\cal B}}
\def\ex{{\rm exp}}
\def\e{\epsilon}
\def\E{{\cal E}}

\def\S{{\cal S}}
\def\Z{{\cal Z}}
\def\tmega{{\bf\Omega}}
\def\tCom{{\rm{\bf Com}}}
\def\tRings{
{\rm{\bf Rings}}}
\def\L{{\cal L}}

\font\mybb=msbm10 at 12pt
\def\bb#1{\hbox{\mybb#1}}
\def\re {\bb{R}}

\font\mybb=msbm10 at 12pt
\def\bb#1{\hbox{\mybb#1}}

\begin{document}

\setcounter{page}{1}
\renewcommand{\thefootnote}{\arabic{footnote}}
\setcounter{footnote}{0}

\begin{titlepage}
\begin{flushright}
ITP-SB-00-40\\
\end{flushright}
\vspace{.5cm}

\begin{center}
{\LARGE Convergence Theorem for Non-commutative Feynman Graphs
and Renormalization
}\\
\vspace{1.1cm}
{\large Iouri Chepelev${}$\footnote{
E-mail: chepelev@insti.physics.sunysb.edu } 
and  Radu Roiban${}$\footnote{ \ E-mail: roiban@insti.physics.sunysb.edu } }\\
\vspace{18pt}

{\it C.N.~Yang Institute for Theoretical Physics}

{\it SUNY at Stony Brook}

{\it  NY11794-3840, USA}
\\
\end{center}
\vskip 0.6 cm

\begin{abstract}
We present a rigorous proof of the convergence theorem for the Feynman 
graphs
in  arbitrary  massive  Euclidean quantum  field theories on
non-commutative  $\re^d$ (NQFT). 
We give  a  detailed  classification  of  divergent  graphs in some
massive  NQFT  and  demonstrate  the  renormalizability
of some of these theories. 
\end{abstract}

\end{titlepage}

\setcounter{footnote}{0}
\noindent

\tableofcontents

\newsection{Introduction}
Recently,   perturbative aspects of non-commutative field theories (NQFT)
have received much attention 
[2-12].\footnote{For the non-perturbative aspects of NQFT see
 ref.\cite{nonpert}.}
  A list of references directly  relevant  for this
work together with a short summary of the corresponding relevant 
results is

\noindent
$\bullet$ Ref.\cite{we}.  Formulates a general convergence theorem for
the non-commutative
Feynman graphs.

\noindent
$\bullet$  Ref.\cite{mrs}.  Finds  explicit  examples  of
divergent  non-planar  graphs  in  the  massive  scalar  NQFT.  Depending
on the order  of  integration  in  the  Feynman  multiple  integrals,
the
divergences
show   up   either  as  infrared (IR)  or  ultraviolet (UV). 
Ref.\cite{mrs}
interprets  these  divergences
as  IR.\footnote{As we will see in section 4,
there  are  certain  types  of  divergent  graphs  with  the  property:
(1) in the corresponding Feynman multiple integrals 
the divergences always
show up as UV, independent of the order of the integration and
(2) the divergences come from the non-planar subgraphs.}

\noindent
$\bullet$ Ref.\cite{arefeva1}. Shows that the massive $\phi^4$ NQFT
is renormalizable up to two loops.
Ref.\cite{arefeva2}. Shows that the massive complex scalar NQFT with
the interaction  term  $\phi^*  \star \phi \star \phi^* \star \phi$ 
is one-loop  renormalizable  and  the  one-loop 
two-point  function  is
 free  of IR  singularities.

In ref.\cite{we}, based on various  consistency 
arguments   
we formulated  a  convergence theorem for the Feynman integrals
 in massive
  scalar NQFT with non-derivative couplings. 
In this paper we give a rigorous proof of the  theorem 
for arbitrary massive NQFT on $\re^d$.

The paper is organized as follows. In section 2 we give a
refined version of 
the heuristic proof of the convergence theorem given in
 ref.\cite{we}.   
The precise statement of the convergence
theorem  and the relevant  definitions  are  given  in  section 3.
In  section  4  we analyze the structure of divergences
in some massive NQFT and give  diagrammatic 
arguments for the renormalizability of some of these
 theories.
In section 5 we cast the proof of the convergence theorem for the
commutative scalar theories
with non-derivative couplings in a form that admits generalization to
the non-commutative case.  
In section 6  we prove the convergence theorem
for the  scalar NQFT with non-derivative
couplings. Using some of the constructions, theorems and lemmas from
section 6, 
in section 7 we  prove  the convergence
theorem for arbitrary NQFT on  $\re^d$ with
massive propagators. 
Section 8 contains some comments. The sections are inter-related as 
follows.

$$
{\rm section}~ 4
$$
$$
\uparrow
$$
$$
{\rm sections}~ 2 ~{\rm and}~ 3
$$
$$
\downarrow
$$
$$
{\rm section}~ 5\rightarrow {\rm section}~ 6 \rightarrow {\rm section}~ 7
$$

\newsection{A heuristic proof of 
the convergence theorem for the non-derivative scalar theories}
In this section we give a 
refined version of the heuristic
proof of the convergence theorem given in
ref.\cite{we}.

\begin{figure}[hbtp]
\begin{center}
\mbox{\epsfxsize=7truecm
\epsffile{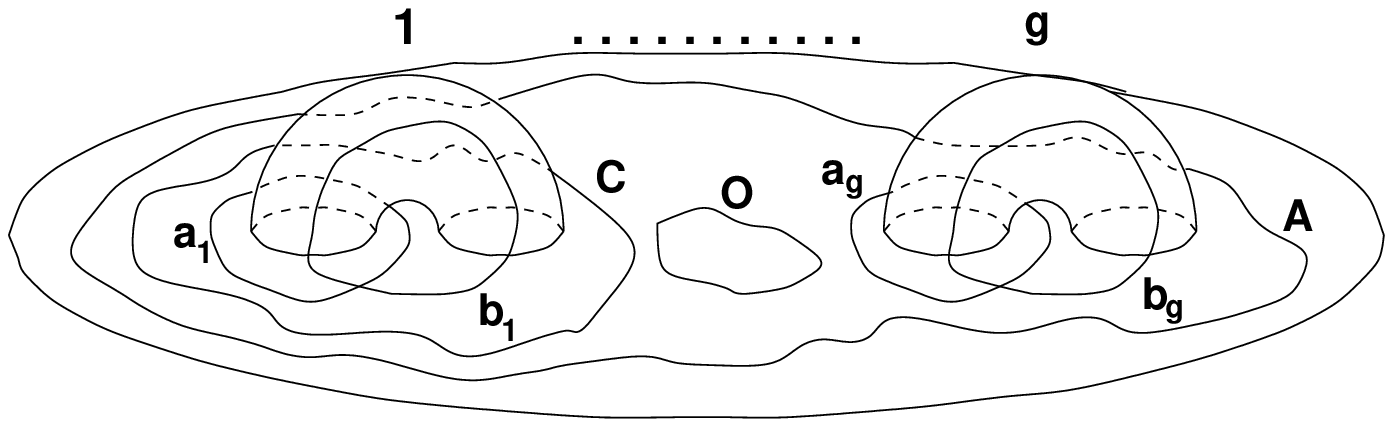}}~~~~
\mbox{\epsfxsize=7truecm
\epsffile{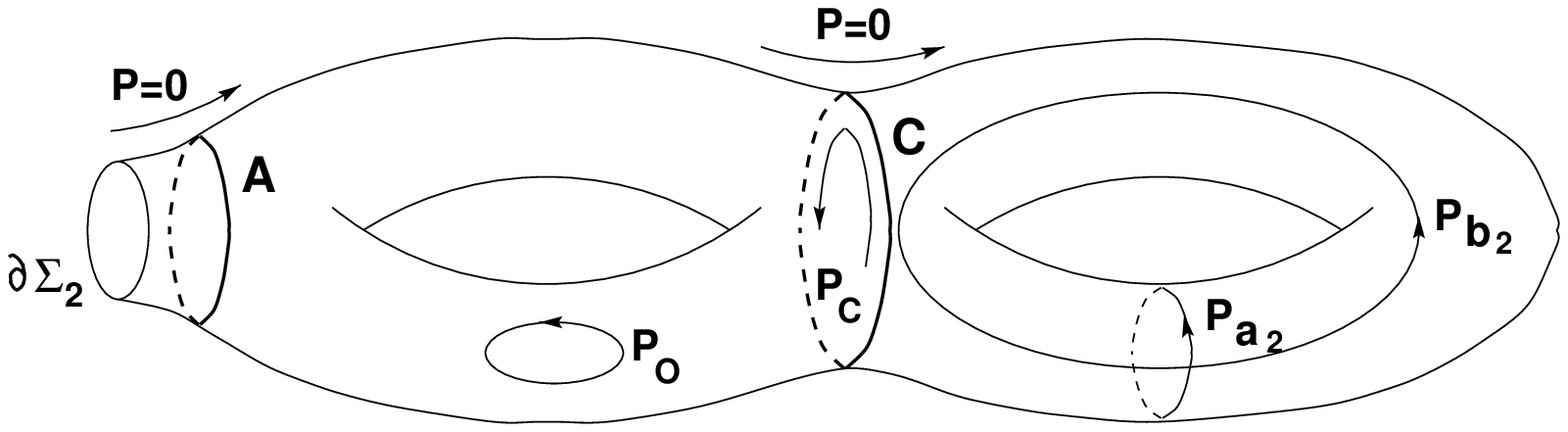}}
\end{center}
\flushleft{~~~~{\bf (a)}A genus $g$ surface with a boundary
~~~~{\bf (b)}Momentum flow on a $g=2$ surface $\Sigma_2$}
\caption{Two-surface with a boundary.}
\label{genushomt}
\end{figure}

\begin{figure}[hbtp]
\begin{center}
\mbox{\epsfxsize=8truecm
\epsffile{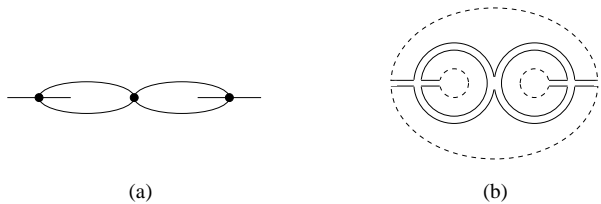}}
\end{center}
\caption{Sphere with three holes.}
\label{threehole}
\end{figure}

Consider a 1PI Feynman graph $G$ in a non-derivative 
massive scalar quantum field theory on non-commutative $\re^d$. 
It can be drawn on
a 2-surface  $\Sigma_g(G)$ of genus $g$ with a boundary 
$\partial \Sigma_g$. 
If $G$ has no external lines,
then $\Sigma_g(G)$ does not have a boundary.
In general $\partial \Sigma_g$   
consists of several  components. 
$\Sigma_g(G)$ for the graph in figure \ref{threehole}(a) has three
holes (boundary components).  This can be  seen from the corresponding
ribbon graph in figure \ref{threehole}(b). The Euler characteristic
$\chi$ is given by
\be
\chi=2-2g-B=C-I+V
\ee
where $g$ is the number of handles, $B$ is the number of holes,
$I$ is the number of edges (including external lines),
$V$ is the number of vertices and $C$ is
the number of closed single lines. For the graph in figure \ref{threehole}(b)
we have $g=0$, $B=3$, $I=8$, $V=3$ and $C=4$.

The non-trivial cycles of $\Sigma_g$ are
$a_1,b_1,\ldots,
a_g,b_g$ (see figure \ref{genushomt}(a)).  Cycles $A$, $C$ and $0$ are
trivial.\footnote{A cycle 
on $\Sigma_g$ is called non-trivial if it is
a non-trivial element of the first homology group $H_1(\Sigma_g)$.
In addition to the  trivial cycles that are contractible to a point,
there are trivial cycles which are not contractible to a point.   
For example, cycles $A$ and $C$  in figure \ref{genushomt} are trivial because 
$A=a_1b_1a_1^{-1}b_1^{-1}\cdots a_gb_ga_g^{-1}b_g^{-1}$ and 
$C=a_1 b_1 a_1^{-1} b_1^{-1}$, i.e. $A$ and $C$ are commutants. 
The fundamental group ${\cal F}(\Sigma_g)$ is a free group generated
by the generators $a_i,b_i$ ($i=1,\ldots,g$) satisfying the relation
$\prod_{i=1}^g a_i b_i a_i^{-1}  b_i^{-1} =1$.  The commutator
group $[{\cal F}, {\cal F}]$ is generated by the products
of commutants $ABA^{-1}B^{-1}$,  $A,B\in {\cal F}$. The first
homology group $H_1(\Sigma_g)$ is  the quotient
$H_1(\Sigma_g)={\cal F}/[{\cal F}, {\cal F}]$.
See
ref.\cite{springer} for the details.}

The  convergence  theorem  of  ref.\cite{we} can  be  stated  as  follows.

\noindent
{\it A 1PI graph $G$ 
is convergent for non-exceptional external momenta 
if for any subgraph  $\gamma\subseteq G$ (possibly disconnected)
at least one of the following conditions is satisfied:

\noindent
(1)
$\omega(\gamma)-c_G(\gamma)d<0$,

\noindent
(2) $j(\gamma)=1$}.

\noindent
where $c_G(\gamma)$ is the number of non-trivial homology cycles of
 $\Sigma(G)$ wrapped by the
subgraph $\gamma$, $j(\gamma)$ is an index which characterizes 
the non-planarity of $\gamma$ with respect to the external lines of graph $G$,
and $\omega(\gamma)$ is the
superficial
degree of divergence of a graph $\gamma$:
\be
\omega(\gamma)=d L(\gamma)-2I(\gamma),
\ee 
where $L$ and $I$ are the number of 
independent  loops and internal lines of $\gamma$ respectively. 
The graph in figure \ref{threehole}  has $j(G)=1$.
The definition of the index $j$ will be given below.

\begin{figure}[hbtp]
\begin{center}
\mbox{\epsfxsize=12truecm
\epsffile{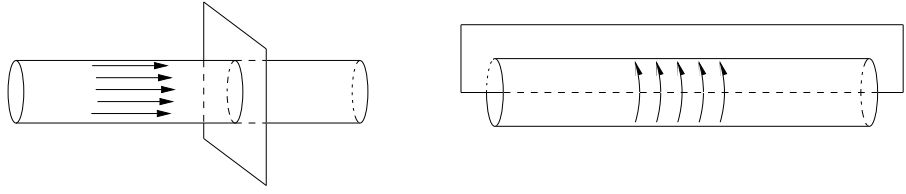}}
\end{center}
\caption{Measurement of momentum flow.}
\label{flow}
\end{figure}
\begin{figure}[hbtp]
\begin{center}
\mbox{\epsfxsize=4truecm
\epsffile{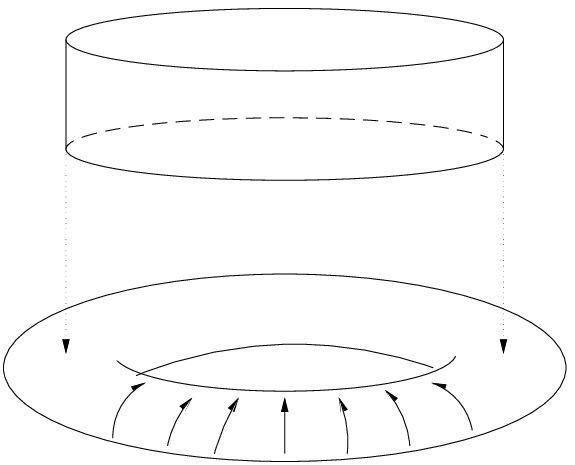}}
\end{center}
\caption{Measurement of momentum flow along $a$-cycle.}
\label{flow2}
\end{figure}

Let $I_{mn}(G)$ be the intersection matrix of the internal lines
of $G$.\footnote{It is constructed as follows. 
On $\Sigma_g(G)$  choose
an  arbitrary  tree $T$ of $G$. Now shrink $T$ to a point. The resulting
graph $G/T$ is a genus $g$ single-vertex graph. Next remove all
the loops  of $G/T$ which wrap homologically  trivial cycles of $\Sigma(G)$.
The $I_{mn}(G)$ can be read-off from  the resulting graph. Note that
$I_{mn}(G)$ depends on the choice of surface $\Sigma(G)$ and tree $T$.
Thus $I_{mn}(G)$ is not an invariant quantity. The procedure for 
constructing $I_{mn}(G)$ described here in Riemann surface-theoretic
terms is equivalent to the one described in refs.\cite{nakayama,filk}
in a different way. 
}
The phase factor associated with the internal lines of the 
Feynman graph $G$ is\cite{nakayama,filk}
\be
{\rm exp}~(i \phi(p))={\rm exp}~\LB i\sum_{m,n}
I_{mn}(G)\Theta_{\mu\nu} p_m^{\mu}p_n^{\nu} \RB
\label{phasefac}
\ee
where $p_m$ are the internal momenta. The matrix
 $I_{mn}(G)$ is not a topologically
invariant quantity. Namely, there may be several different intersection
matrices for a given graph $G$.  
The phase factor \eq{phasefac} gives  
 rise to a unique topology of $G$. Thus \eq{phasefac} should be 
rewritten in more invariant terms. For this purpose let us analyze the 
momentum flow on the surface $\Sigma(G)$. Figure \ref{genushomt}(b)
 illustrates the flow
of momentum on a genus two surface with a boundary. 
 There are  {\it topologically trivial} flows such as $p_0$ and 
{\it topologically non-trivial} flows such as $p_A$, $p_C$, $p_{a_2}$
and $p_{b_2}$. 
Note that the momentum flowing along the cycles is defined as
in figures \ref{flow} and \ref{flow2}.
Since 
the total external momentum flowing into the surface $\Sigma_2$ through 
$\partial\Sigma_2$ is zero,  the net momentum flowing across $A$ and
$C$ is zero (the momenta $p_A$, $p_C$ along $A$ and $C$ are 
nonzero). 
The  phase factor associated with  a graph arises from the {\it
linking} of topologically non-trivial flows. In figure \ref{genushomt}(b), $p_{a_2}$
and $p_{b_2}$  contribute  a phase factor 
$\ex (i\theta_{\mu\nu}p_{a_2}^{\mu}p_{b_2}^{\nu})$.
Cycles $A$ and $C$ do not contribute to the phase factor because
the net momentum flowing across each of these cycles is zero.
Since the cycles $a_2,b_2$ are homologically non-trivial and the
cycles $A$, $C$ and $0$ are homologically trivial, we conclude
that 
\noindent
{\it only the momentum flow along the homologically 
non-trivial cycles contribute to the
phase factor.}

The phase factors are supposed to 
regulate the subgraphs of $G$ and the above argument
suggests that only those subgraphs which wrap homologically non-trivial
cycles are  regulated by the phase factors.
This explains why 
the cycle number $c_G(\gamma)$  in the condition 1 of the 
convergence theorem is defined to be  the number of homologically 
non-trivial cycles wrapped by $\gamma$.

The condition $\omega(G)-c(G) d< 0$ for  $G$ can most
easily be understood from the Schwinger representation of the Feynman
integral $I_G$ for $G$ at zero external momenta. Schematically, this representation
reads
\be
I_G = \int d\alpha_1\cdots d\alpha_I {{\rm exp}( -\sum_l \alpha_l m_l^2 )
\ov (P(G,\theta))^{d\ov 2}}
\label{hlawin}
\ee
where $I$ is the number of internal lines of $G$ and 
\be
P(G,\theta)=\sum_{n=0}^{{c(G)\ov 2}} 
\theta^{2n} P_{2n}(\a),~~~~P_{2n}=\sum_{\{ i_1,i_2,\ldots , i_{L-2n}\} }  \a_{i_1}\a_{i_2}\cdots 
\a_{i_{L-2n}} 
\label{homopoly}
\ee
is a polynomial of degree $c(G)$ in $\theta$.
$c(G)$ is simply twice the
genus of $G$. 
Rescaling 
the Schwinger parameters, $\alpha_i \rightarrow t \alpha_i$,   we
find for $t\sim 0$
\be
I_G \sim t^{I- {d\ov 2} (L-c(G))}
\ee
Thus we see that $I_G$ is overall convergent if  $\omega(G)-c(G) d< 0$.

\begin{figure}[hbtp]
\begin{center}
\mbox{\epsfxsize=6truecm
\epsffile{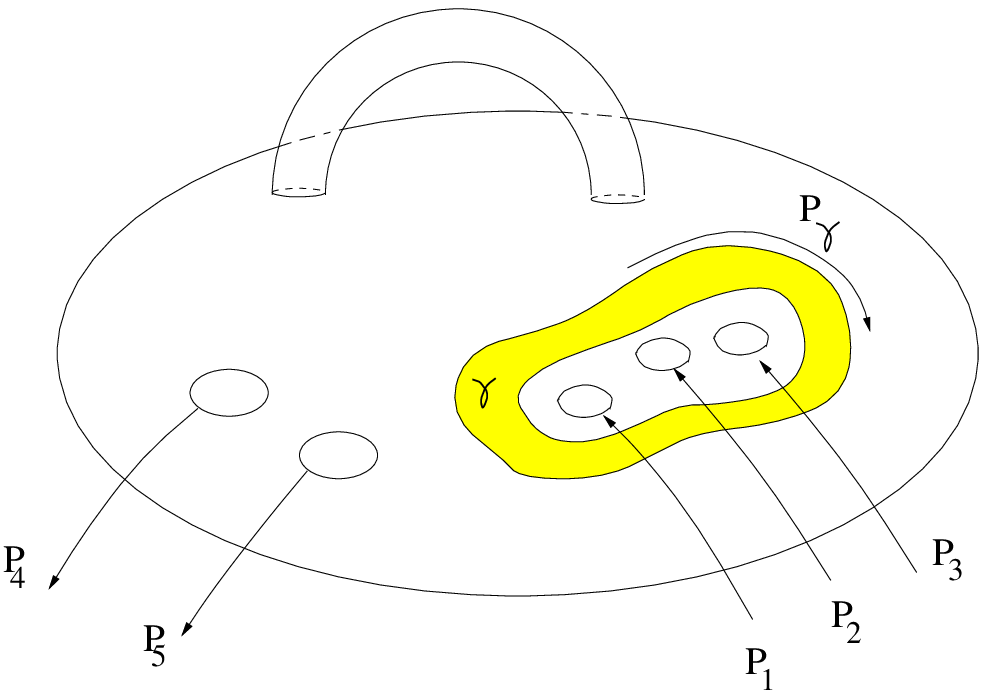}}
\end{center}
\caption{Index $j$.}
\label{jindex}
\end{figure}

The condition  $j(\gamma)=1$  can  be  understood  as  follows.
Consider the subgraph $\gamma$ of 
the $g=1$ graph with five holes shown in figure 
\ref{jindex}.  Let $p_{\gamma}$ be the momentum flowing along
$\gamma$. Let $p_1$, $p_2$ and $p_3$ be the total external
momentum flowing into
the external lines attached to the  holes. The phase factor associated
with the subgraph $\gamma$ is  
\be
{\rm exp}\LB i  (p_1+p_2+p_3) \wedge p_{\gamma}  \RB
\label{jphase}
\ee
where $p\wedge q \equiv p_{\mu}\Theta_{\mu\nu} q_{\nu}$. 
This phase factor will regulate $\gamma$ for the 
arbitrary $\omega(\gamma)$ as long as $p_1+p_2+p_3 \ne 0$.
If a subgraph has at least one hole inside and one hole
outside of it as $\gamma$ from figure \ref{jindex} does, then
$j(\gamma)=1$. Otherwise, $j(\gamma)=0$. A mathematically more precise
definition of $j$ will be given in section 3.
Note that when 
\be
p_1+p_2+p_3=0
\label{exception}
\ee
the phase factor in \eq{jphase} does not regulate $\gamma$. The momenta
satisfying the relation \eq{exception} are called exceptional.

In the Schwinger representation of the Feynman integral the 
condition $j(G)=1$ shows up as follows.
For the non-exceptional
external momenta, there will be a factor
$$
{\rm exp}(-Q(p,\theta))
$$
in the integrand in \eq{hlawin} which will scale like 
$${\rm exp}(-{f(p)\ov t}),~~~f(p)>0$$
if $j(G)=1$
and make the integral overall convergent at $t=0$.

\begin{figure}[hbtp]
\begin{center}
\mbox{\epsfxsize=10truecm
\epsffile{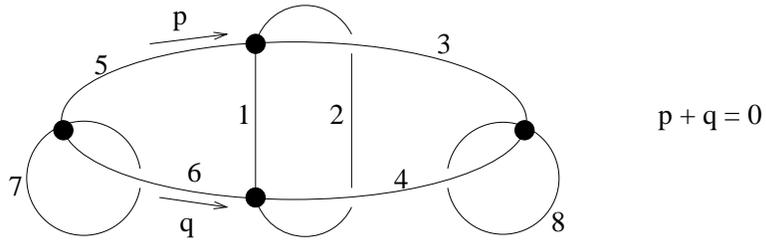}}
\end{center}
\caption{An example of the commutant  $a_1b_1a_1^{-1}b_1^{-1}$.}
\label{commutant}
\end{figure}

\begin{figure}[hbtp]
\begin{center}
\mbox{\epsfxsize=4truecm
\epsffile{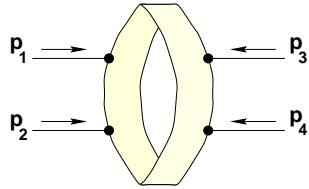}}
\end{center}
\caption{A ring with four external lines.}
\label{generic}
\end{figure}

\begin{figure}[hbtp]
\begin{center}
\mbox{\epsfxsize=6truecm
\epsffile{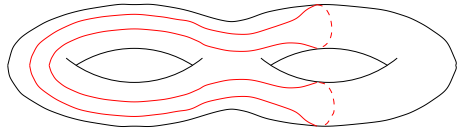}}
\end{center}
\caption{Commutant $b_1 a_2 b_1^{-1} a_2^{-1}$.}
\label{comm1212}
\end{figure}

\begin{figure}[hbtp]
\begin{center}
\mbox{\epsfxsize=12truecm
\epsffile{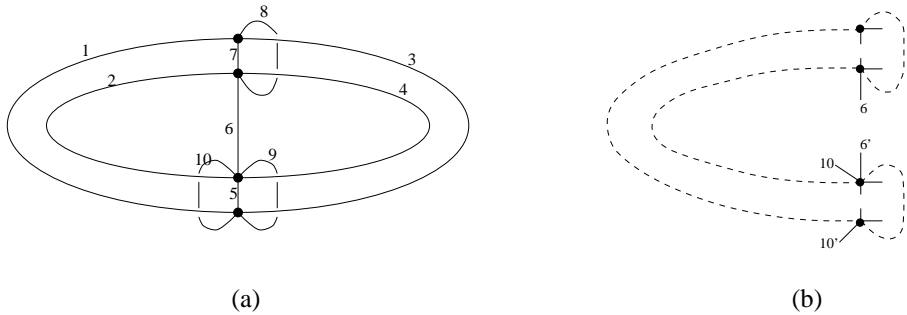}}
\end{center}
\caption{An example of the commutant  $b_1a_2b_1^{-1}a_2^{-1}$.}
\label{commutant1212}
\end{figure}

\noindent
{\bf Remark} (on the commutants). 
The graph  shown in figure \ref{commutant} contains the
 subgraph $\gamma$ formed by the lines 1 and 2 which wraps the cycle
$a_1b_1a_1^{-1}b_1^{-1}$. The latter  cycle is a commutant. 
 From the figure it is clear that
$p+q=0$. The subgraph $\gamma$ considered as a graph on its own is
a sphere with two holes ( see figure  \ref{generic} ). $p+q=0$,
which 
is
the case of the exceptional momenta, means that the total momentum
flowing into the hole is zero. A more sophisticated example of a 
commutant is
given in figure \ref{comm1212}. The graph in figure 
\ref{commutant1212}(a)  contains
a subgraph $\gamma$ formed by the lines 1,2,8 and 9  which wraps
the cycle $b_1 a_2 b_1^{-1} a_2^{-1}$  (see figure \ref{commutant1212}(b)).
As it is clear from figure \ref{commutant1212}(b), $\gamma$ is a sphere with
two holes and the total momentum flowing into the hole is zero.

\newsection{Statement of theorem and definitions}
Let us  give some definitions required for the formulation   
of the convergence theorem. The examples illustrating some of 
these definitions
can be found in ref.\cite{we}. In what follows, unless otherwise
stated, by a graph 
we mean an arbitrary (possibly disconnected) graph.

Let G be a connected NQFT  Feynman
graph. It can be drawn on
 a 2-surface  $\Sigma_g(G)$ of genus $g$ with a boundary 
$\partial \Sigma_g$. 
Let us  define the
cycle number $c_G(\gamma)$ of an arbitrary 
subgraph $\gamma\subseteq G$ with respect to the surface $\Sigma(G)$.

\noindent
{\bf Definition 1.}{\it
The first homology group of  $\Sigma_g(G)$ for the connected 
graph $G$ has the basis 
$\{ a_1,b_1,\ldots,a_g,b_g\}$ (see figure \ref{genushomt}). 
$c_G(\gamma)$ is defined as the number of inequivalent non-trivial cycles
of $\Sigma_g(G)$ spanned by the closed paths in $\gamma$. }

\noindent
{\bf Remark.} One should distinguish between {\it abstract} cycles
which are not wrapped by any loop of the graph and {\it physical} 
cycles which are wrapped by the loops of the graph. To make this point
clear consider a genus $g$ graph $G$. There may be a number of ways of 
drawing $G$ and thus there may be more than one genus $g$ surface
associated with $G$. Let us consider one particular surface $\Sigma_g$.
Let $\{ a_1,b_1,\ldots,a_g,b_g\}$ be the canonical basis of the homology
group associated with $\Sigma_g$. It is clear that  $a_i$, $b_i$ 
 are not necessarily wrapped by any loop of $G$. For example, 
it may well happen
that  $a_1b_1$ and $a_1b_1^{-1}$ are wrapped by loops in $G$, but
$a_1$ and $b_1$ are not. Given a loop ${\cal L}$ in a graph $G$, we
will denote by ${\cal C}({\cal L})$  the corresponding
homology cycle of $\Sigma(G)$. As an example consider the graph in
figure \ref{commutant1212}(a). Let $\{1,3\}$ and $\{2,4\}$ 
 be the subgraphs formed by
the lines 1,3 and 2,4, respectively. Then  
${\cal C}(\{1,3\})={\cal C}(\{ 2,4\})=b_1 b_2$.
Note that ${\cal C}({\cal L})$
is defined to be an {\it element} of the homology group. For example
consider the graph in figure \ref{commutant}. It is clear that
${\cal C}(\{1,3,4\})={\cal C}(\{ 2,3,4\})$,
since  $\{1,3,4\}$  and  $\{2,3,4\}$  differ only
by a commutant, i.e. a trivial element of the homology group.

We will occasionally drop the index $G$ in $c_G$ when the
surface  $\Sigma(G)$ in question is clear from the context.

\noindent
{\bf Remark.} A general  NQFT Feynman graph $G$ with $N_c$ connected
components
can be drawn on  $N_c$ 2-surfaces $\Sigma_{g_1},\ldots,
\Sigma_{g_{N_c}}$.
The cycle number $c_G(\gamma)$ of an arbitrary 
subgraph $\gamma \subseteq G$ is defined as a sum of cycle
numbers of $\gamma$ with respect to the surfaces $\Sigma_{g_1},\ldots,\Sigma_{g_{N_c}}$. In what follows, by the genus $g$ of  a disconnected graph 
we mean the sum of genera of the connected components.

\begin{figure}[hbtp]
\begin{center}
\mbox{\epsfxsize=12truecm
\epsffile{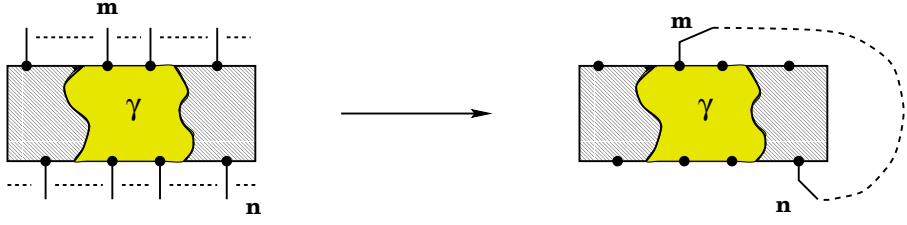}}
\end{center}
\caption{Definition of index $j$.}
\label{defj}
\end{figure}

Let $\E(G)$ be the set of external lines of the graph $G$.
Consider two external lines  $m,n  \in  \E(G)$.
As in figure \ref{defj}
 set the rest of the external momenta of $G$ to zero and
connect 
the lines $m$ and $n$. Denote the resulting graph by $G_{mn}$.
Let $c_{G_{mn}}(\gamma)$ be the cycle number of 
an arbitrary subgraph $\gamma \subset G$  with respect to the 2-surface of
graph $G_{mn}$. There are only two possibilities:
$$c_{G_{mn}}(\gamma)> c_G(\gamma)~~~~~{\rm or}~~~~~
c_{G_{mn}}(\gamma)= c_G(\gamma)$$
The index $j$ of an arbitrary  subgraph
$\gamma \subseteq G$ is defined as 
follows.

\noindent
{\bf Definition 2.}{\it
If there  exists a pair of external lines $m,n$ such that
$c_{G_{mn}}(\gamma)> c_G(\gamma)$, then $j(\gamma)=1$. 
Otherwise, $j(\gamma)=0$.}

From the definition of the index $j$ given above it is easy to see
that if  $\gamma \subseteq \gamma'$  and
$j(\gamma)=1$, then $j(\gamma')=1$.

\noindent
{\bf Remark.} The meaning of $j$ is the following.
Let $\partial \Sigma(G)$ be the boundary of
$\Sigma(G)$. If the number of  components
of $ \partial \Sigma(G)$ is greater than one, then $j(G)=1$. Otherwise,
$j(G)=0$.
The condition $j(\gamma)=1$ for a subgraph $\gamma\subset G$ 
 means
that there is at least one hole inside and one hole outside of 
$\gamma$ (see figures \ref{threehole} and \ref{jindex}.).


\noindent
{\bf Definition 3.}{\it The sets $K_1(G)$ and $K_2(G)$ of the lines
of a graph $G$ are defined as
\be
K_1(G)=\{ l\in G| L(G-l)=L(G)\}
\label{K1set}
\ee
and
\be
K_2(G)=\{ l\in G |  g(G-l) < g(G) \}
\label{K2set}
\ee
}

\noindent
{\bf Remark}.  $K_1(G)\cap K_2(G)=\emptyset$.

We now give the definition of exceptional external momenta
of a 1PI non-commutative graph $G$. Let ${\cal E}(G)$ be the set of
external lines of graph $G$. For a given subgraph $\gamma\subseteq G$, 
we can decompose ${\cal E}(G)$ into
the disjoint subsets ${\cal E}_s(\gamma)~,~~s=1,\dots, n(\gamma)$, 
with the following properties:

\noindent
$\bullet$ For any $s$ we have
$c_{G_{ij}}(\gamma)= c_G(\gamma)$, $\forall i,j \in {\cal E}_s(\gamma)$.

\noindent
$\bullet$ For any $s_1$ and $s_2$ such that $s_1\ne s_2$,
we have  $c_{G_{ij}}(\gamma)> c_G(\gamma)$,
$\forall i \in{\cal E}_{s_1}(\gamma)$,
 $\forall j\in {\cal E}_{s_2}(\gamma)$.

The existence and  uniqueness of the decomposition
\be
{\cal E}(G)=\bigcup_{s=1}^{n(\gamma)} {\cal E}_s(\gamma)
\label{decompos}
\ee
for a given $\gamma$
is proven in lemma \ref{lem4}. The meaning of this decomposition is
the following. $n(G)$ is simply the number of  components
of $\partial\Sigma(G)$. In other words it is the number of holes
of $\Sigma(G)$. The meaning of $n(\gamma)$ for the subgraph $\gamma$
 can 
be understood from the figure~\ref{jindex}.

\noindent
{\bf Definition 4.}{\it
The  external momenta $\{ p_i \}$ of a 1PI graph $G$ are called
exceptional if for at least one $\gamma\subseteq G$ with $n(\gamma)=2$ in
\eq{decompos}
we have}
\be
\sum_{i\in {\cal E}_1} p_i =\sum_{i\in {\cal E}_2} p_i= 0
\label{deffour}
\ee

\noindent
{\bf Remark.} The meaning of the exceptional momenta is
the following. Consider the  subgraph $\gamma \subset G$ in
figure \ref{jindex}. It separates the holes of $\Sigma(G)$ into two sets.
The external momenta are exceptional if the sum of momenta flowing
into  the holes inside $\gamma$ ( or equivalently, outside of $\gamma$ )
is zero. For the momenta to be exceptional, it is not required to have
the momentum flowing into each hole to be zero separately.

Let us now give a generalization of the usual definition of the 
superficial degree of divergence to the non-commutative 
theories. We will consider the class of non-commutative field theories
with massive propagators. Examples of this class are
massive scalar theories with arbitrary couplings ( with or
without derivatives ) and  massive spinor theories.
The factor in the Feynman integral 
associated with a line $l$ in a Feynman graph of such theories
is of the form
$$
{P(q_l)\ov q_l^2 +m_l^2}
$$
where $P(q_l)$ is a polynomial of degree ${\rm deg}(l)$ in 
the momentum $q_l$ associated with the line $l$. Let us define 
an index ${\rm ind}_{K_2}(l)$ associated with a line $l\in G$ as
\be
{\rm ind}_{K_2}(l) = \left\{
\begin{array}{cl}
1 & {\rm ~if~}l\notin K_2\\
0 &{\rm ~if~}l\in K_2
\end{array} \right.
\label{indK2}
\ee

\noindent
{\bf Definition 5.}{\it
For an arbitrary subgraph $\gamma \subseteq G$ of a graph $G$,
the degree of divergence $\omega(\gamma)$ is defined as
\be
\omega(\gamma)= d L(\gamma)-2I(\gamma) + \sum_{l\in \gamma} {\rm ind}_{K_2}(l)
{\rm deg}(l)
\label{newomega}
\ee
where $L(\gamma)$ and $I(\gamma)$ are the number of 
independent  loops and internal lines of $\gamma$, respectively.}

The convergence theorem (Theorem 5) can be stated as follows.

{\it  Let $I_G$ be the Feynman integral  
for a 1PI graph $G$ in a 
field theory over non-commutative $\re^d$ with the massive propagators.
There are three cases:

\noindent
(I) If $j(G)=0$, then $I_G$ is convergent if 
   $\omega(\gamma)-c_G(\gamma)d<0$ for all $\gamma\subseteq G$.

\noindent
(II) If $j(G)=1$ and the external momenta are non-exceptional, then
$I_G$ is convergent  if 
for any subgraph $\gamma\subseteq G$ 
at least one of the following conditions is satisfied:(1)
$\omega(\gamma)-c_G(\gamma)d<0$, (2) $j(\gamma)=1$.

\noindent
(III) If $j(G)=1$ and the external momenta are exceptional, then
$I_G$ is convergent if
$\omega(\gamma)- c_G(\gamma)d<0$ for all $\gamma\subseteq G$.
}

When ${\rm deg}(l)=0, \forall l\in G$, i.e. for the theories 
with non-derivative couplings, the above theorem reduces to the
one we conjectured in ref.\cite{we}.

\begin{figure}[hbtp]
\begin{center}
\mbox{\epsfxsize=2truecm
\epsffile{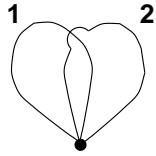}}
\end{center}
\caption{Illustration for a peculiar example.}
\label{exampleps}
\end{figure}

Let  us   give  a  simple  example  illustrating  a  peculiar 
convergence 
property of diagrams with the derivative couplings
related to the definition \eq{newomega} of $\omega$. Consider
the graph $G$ in figure \ref{exampleps}. Suppose that there is a factor of 
$p_2^n$
in the numerator of the Feynman integrand
for  $G$. Then

\be
I_G= {\rm const} \int_0^{\infty} d\a_1 d\a_2 
{\ex (-\a_1 m_1^2-\a_2 m_2^2 )\ov [\a_1\a_2 + 
\theta^2]^{d\ov 2}}
{\partial^n \ov \partial q^n} \ex [- {\a_1 q^2 \ov \a_1\a_2 + 
\theta^2} ]\bigg|_{q=0} 
\ee
( For simplicity we assumed that the absolute values of the 
eigenvalues of $\Theta$ are
all equal).
It is easy to see that this integral is convergent for any $n$. 
This property is directly related  to the fact that
line 2 of $G$ belongs to the set $K_2(G)$ ( see definition 3 ).

\newsection{Renormalization}
\subsection{ General discussion.}

Let us recall the basic idea behind the subtraction formula of 
ref.\cite{we}. 
The vertices in the non-commutative Feynman graphs come from 
the interaction  term $\int {\cal L}_{int}$. The interaction term has only
cyclic symmetry. Thus one may think of an n-vertex as a rigid ribbon
vertex.
 The topology of a Feynman graph in 
NQFT  arises  from  {\it the way}
 the legs of the $n$-vertices are {\it joined} together
by the propagators. Locally around the center of a vertex the graph
is planar. Thus the  Lagrangian is capable of generating only planar
vertices.

To illustrate the idea, consider the $U(N)$ real scalar QFT in $d=4$ with the
tree level Lagrangian
\be
{\cal L}=\Tr [(\partial_{\mu}\phi)^2+m^2\phi^2+ g~ \phi^4 ]
\label{phifour}
\ee
The  Lagrangian  \eq{phifour} is invariant under the global symmetry
$\phi\rightarrow U^{-1}\phi U$. It is clear that \eq{phifour} is not
the most general $U(N)$ invariant Lagrangian. The 
quantum corrections will induce the terms of the following forms.

\noindent
$\bullet$
The overall divergent diagrams 
of the type   ``sphere with two holes''   will induce
the terms
$(\Tr \partial_{\mu} \phi)^2$, $(\Tr \phi)^2$, 
$(\Tr \phi)(\Tr \phi^3)$ and  $(\Tr \phi^2)^2$.

\noindent
$\bullet$
The overall divergent diagrams 
of the type  ``sphere with three holes''   will induce
the term
$(\Tr \phi)^2 (\Tr \phi^2)$.

\noindent
$\bullet$
The overall divergent diagrams 
of the type   ``sphere with four holes''   will induce
the term 
$(\Tr \phi)^4$.

Thus we see that $\Tr \phi^4$ is capable of subtracting the divergences
only
from the graphs of the type ``sphere with one hole'' (=planar graphs). 
At first sight it
 seems that the same is  true for  $\int \phi\star\phi\star\phi\star\phi$ of
the NQFT, since $\int$ can be thought of as a trace.

The subtraction of the planar divergences which occur in a non-commutative
graph is most natural in BPHZ approach. In the latter approach one
does not introduce a regulator, but rather deals directly with the
integrand of the Feynman integral. One subtracts the first few terms
of the Taylor series of the integrand as a function the external
momenta. 
The topology of a non-commutative graph arises from the phase
factors coming from the interaction vertices.  Thus to keep the topology
of the graph intact, one should not act with the Taylor expansion
operator on the phase factors.

Thus, if the planar subgraphs of a non-commutative graph $G$
are the only subgraphs which violate the conditions of
the convergence theorem, then these divergences can be subtracted using
the counter-terms of the form which occur in the original Lagrangian.
This class of graphs was called class $\Omega_d$ in ref.\cite{we}.
Here we will simply call it class $\tmega$, keeping in mind that
$\tmega$ refers to a particular theory in a particular dimension.

\begin{figure}[hbtp]
\begin{center}
\mbox{\epsfxsize=10truecm
\epsffile{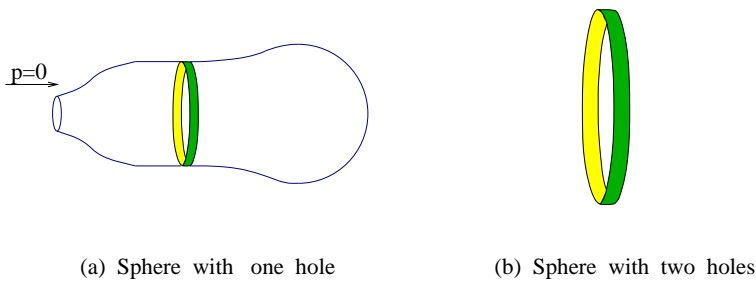}}
\end{center}
\caption{A baloon.}
\label{baloon}
\end{figure}

Note that a planar graph (=sphere with one hole) may contain a divergent
subgraph which is a sphere with more than one hole. Consider the 
sphere with a hole shown
in figure \ref{baloon}(a). Since the total external 
momentum flowing into the hole is zero, the momentum flowing
across the ring subgraph (=sphere with two holes) is zero. Thus
the phase factors will not regulate this ring subgraph. More
technically, this  follows from the fact that $c(\gamma)=0$ for the ring
subgraph $\gamma$ of the sphere with a hole.

\begin{figure}[hbtp]
\begin{center}
\mbox{\epsfxsize=6truecm
\epsffile{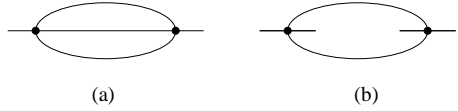}}
\end{center}
\caption{A non-planar subgraph of a planar graph.}
\label{planar}
\end{figure}

Consider $\phi^4$ theory
in $d=4$.
The  graph in figure \ref{planar}(a) contains a subgraph of the type
``sphere with two holes'' figure \ref{planar}(b).\footnote{This
particular graph was given in ref.\cite{arefeva1}.}

\begin{figure}[hbtp]
\begin{center}
\mbox{\epsfxsize=6truecm
\epsffile{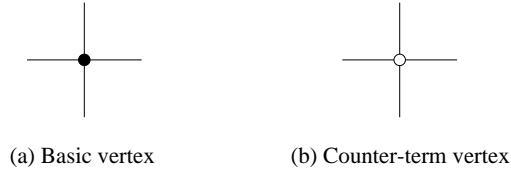}}
\end{center}
\caption{Vertices}
\label{vertex}
\end{figure}

\begin{figure}[hbtp]
\begin{center}
\mbox{\epsfxsize=10truecm
\epsffile{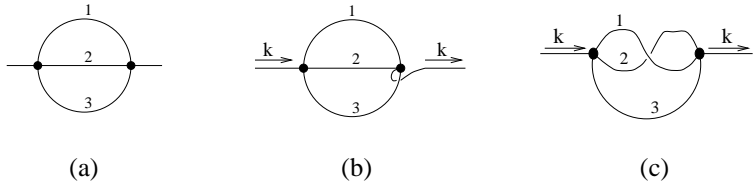}}
\end{center}
\caption{Sunset diagrams.}
\label{sunset}
\end{figure}

\begin{figure}[hbtp]
\begin{center}
\mbox{\epsfxsize=10truecm
\epsffile{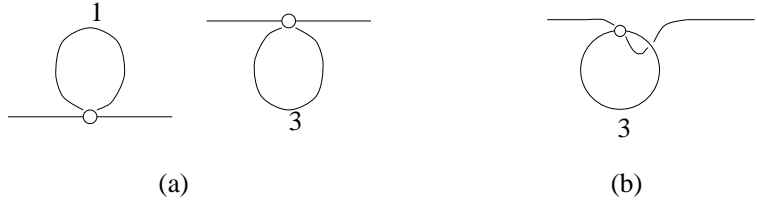}}
\end{center}
\caption{Counter-term graphs.}
\label{counter}
\end{figure}

Let us analyze  the sunset diagrams in figure 
\ref{sunset}.
The planar sunset diagram figure \ref{sunset}(a) is overall divergent
and contains three divergent subgraphs: 1-2, 1-3 and 2-3. 
The overall divergence as well as the divergences of 1-2 and 2-3 can
be subtracted using the planar vertices (see figure \ref{counter}(a)).
The basic and the counter-term vertices 
\be
g~ \phi\star\phi\star\phi\star\phi,~~~~~~\delta g(\Lambda)~
\phi\star\phi\star\phi\star\phi
\ee
are shown in figure \ref{vertex}.
But the subgraph 1-3 seems to be quite problematic because there 
is no
corresponding counter-term graph. By analogy with the commutative
$U(N)$ $\phi^4$ theory one expects counter term of the type
$(\Tr \phi^2)^2$. Under  the identification $\int d^4x \leftrightarrow 
{\rm Tr}$ we seem to need
\be
\LB\int d^4x \phi^2 \RB^2
\ee
counter term, which is catastrophic. Ref.\cite{arefeva1} 
worked with the single line notation for the Feynman graphs and used
the 
cosines for the interaction vertices to show that
$\phi^4$ NQFT is 2-loop renormalizable.  So what is wrong with 
 our argument  regarding the planar sunset diagram? The 
resolution of this puzzle will be given shortly.

Consider now the non-planar sunset diagram in figure \ref{sunset}(b).  
From the convergence theorem it follows that
for $k\ne 0$
the non-planar sunset diagram
figure \ref{sunset}(b) is overall convergent and the only divergent
subgraph is 1-2. The whole graph and the 
subgraphs 1-3 and 2-3 have $j(\gamma)=1$.
The corresponding counter-term graph is shown in
figure \ref{counter}(b).

The non-planar sunset diagram figure \ref{sunset}(c)
is convergent for any $k$ because all subgraphs satisfy the condition
$\omega(\gamma)-4 c(\gamma)<0$.

Now let us return to our puzzle. In the dimensional regularization
the divergent part of the
Feynman integral for the planar
sunset diagram is of the form 
\be
c_1~ {m^2\ov \epsilon^2}+ c_2~ {m^2\ov \epsilon}
+c_3 ~ {m^2\ov \epsilon}{\rm ln}{m^2\ov \mu^2}
+c_4~{k^2\ov \epsilon}
\label{ramsunset}
\ee
where $c_1$, $c_2$, $c_3$ and $c_4$ are some numerical constants and
the overall factor $g^2$ is omitted.
In the commutative $\phi^4$ theory one subtracts the divergences from the
subgraphs
1-2, 1-3 and 2-3  by a single counter term graph in figure \ref{counter}(a)
which comes with the factor 3:
\be
- 3 \LB c_5 {m^2\ov \e^2}+ c_6~ {m^2 \ov \e}+ {c_3\ov 3}~ {m^2\ov \e}
{\rm ln}{m^2\ov \mu^2}\RB
\label{sunsetdiv}
\ee
where $c_5$ and $c_6$ are some  numerical constants.
The remaining divergence 
\be
(c_1-3c_5)~ {m^2\ov \epsilon^2}+ (c_2-3c_6)~{m^2\ov \epsilon}+
 c_4~{k^2\ov \epsilon}
\label{phiaref}
\ee
can be absorbed in the 2-loop wave function and mass renormalization.

The story is different in
 the non-commutative $\phi^4$ theory. There are only two counter-term
graphs figure \ref{counter}(a)
each coming with the factor one. Instead of \eq{phiaref} one has
\be
(c_1-2c_5) ~ {m^2\ov \epsilon^2}+ (c_2-2c_6)~{m^2\ov \epsilon}+
{c_3\ov 3}~ {m^2\ov \e}
{\rm ln}{m^2\ov \mu^2}+ c_4~{k^2\ov \epsilon}
\ee
But the extra divergence is independent of $k$ and can be absorbed
into the renormalization of mass. 
Thus our puzzle is resolved.

In addition to the graphs of type $\tmega$, there are two other
types of graphs: ${\rm{\bf Com}}$ (commutant) 
and ${\rm{\bf Rings}}$. The definition
of ${\rm{\bf Com}}$  follows.

\noindent
{\bf Definition 6.}~{\it  A Feynman graph in a NQFT is from the class
$\tCom$ if it contains subgraphs with $\omega \ge 0$ which wrap
homologically trivial, but topologically non-trivial cycles 
(e.g. cycle
C in figure \ref{genushomt}).}

\begin{figure}[hbtp]
\begin{center}
\mbox{\epsfxsize=14truecm
\epsffile{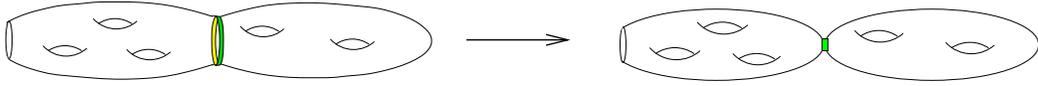}}
\end{center}
\caption{Shrinking (=UV diverging) commutant.}
\label{tadpole}
\end{figure}

\begin{figure}[hbtp]
\begin{center}
\mbox{\epsfxsize=8truecm
\epsffile{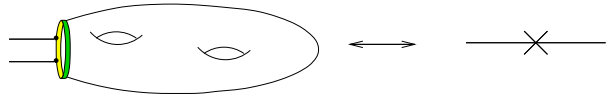}}
\end{center}
\caption{Counter-term}
\label{tadcount}
\end{figure}

\begin{figure}[hbtp]
\begin{center}
\mbox{\epsfxsize=11truecm
\epsffile{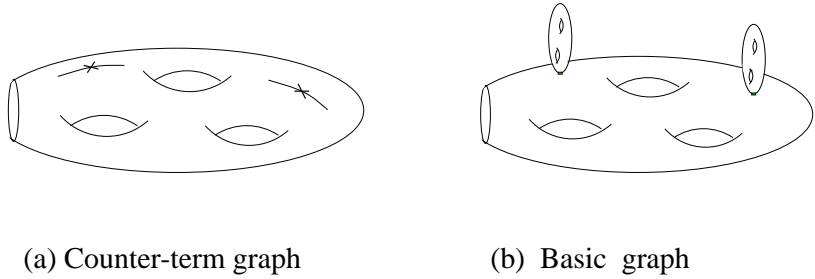}}
\end{center}
\caption{Counter-term and basic graphs.}
\label{gencounter}
\end{figure}

Consider a subgraph $\gamma$ which wraps
 the cycle $C$ in figure \ref{genushomt}(b).
In $\phi^4$ theory in $d=4$ we have $\omega(G)=4-E$ for any 1PI graph
$G$ with $E$
 external lines. We are interested in the subgraphs with
$\omega \ge 0$.
Thus for the graph in figure \ref{genushomt}(b)
to be 1PI, the subgraph $\gamma$
 has to be of the
form shown in figure \ref{generic}. In the $\phi^4$ NQFT the graph
in figure \ref{generic} will be overall convergent as long as 
$p_1+p_2\ne 0$. But if the graph \ref{generic} wraps the $C$ cycle
in figure \ref{genushomt}(b), the momentum conservation will
enforce $p_1+p_2=0$. Define $k=p_1=-p_2$.  
 Figure \ref{tadpole} illustrates the shrinking (=UV divergence)
of the subgraph wrapping the commutant. From the figure \ref{tadpole}
it is easy to see that the corresponding
divergent piece of the  
Feynman integral will be independent of the momentum $k$.
Thus we can absorb the divergence into the mass renormalization
as  in  figure  \ref{tadcount}. The  basic  and  counter-term 
 graphs
are  illustrated in figure \ref{gencounter}. 
The details of this 
procedure will be discussed in section 4.4 on the example of
 $\phi^6$ 
NQFT in $d=2$.

\begin{figure}[hbtp]
\begin{center}
\mbox{\epsfxsize=5truecm
\epsffile{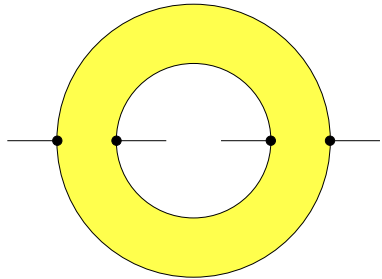}}
\end{center}
\caption{Two-two ring.}
\label{twotwoh}
\end{figure}

\begin{figure}[hbtp]
\begin{center}
\mbox{\epsfxsize=10truecm
\epsffile{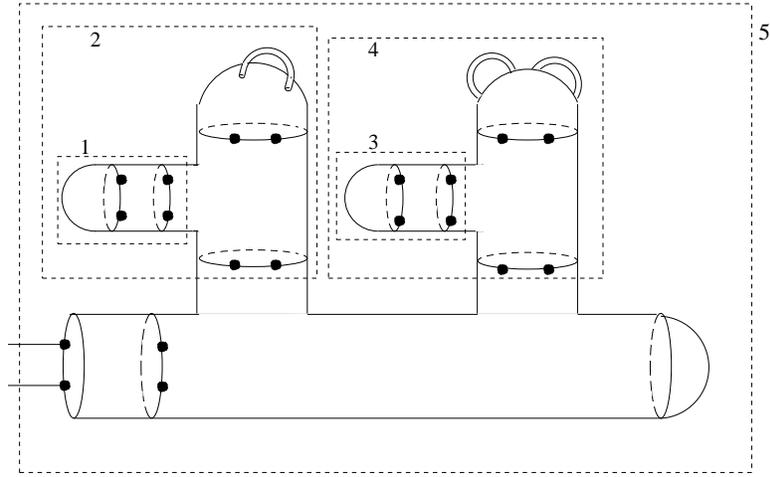}}
\end{center}
\caption{Nested and disjoint boxes.}
\label{submarine}
\end{figure}

\noindent
{\bf Remark.} Let us define the two-two rings in $\phi^4$ NQFT 
to be the 1PI graphs of the type ``sphere with two holes'' 
with two external lines on one hole and two on the other hole (see
figure \ref{twotwoh}). 
It  is  not  difficult to see that 
the  boxes  which enclose the subgraphs containing the two-two rings
are  either disjoint or  nested (see figure \ref{submarine})\footnote
{More precisely, the boxes enclose the two-two ring and the ``dead end''
part of the graph as can be seen from the figure \ref{submarine}.}. 
For  example, the subtraction of the graph 
in  figure \ref{submarine} goes as follows. All
 planar
divergent subgraphs are  subtracted  first 
using the recursive formula of ref.\cite{we}.
 Since 
$${\rm box}~1~\subset {\rm box}~2~\subset {\rm box}~5~~~~~~~{\rm and}
~~~~~~~{\rm box}~3~\subset {\rm box}~4~\subset {\rm box}~5,
$$
and  box 2  and  box  4  do  not  overlap,  box 1 and box 3 are subtracted
first  and  then box 2 and box 4, and finally box 5.

Note that in $\phi^4$ theory 1PI graphs with  more than four
external lines
have $\omega < 0$. This means that spheres with two holes and six external
lines (four lines on one boundary and two lines on the other boundary) are
overall convergent. Thus the counter-terms for the graphs of the type
in figure \ref{tadcount} but with four external lines are not needed. 
We thus conclude that $\delta g$ is independent of $\theta$.

From the above discussion it follows that the $\beta$-function 
\be
\beta = \mu {dg(\mu)\ov d\mu}
\ee
will be independent of $\theta$
whereas the   renormalization   group   coefficient 
\footnote{In the commutative QFT
in the minimal subtraction scheme,
$\beta$ and $\gamma_m$ depend only on $g$.}
\be
\gamma_m = - m^{-2} \mu {d m^2(\mu)\ov d\mu}
\ee
will be a non-trivial function of $\theta$.

\begin{figure}[hbtp]
\begin{center}
\mbox{\epsfxsize=3truecm
\epsffile{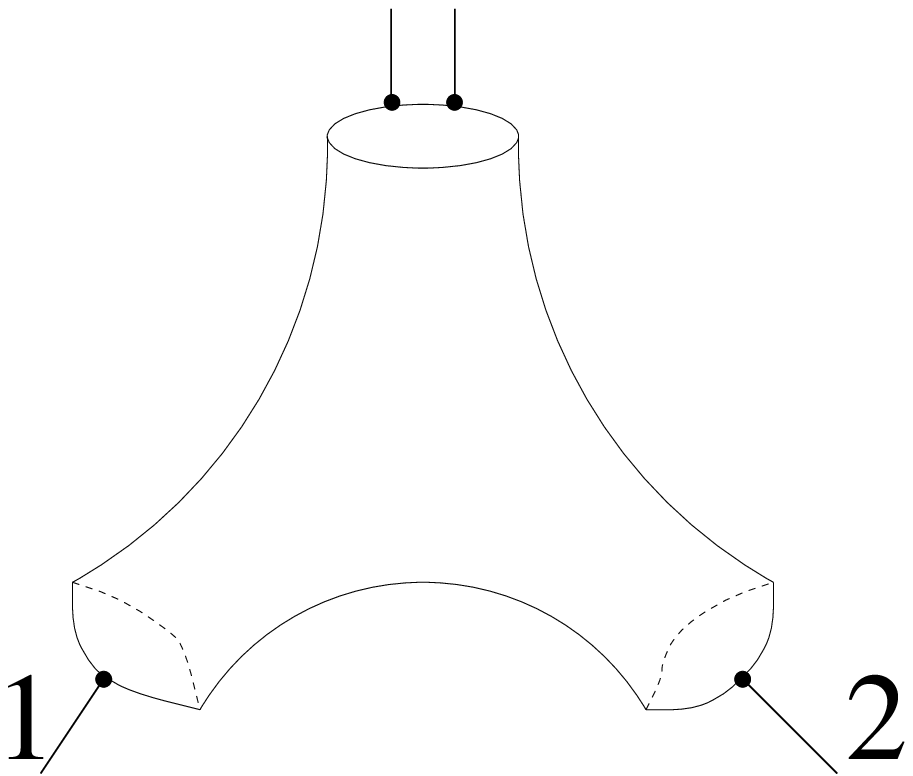}}
\end{center}
\caption{Pants}
\label{pants}
\end{figure}

\begin{figure}[hbtp]
\begin{center}
\mbox{\epsfxsize=7truecm
\epsffile{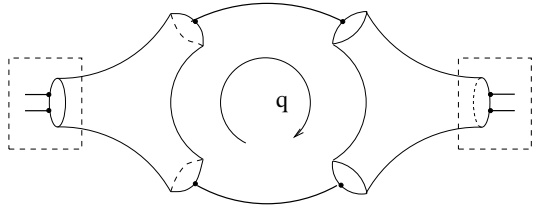}}
\end{center}
\caption{A pair of pants}
\label{twopant}
\end{figure}

We have seen that spheres with two holes are overall convergent
at non-exceptional external momenta, but can  diverge inside
a bigger graph. 
One may wonder if the same is true for 
the graphs of the type ``sphere with $n>2$ holes''. Consider
the sphere with three  holes (``pants'') and  four  external  lines shown in
figure \ref{pants}. If it is part of a bigger 1PI graph $G$ and the
external lines $1$ and $2$ are internal lines of $G$, then  it
has to be joined to the rest of the diagram as in figure \ref{twopant}.
But then there will be virtual momentum $q$ circling as in figure
\ref{twopant}. The dotted boxes in figure \ref{twopant} indicate the
subgraphs of the type sphere with two holes. Thus in an obvious 
sense the case of ``spheres with $n>2$ holes'' reduces to that of
``spheres with two holes''.

We have been discussing the spheres with two holes which wrap
commutants of a very specific type, namely, those which ``separate the
handles'' (see, for example,
 the commutant $C$ in figure \ref{genushomt}(b)).
There are actually some other types of commutants as, for example,
 in figure
\ref{comm1212}. Should this type of commutants
be considered in the divergence analysis? The answer is no.
For the theories of interest,
$\omega$ for such a subgraph $\gamma$ will be negative because  
$\gamma$ will have
too many external lines (see figure \ref{commutant1212}(b)).

We now define a new class of graphs.

\noindent
{\bf Definition 7.}~{\it A Feynman graph in a NQFT is from the class
$\tRings$ if it contains subgraphs which wrap a homologically non-trivial
cycle and satisfy the condition $\omega(\gamma)- d <0$, but the disjoint
union of these subgraphs violate the condition $\omega(\gamma)- d <0$.}

\begin{figure}[hbtp]
\begin{center}
\mbox{\epsfxsize=10truecm
\epsffile{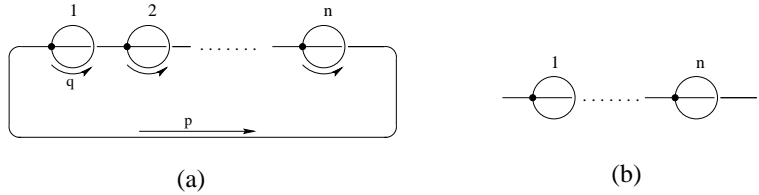}}
\end{center}
\caption{A chain of rings in $\phi^4$ theory.}
\label{phi4}
\end{figure}

\begin{figure}[hbtp]
\begin{center}
\mbox{\epsfxsize=8truecm
\epsffile{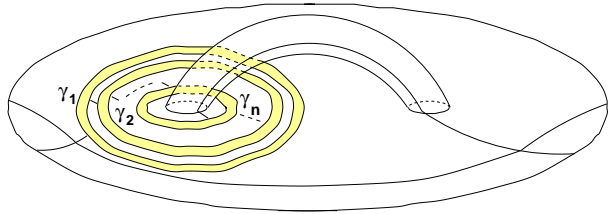}}
\end{center}
\caption{A diagram from the class $\tRings$ in $\phi^4$, $d=4$ theory.}
\label{div2}
\end{figure}

\noindent
{\bf Remark.} A particular example of  graph from $\tRings$ is shown
in figure \ref{phi4}(a). Ref.\cite{mrs} first noted that the latter graph
for $n\ge 2$
is divergent in $d=4$ by the explicit calculation. The Feynman multiple
integral $\int d^4q d^4p \cdots$ is divergent. If the integration over
$q$ is performed first, then the integral  $\int d^4p$ diverges at $p=0$
(IR divergence). If the integration over $p$  is performed first, then
the integral  $\int d^4q$ diverges at $q=\infty$ (UV divergence). 
The general graph in figure
\ref{div2} is from the class  $\tRings$.

Now let us ask the following question. Let $G$ be a 1PI graph  in 
$\phi^4$ NQFT in $d=4$. Is it possible to have a 1PI subgraph
$\gamma\subset G$  with $c_G(\gamma)\ge 1$ which would violate
the convergence condition $\omega(\gamma)-c_G(\gamma) d < 0$ ?

The answer is no. The reason is simple: for any 1PI graph which is not a 
vacuum  bubble
in this theory we 
have  $\omega(\gamma) < d=4$.

We  thus  conclude  that  any  subgraph $\gamma$  with  
$c_G(\gamma)\ge 1$  which  violates  the 
condition $\omega(\gamma)-c_G(\gamma) d < 0$ has to be a disjoint
union of 1PI graphs with $c\ge 1$. This is true for any theory
in which $\omega(G) < d$ for any 1PI graph $G$ with external lines. 
We now show that only $c=1$ disjoint graphs can violate the convergence
condition. This follows from the following obvious  statement:

\noindent
 If 

\noindent
(1) $\gamma_1$ and $\gamma_2$ are
1PI subgraphs of $G$, and

\noindent
(2)  $\gamma_1$ and $\gamma_2$ wrap the same cycles, and 
$c_G(\gamma_1)=c_G(\gamma_2)>1$, 

\noindent
then $\gamma_1 \cup \gamma_2$ is 1PI. Thus $\gamma_1 \cup \gamma_2$
is connected.

\begin{figure}[hbtp]
\begin{center}
\mbox{\epsfxsize=10truecm
\epsffile{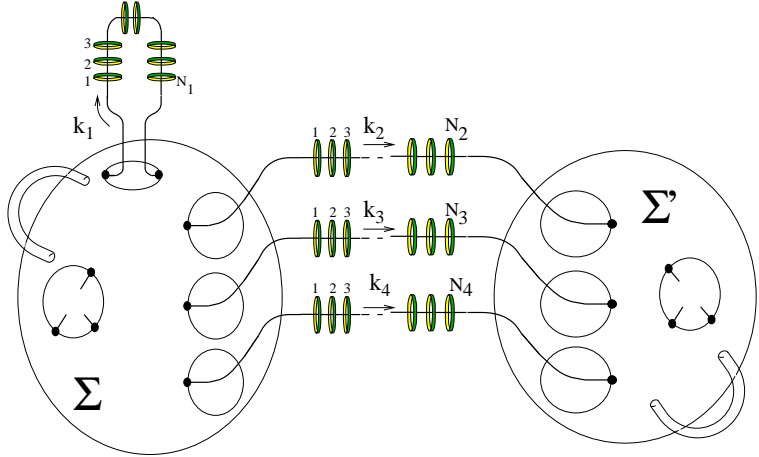}}
\end{center}
\caption{A generic structure of $\tRings$ graphs.}
\label{last}
\end{figure}

Figure \ref{last} illustrates a generic structure of 1PI 
$\tRings$ graphs
in 
  $\phi^4$  NQFT.  Note that the rings do not overlap\footnote{
A ring has only two external vertices which are reserved for
the joining   to the rest of the rings wrapping a given cycle.
Thus a ring wrapping a given cycle cannot overlap a ring wrapping
a different cycle. Hence the structure shown in figure \ref{last}.}.
The sub-surfaces $\Sigma$ and $\Sigma'$ do not contain any ring.
In  general,  there  will  be  $n$  sub-surfaces joined together in
all  possible  ways  by  the  ``tubes''  containing  rings.

Ref.\cite{mrs} suggests the following procedure to deal
with the divergence of the graph \ref{phi4}(a) with $n\ge 2$. The connected
diagram shown in figure  \ref{phi4}(b) is convergent for the non-zero 
external momentum $k$.  It is divergent at $k=0$.
A power counting
argument  shows that for sufficiently small $k$ the $n=1$ approximation
is not adequate and $n\ge 2$ diagrams become important. Thus
one has to sum over  $n$.  The resulting integral over $k$ 
is convergent at
$k=0$ (see ref.\cite{mrs} for the details).

\begin{figure}[hbtp]
\begin{center}
\mbox{\epsfxsize=13truecm
\epsffile{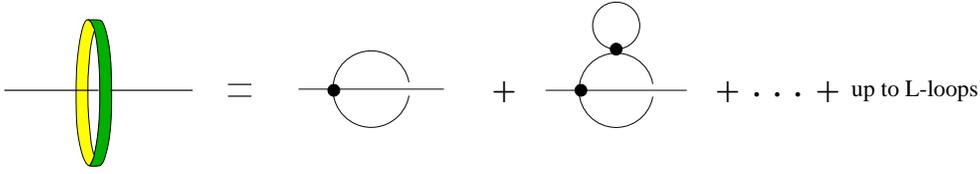}}
\end{center}
\caption{A sum of 1PI rings.}
\label{last1}
\end{figure}

Now let us denote by $\Pi_L (k)$ the sum of contributions
(after the subtraction  of the planar subdivergences) of 
up to $L$-loop 1PI diagrams
on the r.h.s. of the equation in figure \ref{last1}. 
It is not difficult
to see that  the  loop  corrections  within  an
individual  ring  are  not  important  at  small  $g$.
Thus in computing a Green's function in perturbation theory
for a given accuracy, it is enough to have $L < L_{max}$ for
some given $L_{max}$. The surfaces $\Sigma$ and $\Sigma'$ in 
figure \ref{last} also have the number of loops bounded from above
by $L_{max}$.

Now consider the graph in figure \ref{last}. The non-overlapping 
structure of the rings allows  to re-group the set of graphs as in
figure \ref{last}, where the ring is given by figure \ref{last1}. 
Denoting by $k_1,\ldots,k_4$ the momenta flowing along the cycles
to which the rings are attached, we have, after the summation 
$
\sum_{N_1,N_2,N_3,N_4=1}^{\infty}
$,
the product
\be
{\Pi_{L}(k_1)\ov (k_1^2+m^2) (k_1^2+m^2 - \Pi_{L}(k_1))}\cdots
{\Pi_{L}(k_4)\ov
(k_4^2+m^2)(k_4^2+m^2 - \Pi_{L}(k_4))}
\label{Piprod}
\ee
In the limit $k\rightarrow 0$ we have  
$\Pi_L(k)\rightarrow \infty$ and thus \eq{Piprod} reduces to
\be
{1\ov (k_1^2+m^2)(k_2^2+m^2)(k_3^2+m^2)(k_4^2+m^2)
}
\label{Piprodze}
\ee
Suppose that the sub-surfaces $\Sigma$ and $\Sigma'$ in figure \ref{last}
are convergent. Since $\Sigma$ and $\Sigma'$ do not contain rings, 
this can always be achieved by the subtraction procedure discussed 
earlier.
Now  we  join  $\Sigma$  and  $\Sigma'$  using  the 
resummed  rings. Since $\Pi_{L}(k)$ is exponentially vanishing
at $k\rightarrow \infty$, the integrals 
\be
\int d^4k_1 \int d^4k_2 \int d^4k_3 \int d^4k_4 \cdots
\label{kkkk}
\ee
are convergent at $k_i\rightarrow\infty$.

Thus  only  the 
 convergence at $k=0$ is questionable. From \eq{Piprodze}
it  is  clear that the resummed rings are harmless at $k\rightarrow 0$. 
Thus the only
possible divergence may come from the surfaces $\Sigma$ and $\Sigma'$.
Recall that $\Sigma$ and $\Sigma'$ may diverge at the exceptional
external momenta. When we join  $\Sigma$ and $\Sigma'$ using the 
resummed rings, we necessarily make some of the external momenta
(which actually become internal lines in the resulting graph)
exceptional. But these momenta are integrated over (we are talking
about the momenta $k_1$,...,$k_4$ in figure \ref{last}). Since
$\Sigma$ and $\Sigma'$ do not contains rings, we conclude that the
integral \eq{kkkk} is convergent at $k_i\rightarrow 0$. Thus
at least at
 a heuristic level it is clear that the divergences
of  $\tRings$  graphs can be dealt with.

\subsection{Wess-Zumino model}
In ref.\cite{we} we suggested that the non-commutative version
of the Wess-Zumino model (in $d=4$) is  renormalizable  to  all  orders
in the
perturbation theory. The argument given in ref.\cite{we}
is short and precise: 
$\omega \le 0$ for 1PI supergraphs. Thus there are no $\tRings$ 
graphs\footnote{
Refs. \cite{mrs, susskind, arefeva3, gomes} subsequently
studied one-loop diagrams in supersymmetric NQFT
and found 
that IR poles typical of NQFT are  absent (more precisely, there are
only logarithmic singularities).}.

The argument given in ref.\cite{we} is 
not complete  because

\noindent
(1)
The subtraction procedure in
BPHZ scheme necessarily introduces
 the propagators of the form $P(k)/(k^2+m^2)$,
where $P(k)$ is a polynomial in $k$.  
The convergence theorem for the NQFT 
with the propagators of this form was neither  proven nor argued to be
true in ref.\cite{we}, but it was used in the argument for the
renormalizability of Wess-Zumino model.

\noindent
(2)
The convergence theorem was formulated for the ordinary graphs,
but used for the supergraphs.

\noindent
(3)
The  $\tRings$ graphs were shown to be absent simply because
there are at most the logarithmic divergences.\footnote{This can most
easily be seen in the superspace language.}.  
But it was not shown that  $\tCom$ graphs are absent.

Point (1)   is dealt with  in   section 7.
Point (2) can be dealt with as follows. The star product in the
Wess-Zumino action does not depend on the Grassmann variables $\theta$
and so the super Feynman graphs are the ribbon graphs. As in the
commutative case, there will be $\theta$'s associated with each vertex.
In the language of the Feynman integrals this means that only
the phase factors will carry the information about the topology
of the graph and these phase factors are the same as in non-commutative
$\phi^3$ theory. Apart from the phase factors, the rest of the Feynman
integrand is identical to that of the commutative Wess-Zumino model.
For further discussions of the superfield formalism in
SUSY NQFT see ref.\cite{susy}.

The fact that there are no  1PI  $\tCom$ graphs can  easily
be seen as follows. Let us look at the genus two graph shown in
figure \ref{genushomt}(b). 
In order for the graph to be 1PI, any subgraph which wraps the $C$ cycle
should have at least two external legs on each hole as in
the figure \ref{generic}. But such graphs have $\omega<0$. Thus
there  are  no  $\tCom$  graphs. Note that  by
a  graph  we  mean  a  supergraph.

\begin{figure}[hbtp]
\begin{center}
\mbox{\epsfxsize=10truecm
\epsffile{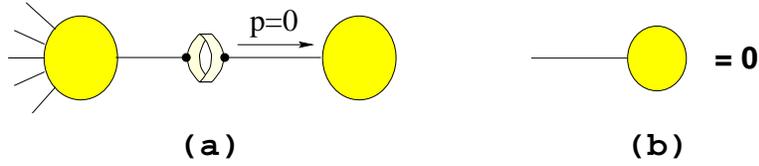}}
\end{center}
\caption{The supergraphs which are identically zero in Wess-Zumino model.}
\label{impossible}
\end{figure}

Thus the divergences from all 1PI supergraphs 
(with the non-exceptional external momenta)
in the Wess-Zumino model
can be removed. What about the non-1PI connected graphs? In principle,
the non-1PI graphs of the type shown in figure \ref{impossible}(a)
are problematic. But in this theory  the graphs with one external
line are identically zero. Thus 
the general discussion of this section suggests that the Wess-Zumino
model should be renormalizable to all loop 
orders.\footnote{A different approach  for the renormalization of
non-commutative Wess-Zumino model was suggested in ref.\cite{gomes}.}

\subsection{$\phi^4$ in $d=2$}
The corresponding commutative theory  is super-renormalizable.
The superficial degree of divergence of a 1PI 
graph with $V$ vertices in this
theory is
$\omega=2-2V$. So $\omega$ is at most 0.
There are no 1PI  $\tRings$ and $\tCom$ graphs.
Thus the divergences from  all 1PI graphs can be removed by the
introduction of
the planar counter-terms in the Lagrangian.

\begin{figure}[hbtp]
\begin{center}
\mbox{\epsfxsize=6truecm
\epsffile{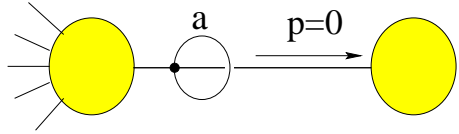}}
\end{center}
\caption{A divergent non-1PI graph.}
\label{connected}
\end{figure}

At first sight
the situation seems to be  different with the non-1PI graphs.
There seems to be a  specific type of 
non-1PI connected graphs which  contain the nonplanar divergences.
The graph in figure \ref{connected}  contains a non-planar divergent
subgraph a. 
The momentum flowing across this subgraph is zero by the 
momentum
conservation.  
But the 
graphs of the type shown in figure \ref{connected} are 
impossible in $\phi^4$ theory because there are no graphs with one
external leg in this theory. The latter statement follows from the
$\phi \rightarrow -\phi$ symmetry of the action or, equivalently,
from the relation $E+2I=4V$.
Thus  $\phi^4$ theory in $d=2$ is 
renormalizable to all orders in the pertubation theory. 
By this we mean that
the divergences from all connected Green's functions at non-exceptional
external momenta can be removed in the counter-term approach.

\subsection{$\phi^6$  in  $d=2$}
The quantum corrections  to  $\phi^6$  action  will  induce $\phi^4$ terms.
Thus we consider the combined $\phi^6$ plus $\phi^4$ 
 action in $d=2-2\epsilon$:
\bea
S&=&\int d^{2-2\epsilon}x~[ {1\ov 2} (\partial_{\mu}\phi)^2+{1\ov 2}(m^2
+ \delta m^2)
\phi^2 
+{1\ov 4} (g_4+\delta g_4)~\mu^{2\epsilon}~ 
( \phi\star\phi\star\phi\star\phi) \non
&& + {1\ov 6}
g_6~\mu^{4\epsilon}~( \phi\star\phi\star\phi\star\phi
\star\phi\star\phi)]
\label{phi6star}
\eea

\begin{figure}[hbtp]
\begin{center}
\mbox{\epsfxsize=11truecm
\epsffile{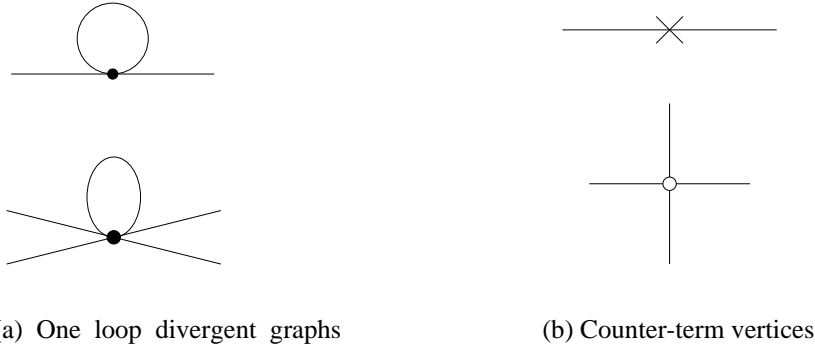}}
\end{center}
\caption{One loop divergent graphs in $\phi^6$ theory in $d=2$. }
\label{basicphi6}
\end{figure}

For  convenience let us define
\bea
\mu^{2\epsilon}~\int {d^{2-2\epsilon}q\ov (2\pi)^{2-2\epsilon}}
{1\ov (q^2+m^2)}&=&
{1\ov 4\pi}~[{1\ov\e}+\psi(1)-{\rm ln}{m^2\ov 4\pi \mu^2}
+{\cal O}(\e)] \non
&\equiv& {c_0\ov \e}+ c_1 + \e ~ c_2 +\cdots
\eea
We will show that the theory with the action \eq{phi6star} is finite
in the $\e \rightarrow 0$ limit,
provided that we choose  appropriate  $\delta m^2$ and $\delta g_4$:
\be
\delta g_4 \equiv - {4c_0g_6\ov \e}
\ee
and
\bea
\delta m^2&=&-{2c_0g_4\ov \e}+{3c_0^2 g_6\ov \e^2}-{2c_0 c_1 g_6\ov \e}
\non
&&+{2c_0 g_4 g_6 m^{-2} \ov \e}  f(\theta m^2) + (n>3)~{\rm loop~corrections}
\label{deltam2}
\eea
where $f$ is given by
\be
{1\ov m^2}~f(\theta m^2)\equiv 
 \int {d^2q\ov (2\pi)^2}{d^2p\ov (2\pi)^2} {{\rm exp}(i p\wedge q)
\ov (p^2+m^2)^2 (q^2+m^2)}
\label{phi6f}
\ee
Note that $\delta m^2$ receives contributions from all loop orders.

\begin{figure}[hbtp]
\begin{center}
\mbox{\epsfxsize=11truecm
\epsffile{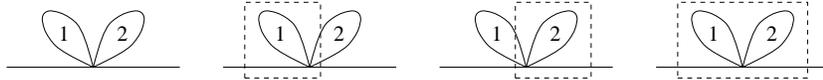}}
\end{center}
\caption{Subtraction in the commutative theory.}
\label{subtract}
\end{figure}

The one loop divergent graphs and the corresponding
mass and coupling constant  vertices 
are listed in figure \ref{basicphi6}. The one loop mass and coupling
constant corrections are
\be
\delta m^2=-{2c_0g_4\ov \e}
\ee
and
\be
\delta g_4 = - {4c_0g_6\ov \e}
\ee
The one loop coupling constant counter-term $\delta g_4$
does not receive corrections from the higher loops.

\begin{figure}[hbtp]
\begin{center}
\mbox{\epsfxsize=11truecm
\epsffile{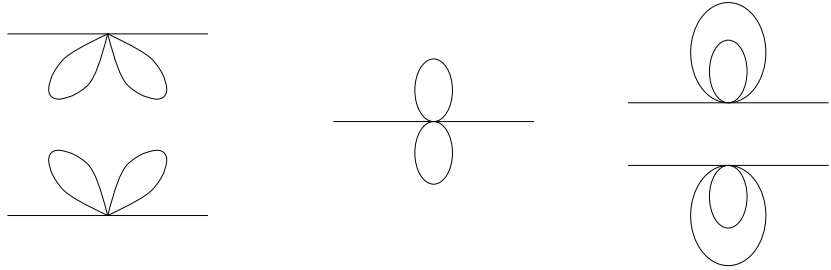}}
\end{center}
\caption{Diagrams contributing to the 2-loop mass renormalization.}
\label{bunch}
\end{figure}

\begin{figure}[hbtp]
\begin{center}
\mbox{\epsfxsize=7truecm
\epsffile{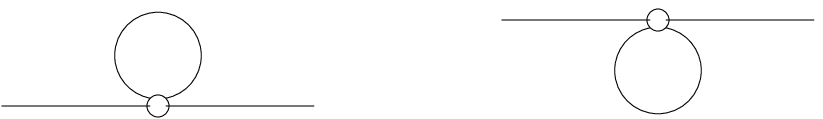}}
\end{center}
\caption{Counter-term graphs.}
\label{phi6new}
\end{figure}

\begin{figure}[hbtp]
\begin{center}
\mbox{\epsfxsize=11truecm
\epsffile{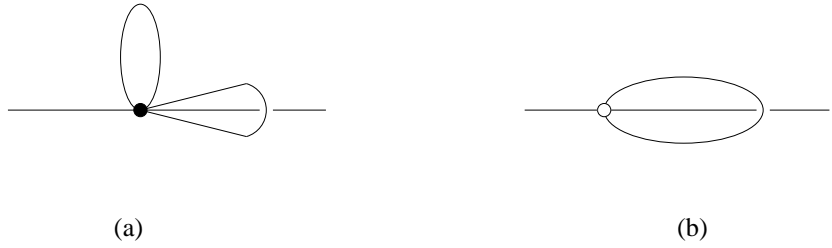}}
\end{center}
\caption{Basic and counter-term graph. }
\label{phi6hook}
\end{figure}

Consider the two loop graph in figure \ref{subtract}. In the commutative
theory it comes with the factor of $15$. The subtractions are
done as in figure \ref{subtract}. In the non-commutative theory there
are $15$ different diagrams each with the factor one. Some
of these
graphs are shown in figures \ref{bunch} and \ref{phi6hook}(a). The 
counter-term graph for the graph \ref{phi6hook}(a) is shown in
figure \ref{phi6hook}(b).  The graphs in figure \ref{bunch} are the only
graphs which contribute to two loop mass counter-term. Two of the 
graphs in figure \ref{bunch} contain divergent non-planar one loop
subgraph
of the type ``sphere with two holes''. As explained in section 4.1., the
corresponding divergence can be
 absorbed in the two loop mass counter-term.  The sum of
 the contributions
from figures \ref{bunch} and \ref{phi6new} reads
$$
-5 g_6 ({c_0\ov \e}+c_1+\cdots)^2 - 2 \delta g_4 ({c_0\ov \e}+c_1+\cdots)
$$
\be
={3c_0^2 g_6\ov \e^2}-{2c_0 c_1 g_6\ov \e}+{\rm finite}
\ee
Thus the mass counter-term up to two loops reads
\be
\delta m^2=-{2c_0g_4\ov \e}+{3c_0^2 g_6\ov \e^2}-{2c_0 c_1 g_6\ov \e}
\ee

\begin{figure}[hbtp]
\begin{center}
\mbox{\epsfxsize=11truecm
\epsffile{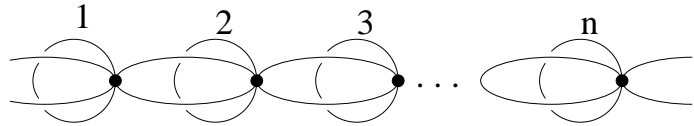}}
\end{center}
\caption{A chain of rings in $\phi^6$ theory.}
\label{phi6}
\end{figure}

In this theory there are no 1PI  $\tRings$ graphs since $\omega=2-2V$
is at most zero. 
But there are  1PI  $\tCom$  graphs. 
In order for  a graph to wrap a commutant ( see cycle C in figure
\ref{genushomt} ), it should be ``a sphere with two holes''.
It is easy
to see that the only such graphs (with four external lines)
which are divergent at the vanishing
total momentum flowing into each hole are the ones
shown in figure \ref{phi6}.

\begin{figure}[hbtp]
\begin{center}
\mbox{\epsfxsize=6truecm
\epsffile{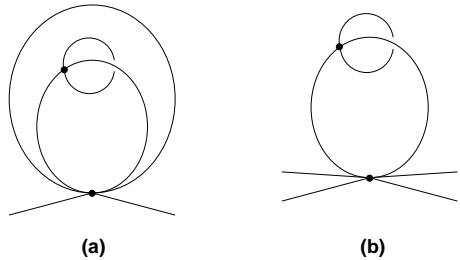}}
\end{center}
\caption{The first non-trivial diagram in $\phi^6$ NQFT in $d=2$.}
\label{phi6mass}
\end{figure}

The first non-trivial $\theta$ dependent contribution to $\delta m^2$
comes from the 3-loop graph in figure \ref{phi6mass}(a). The Feynman
integral for the subgraph in  figure \ref{phi6mass}(b) is given by 
\eq{phi6f}. Thus $\delta m^2$ is given up to 3-loops by
\eq{deltam2}.

\begin{figure}[hbtp]
\begin{center}
\mbox{\epsfxsize=11truecm
\epsffile{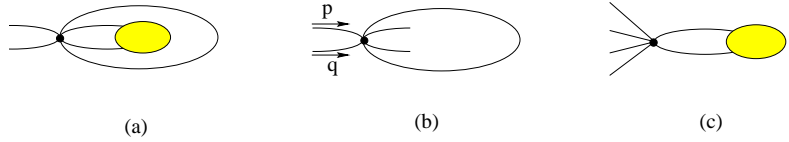}}
\end{center}
\caption{The graphs which contain non-planar subgraphs.}
\label{phi6plan}
\end{figure}

Now consider the graph  in figure \ref{phi6plan}(a).
It  contains a divergent non-planar subgraph \ref{phi6plan}(b). It is 
easy to see that the divergence is independent of the external momentum.
As in section 4.1 the divergence of the graph \ref{phi6plan}(a)
is absorbed in the mass renormalization. If the Feynman integral corresponding
to the subgraph shown in figure  \ref{phi6plan}(c) is convergent, 
the contribution of figure \ref{phi6plan}(a) to 
the mass counter-term will be 
\be 
{1\ov \epsilon} g_4^{V_4} g_6^{V_6} m^{\omega(G)} F(\theta m^2)
\ee
for a non-trivial function $F$ if the graph \ref{phi6plan}(c) is non-planar,
where $V_4$ and $V_6$ are the number of $\phi^4$ and $\phi^6$ vertices, 
respectively.

\noindent
{\bf Remark.} The analysis of this subsection can readily be extended
to $\phi^{2n}$ NQFT in $d=2$. 
It  is  not  difficult  to  see  that 
the  counter-terms  $\delta  m^2$, 
$\delta  g_4$,  $\delta g_6$,\ldots, $\delta g_{2n-4}$,
 will
depend non-trivially on $\theta$.

\subsection{$\phi^* \star \phi \star \phi^* \star \phi$  in $d=4$}
This theory is interesting because $\tRings$  graphs are absent
for a purely graph-theoretic reason. This should be contrasted with
what happens in supersymmetric non-commutative theories in general
and 
non-commutative Wess-Zumino model in particular.
In SUSY theories $\tRings$  graphs are absent because $\omega \le 0$
for the supergraphs.

\begin{figure}[hbtp]
\begin{center}
\mbox{\epsfxsize=8truecm
\epsffile{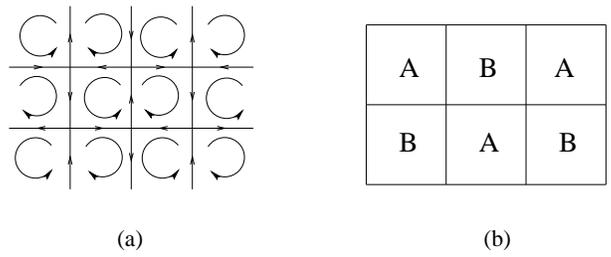}}
\end{center}
\caption{A graph in $\phi^* \star \phi \star \phi^* \star \phi$  theory and
two color problem.}
\label{twocolor}
\end{figure}

A general 1PI graph in this theory can be drawn on a genus $g$ surface
with a  number of holes. The only
difference from a graph in the real $\phi^4$ theory is that the lines
in  the  complex  $\phi^4$  theory carry arrows whereas the lines in the real
$\phi^4$ theory do not carry them.
In other words the 2-surfaces in the complex $\phi^4$ theory are
{\it decorated} whereas those in the real $\phi^4$ theory are not.
Let us consider the piece of a
 graph in 
$\phi^* \star \phi \star \phi^* \star \phi$ theory shown in 
figure \ref{twocolor}(a). Associating the clockwise direction of
the arrows with the color $A$ and the counter-clockwise direction
with the color $B$, we can equivalently draw the graph as in 
figure \ref{twocolor}(b). 
Let $G$ be a  graph in the real $\phi^4$ NQFT and 
$\Sigma(G)$ the associated surface. Then it is easy
to see that $G$ is also a graph in the complex $\phi^4$ theory
if and only if the plaquettes of $\Sigma(G)$ can be painted in two
colors as in figure \ref{twocolor}(b).
According to a well-known coloring theorem, any planar graph in
the real $\phi^4$ NQFT admits coloring and thus it is a graph
of the complex $\phi^4$ theory.

\begin{figure}[hbtp]
\begin{center}
\mbox{\epsfxsize=8truecm
\epsffile{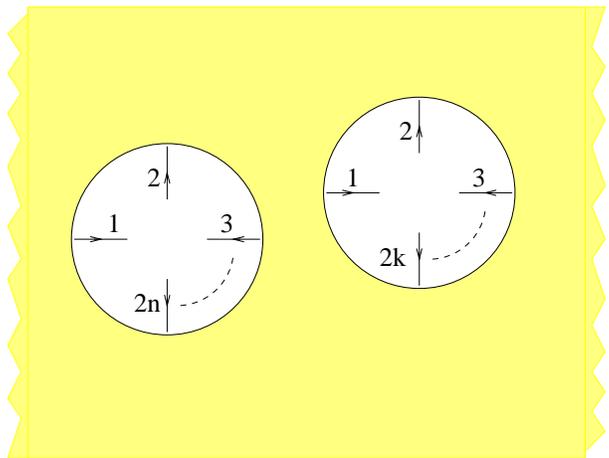}}
\end{center}
\caption{Two holes.}
\label{complexhole}
\end{figure}

\begin{figure}[hbtp]
\begin{center}
\mbox{\epsfxsize=7truecm
\epsffile{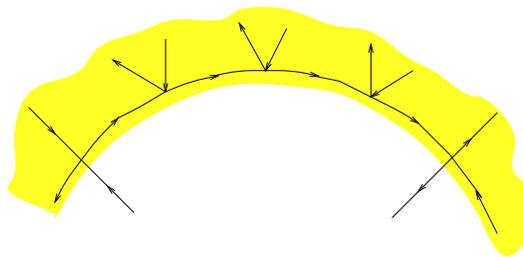}}
\end{center}
\caption{Local structure at the boundary.}
\label{complexlocal}
\end{figure}

Let  $\Sigma(G)$ 
be  the  surface  associated  with  a  graph  $G$  in 
$\phi^* \star \phi \star  \phi^*  \star  \phi$  theory. 
As in the figure \ref{complexhole} the
total  number  of   external   lines   attached   to   each  hole  of 
$\Sigma(G)$ 
is  always  an  even  number  and  the  arrows  carried  by  these
 lines  always
appear in the  alternating  order  in-out-in-out-...  .
This statement  is  evident  from  the
figure  \ref{complexlocal},  where the local structure at the
boundary  of  a  hole  is  depicted.

\begin{figure}[hbtp]
\begin{center}
\mbox{\epsfxsize=5truecm
\epsffile{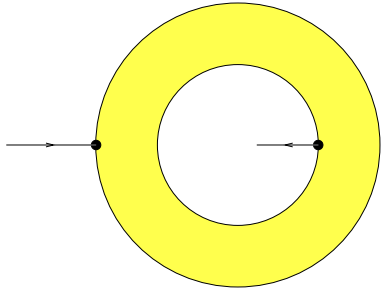}}
\end{center}
\caption{Impossible ring diagram.}
\label{ringps}
\end{figure}

This means  that  there are no 1PI graphs of the
type ``sphere with two holes'' and two external lines shown in
figure
\ref{ringps} in the $\phi^* \star \phi \star \phi^* \star \phi$  theory.
Thus there  are  no   $\tRings$ graphs.

It is also clear that the graphs of the type ``sphere with two holes''
with four external lines
shown in figure \ref{generic} are permissible in this theory. Thus
there will be  $\tCom$ graphs in this theory. There will
also be the planar graphs which contain divergent non-planar subgraphs
of the type shown in figure \ref{generic}.

\begin{figure}[hbtp]
\begin{center}
\mbox{\epsfxsize=7truecm
\epsffile{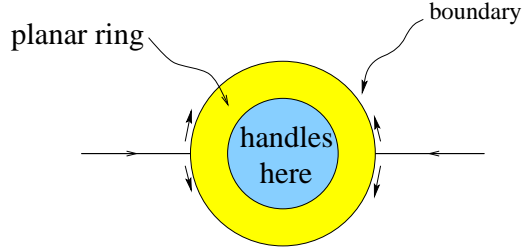}}
\end{center}
\caption{Impossible diagram.}
\label{twopoint}
\end{figure}

The external lines of a planar 1PI graph with four external lines
are necessarily in the order   $\phi^* - \phi - \phi^* - \phi$
(see figure \ref{complexhole}).
Thus the overall divergences 
from the planar  1PI graphs with four external lines can be removed
by the coupling constant counter-term 
$$
\delta g~ \phi^* \star \phi \star \phi^* \star \phi
$$

Let us now consider the counter-term of the type shown in figure
\ref{tadcount}. From the figures \ref{complexhole} and
\ref{twopoint}, it is easy
to see that $\phi^2$ and $\phi^{*2}$ terms cannot be induced
in $\phi^* \star \phi \star \phi^* \star \phi$  theory.
Thus the divergences from the diagrams of the type shown in 
\ref{tadcount} can be subtracted using mass counter-term
$\delta m^2 \phi^* \phi$.

\begin{figure}[hbtp]
\begin{center}
\mbox{\epsfxsize=7truecm
\epsffile{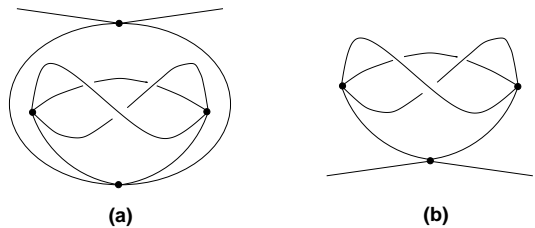}}
\end{center}
\caption{The first non-trivial diagram in $\phi^* \star \phi 
\star \phi^* \star \phi$ NQFT in $d=4$.}
\label{beatle}
\end{figure}

It can be shown diagrammatically  that
the mass counter-term 
$\delta m^2$ is independent of $\theta$ up to 3 loop order.
The first non-trivial $\theta$ dependent contribution 
to $\delta m^2$ comes from the 4 loop diagram shown in figure \ref{beatle}(a).
The Feynman integral corresponding to the graph in figure \ref{beatle}(b)
is convergent. On dimensional grounds it is of the form
$m^2 f(\theta m^2)$, where $f$ is a non-trivial function.
Thus the contribution to $\delta m^2$ is 
\be
{{\rm const}\ov \e} ~ m^2 f(\theta m^2)
\ee

The coupling constant counter-term
$\delta g$  does not depend on $\theta$ as follows from the
discussion in section 4.1.

From
the general discussion of this section it is quite clear that
$\phi^* \star \phi \star \phi^* \star \phi$  theory should
be renormalizable
to all orders in the perturbation theory.





\subsection{A comment on non-commutative 
super Yang-Mills theories (NSYM).}
The bosonic non-commutative YM theory contains 
$\tCom$ graphs. As we discussed before, in the massive theories
the divergences from this type of  graphs can readily be removed by
the mass renormalization.  
In the commutative YM theory  tadpole type graphs
vanish.  Recall that in the dimensional regularization 
$$
\int d^dp (p^2)^{\alpha}=0
$$ 
Due to the presence of the scale $\theta$ in
 the non-commutative case, the situation with the
tadpoles is  somewhat involved.
Thus $\tCom$  graphs are quite problematic in
  the non-commutative YM. 
A priori there is no reason for the cancellation of divergences
in $\tCom$ graphs in non-commutative YM.

Ref.\cite{susskind} reports that
there are quadratic divergences in non-commutative
YM. Thus there are  $\tRings$  graphs in this theory.

How about NSYM? In these theories 
$\omega \le 0$    and  therefore
$\tRings$  graphs are absent. It is easy to see that
in general
there will be $\tCom$  graphs. But it is likely
that $N=4$  NSYM
theory is an exception in this respect and there are no
 $\tCom$ graphs in this theory. The ref.\cite{susskind}
computes one-loop diagrams in $N=4$ NSYM and shows that
IR singularities are absent. At one loop IR divergences come from
the spheres with two holes. In the theories with $N<4$ SUSY, there
are logarithmic IR singularities.

The analysis of divergences given in this paper is based on
the convergence theorem for the massive NQFT. 
The convergence properties of the Feynman integrals
in   massless NQFT are much more
involved (see the first comment in section 8). 
Thus, NYM theories are expected to have rich divergence structures.

\newsection{Convergence theorem for non-derivative scalar commutative
theories}
To make the proofs given in sections 6 and 7 easier to follow,
we urge the reader to follow through  the proof of 
the theorem for commutative scalar theories  given 
in this section.

In a commutative scalar theory with non-derivative couplings  the 
parametric  representation of a Feynman integral for an arbitrary
graph $G$ reads\footnote{In the  commutative case
one usually considers \eq{comfeyn} only for 1PI graphs.
For the purpose of the proof of the convergence theorem 
in the non-commutative case,
we will find it very useful to consider arbitrary disconnected graphs. It is
easy to see that \eq{comfeyn} holds for general  disconnected graphs. 
} 
\be
I_G(p)= \int_0^{\infty} d\a_1 \cdots d\a_I  {\ex -[\sum \a_l  m_l^2+ Q(p,\a)]
\ov  P(G)^{d\ov 2} }
\label{comfeyn}
\ee
where $Q$ is a quadratic, positive definite form in the external 
momenta $p$ and a homogeneous rational fraction of degree 1 in the $\a$. Hence,
only the convergence at $\a = 0$ is questionable. The exponent being 
bounded at the origin plays no role. The polynomial $P(G)$ is a sum
of monomials of degree $L$. Let $G_i,~~i=1,\ldots,n$ be 1PI components
of the graph $G$. Then
\be
P(G)=\prod_i P(G_i)
\ee
As in ref.\cite{itz}, let us  divide  the integration domain into 
sectors
\be
0 \le \a_{\pi_1} \le \a_{\pi_2} \le \cdots \le \a_{\pi_I}
\ee
where $\pi$ is a permutation of $(1,2,\ldots,I)$. We shall prove the
convergence of the integral sector by sector. To each sector 
corresponds a family of nested subsets $\gamma_l$ of lines of $G$:
\be
\gamma_1 \subset \gamma_2 \subset \cdots \subset \gamma_l \subset
\cdots \subset \gamma_I \equiv G
\label{nested}
\ee
where $\gamma_l$ contains the lines pertaining to 
$(\a_{\pi_1},\ldots,\a_{\pi_l})$. In a given sector specified by $\pi$,
we perform a change of variables
\be
\begin{array}{ccccccccc}
\a_{\pi_1}&=&\b_1^2&\b_2^2& \cdots &\b_s^2&\cdots&\b_{I-1}^2&\b_I^2 \\
\a_{\pi_2}&=&& \b_2^2 & \cdots &\b_s^2&\cdots&\b_{I-1}^2&\b_I^2 \\
~&\vdots& &&&&&& \\
\a_{\pi_s}&=&&&&\b_s^2&\cdots&\b_{I-1}^2&\b_I^2 \\
~&\vdots& &&&&&& \\
\a_{\pi_{I-1}}&=&&&&&&\b_{I-1}^2& \b_I^2 \\
\a_{\pi_I}&=&&&&&& &\b_I^2
\end{array}
\la{permute}
\ee
the jacobian of which is
\be
{D(\a_{\pi_1},\ldots,\a_{\pi_I}) \ov D(\b_1,\ldots,\b_I)} = 2^I \b_1\b_2^3
\cdots \b_I^{2 I-1}
\label{jacob}
\ee
In these $\b$ variables the integration domain reads
\be
0 \le \b_I \le \infty~~~~{\rm and}~~~~0 \le \b_l \le 1~~~~{\rm for}~~1\le l
\le I-1
\ee
Using the well-known  graph-theoretic 
formula for $P(G)$ one may prove the following relation \cite{itz}
\be
P(G) = \b_1^{2L(\gamma_1)} \b_2^{2L(\gamma_2)}\cdots
\b_I^{2L(\gamma_I)} [1+{\cal O}(\b^2)]
\label{mrel}
\ee
The integrand is majorized by the factor
\be
{ \b_1\b_2^3
\cdots \b_I^{2 I-1} \ov \b_1^{d L(\gamma_1)} \b_2^{d L(\gamma_2)}
\cdots \b_I^{d L(\gamma_I)}}
=\prod_{l=1}^I \b_l^{-d L(\gamma_l)+2l-1} = \prod_{l=1}^I \b_l^{-\omega_l-1}
\ee
and since $\omega_l < 0$, the integral
$
\int_0 \prod d\b_l  \b_l^{-\omega_l-1}
$
is convergent at the origin.

For the purpose of generalization to non-commutative theories,
we will give an alternative  proof of the relation \eq{mrel} which does 
not require
the use of the graph-theoretic formula. The only properties of $P(G)$ which we 
will need
for the proof are

\noindent
$\bullet$
$P(G)$ is a homogeneous polynomial in $\alpha$ of degree $L(G)$

\noindent
$\bullet$
\be
P(G) = \left\{
\begin{array}{cl}
P(G-l) & {\rm ~if~}L(G-l)=L(G)\\
\a_l P(G-l) + X &{\rm ~if~}L(G-l)=L(G)-1
\end{array} \right.
\label{pofg}
\ee
where $G-l$ is the graph $G$ with the line $l$ removed, 
$X$  is a sum of monomials of 
degree $L$ in $\a$ and it does not depend on $\a_l$.
For a 1PI graph $G$, one can  have only the case $L(G-l)=L(G)-1$ in \eq{pofg}.

It is easy to check that  \eq{mrel}  
holds for
all one loop graphs. Assuming that \eq{mrel} holds for 
all of the graphs which have at most 
$L$ loops, let us prove that \eq{mrel} holds for all $L+1$ loop 
graphs.
Let $G$ be a general $L+1$ loop graph with $I$ lines.
Let us perform a change of
variables \eq{permute}  in $P(G)$.

Let $n \ge 1$ be an integer such that the removal of any one of the lines
$\pi_I, \pi_{I-1},
\ldots, \pi_{n+1}$ from the graph $G$ does not change the number of loops
and the removal of line $\pi_n$ from $G$ changes the number of loops. That is
$$
L(G-\pi_n)=L(G)-1
$$
\be
L(G-\pi_i)=L(G),~~~~n+1 \le i \le I
\label{lminusone}
\ee
From \eq{pofg} it follows that
$P(G)$
does not depend on $\alpha_{\pi_I},
\alpha_{\pi_{I-1}},
\ldots, \a_{\pi_{n+1}}$.
Since $P(G)$ is a homogeneous polynomial of order $L(G)$ in $\a$, we
have
\be
P(G)=(\beta_{n}\beta_{n+1}\cdots\beta_I)^{2L(G)}
P(G | \beta_{n}=1,\ldots,\beta_I=1)
\ee
From \eq{pofg} we have
\bea
&&P(G | \beta_{n}=1,\ldots,\beta_I=1) \non
&&~~= \beta_{n-1}^{2L(G-\pi_n)} [ P(G-\pi_n | \beta_{n-1}=1,\ldots,\beta_I=1 ) +
{\cal O}(\beta_{n-1}^2) ]
\label{hla1}
\eea
Now since $G-\pi_n$ is an $L$ loop graph, we have by  inductive assumption
and the fact that $P(G-\pi_n)$ does not depend on 
$\a_{\pi_{n+1}},\ldots,\a_{\pi_I}$,
\footnote{The lines of $G-\pi_n$ are: $\pi_1,\ldots,\pi_{n-1},\pi_{n+1},
\pi_{n+2},\ldots,\pi_I$.
Since the total number of lines in $G-\pi_n$ is $I-1$, one should actually
have a different notation in the argument of $P$ in \eq{hla2}:
${\hat \beta}_1=\beta_1,\ldots,{\hat \beta}_{n-1}=\beta_{n-1}$,
${\hat \beta}_n = \beta_{n+1},\ldots, {\hat \beta}_{I-1}=\beta_I$.
}
\be
P(G-\pi_n | \beta_n=1,\ldots,
\beta_I=1)= \LB \prod_{i=1}^{n-1} \beta_i^{2L(\gamma_i)}\RB
 [1+{\cal O}(\beta^2)]
\label{hla2}
\ee
Using \eq{hla1}, \eq{hla2} and relations
\be
L(G)\equiv L(\gamma_I)=L(\gamma_{I-1})=\cdots = L(\gamma_n),~~~~~~~L(G-\pi_n)=
L(\gamma_{n-1})
\ee
we find \eq{mrel}.




\newsection{Convergence theorem for non-commutative 
theories--Non-derivative case}
\subsection{Outline of the proof}
Let us recall 
the 
parametric  representation of a Feynman integral for a 
 graph $G$ 
in the  scalar  field theory over non-commutative $\re^d$ 
without derivative couplings\cite{we}. It reads
\be
I_G(p)= \int_0^{\infty} d\a_1 \cdots d\a_I  {\ex -[\sum \a_l  m_l^2+ 
Q(p,\a,\Theta)+i {\tilde Q}(p,\a,\Theta)]
\ov \prod_{i=1}^{d/2} P(G,\theta_i)  }
\label{nfeyn}
\ee
where
\begin{equation}\label{stc}
 \Theta=\left(
 \begin{array}{ccccc}
 0 & -\theta_1 & & & \\
 \theta_1 &0& & & \\
 & &\ddots & & \\
 &&&0 & -\theta_{\frac{d}{2}}  \\
 &&&\theta_{\frac{d}{2}} &0 \\
 \end{array}
 \right),
 \end{equation}
is $d\times d$ anti-symmetric matrix in the Jordan form, 
$$
P(G,\theta) = \det\A\det\B
$$
\be
\A_{mn}(\theta)\equiv \a_m \delta_{mn}  -  \theta I_{mn} 
,~~~~\B_{v\tilde v}(\theta)=\e_{vm}(\A^{-1})_{mn}
\e_{\tilde v n},~~v,\tilde v=1,\ldots,V-1
\label{ptwog}
\ee
is the determinant which comes from the integration over momenta,
$Q(p,\a,\Theta)$ and  ${\tilde Q}(p,\a,\Theta)$ are real-valued
functions.\footnote{
The graph-theoretic formula for $P(G,\theta)$ was given in ref.\cite{we,mrs}.
$Q$ and ${\tilde Q}$ can be expressed in terms of $P(G,\theta)$ \cite{mrs}.}

Let us briefly outline the proof of the convergence theorem. In
\eq{nfeyn}, only the convergence at the origin $\a = 0$ is questionable.
Thus we  consider the following part of the integrand in \eq{nfeyn}
\be
{\ex [- 
Q(p,\a,\Theta)]
\ov \prod_{i=1}^{d/2} P(G,\theta_i)  }
\label{eqone}
\ee
Let us perform the change of variables \eq{permute} in the above equation.
According to  theorem 1 to be proven below, the 
following relation
holds  around the origin $\beta =0$ 
\be
P(G,\theta)= \theta^{ c(G)} 
\LB \prod_{i=1}^I\beta_i^{2(L(\gamma_i)-c_G(\gamma_i))}\RB [1+{\cal O}(\beta^2)]
\label{eqtwo}
\ee
where $c(G)=2g$ is twice the genus of the graph $G$.

In theorem 2 it will be shown that 
\be
Q(p,\a,\Theta)= \LB \prod_{i=1}^I \beta_i^{-2j(\gamma_i)} \RB 
                  [ f_G(\pi, p, \Theta)  + {\cal O}(\beta^2)]
\label{eqthree}
\ee
where $f_G$ is a function which depends on external momenta $p$ 
of graph $G$, $\Theta$ and  the permutation $\pi$ from \eq{permute}.
The function $f_G$ is non-negative $f_G \ge 0$. For the non-exceptional
external momenta $p$ it is strictly positive: $f_G > 0$.

Combining \eq{eqone}, \eq{eqtwo}, \eq{eqthree} and \eq{jacob}, it is
easy to see that the integrand of \eq{nfeyn} is majorized by the
factor
\be
{\ex \LB - f_G(\pi, p, \Theta) \prod_{i=1}^I \beta_i^{-2j(\gamma_i)} \RB}
\prod_{l=1}^I \beta_l^{-[\omega(\gamma_l)- c_G(\gamma_l) d]-1} 
\label{eqfour}
\ee
The convergence theorem follows
from  \eq{eqfour}.

\subsection{Lemmas and theorems}
In this section we prove four lemmas and three theorems which are
inter-related as follows:
$$
~~~~~~~~~~~~~~~~~~{\rm Lemma}~\ref{lem3}
$$
$$
~~~~~~~~~~~~~~~~~~\downarrow
$$
$$
{\rm Lemma}~\ref{lem1}~\rightarrow{\rm Lemma}~\ref{lem2}~\rightarrow{\rm 
Theorem}~\ref{ther1}~\rightarrow{\rm Theorem}~\ref{ther2}~
$$
$$
~~~~~~~~~~~~~~~~~~~~~~~~~~~~~~\searrow~~~~~~~~\swarrow
$$
$$
~~~~~~~~~~~~~~~~{\rm Lemma}~\ref{lem4}~\rightarrow{\rm Theorem~\ref{ther3}~}
$$

In lemma \ref{lem1}
 we  prove a non-commutative analog of relation \eq{pofg}.
For this purpose we need the following representation of $P(G,\theta)$
from \eq{ptwog}. 
Using the 
Cauchy-Binet theorem and the Jacobi ratio theorem, $P(G,\theta)$  can be expressed 
as a double sum over
all possible trees ${\cal T}(G)$ of the graph $G$ as follows
\footnote{
 For an analogous representation in the
commutative case see  ref.\cite{nakanishi}. Note that this representation
assumes that  $G$ has no tadpoles.
 The graphs containing tadpoles
can be treated as follows. If a line $l$ begins and ends on the same vertex
$v$, replace it by two lines $l$ and $l'$ forming a loop. Call the resulting
graph $G'$. The original
graph $G$ can be obtained from $G'$ by   shrinking
line $l'$. Now in the sum over trees \eq{eq:sumtrees} the lines $l$ and $l'$
will enter symmetrically. In other words, the associated Schwinger parameters
will always come in the combination $(\a_l +\a_{l'})$. This means that the
relations which follow from \eq{eq:sumtrees} are also true  for the graphs
containing tadpoles. In particular, lemma \ref{lem1}
 holds for the graphs with
tadpoles as well as those without.}  
\be
P(G,\theta)=\sum\limits_{T,T'\in{\cal T}(G)} (-)^{T+T'} \det\e[T]\,\,
\det\e[T']\,\det\A [T^*| {T'}^*]
\label{eq:sumtrees}
\ee
where $\e[T]$  denotes the minor of the incidence matrix $\e$ 
corresponding to the lines 
in the tree $T$, while $T^*$ denotes the complement of the tree $T$, 
i.e. $T^* = G\setminus T$. The quantity $(-)^{T+T'}$ is defined as follows.
If  
$$T=\{ i_1,i_1,...,i_{V-1}\},~~~~~T'=\{ j_1,j_1,...,j_{V-1}\}
$$ 
then\footnote{It is assumed that $i_1< i_2 < \cdots < i_{V-1}$ and
$j_1 < \cdots < j_{V-1}$.}
\be
(-)^{T+T'}\equiv (-)^{\sum_{k=1}^{V-1} i_k+j_k}
\ee

\newtheorem{lm}{Lemma}[section]
\begin{lm}
 If $G$ is an $L$ loop graph and $l$ is one of its lines,
then
\be
P(G,\theta) = \left\{
\begin{array}{cl}
P(G-l,\theta) & {\rm ~if~}L(G-l)=L(G) \\
 \alpha_l\,P(G-l,\theta) + X(G|l) &{\rm ~if~}L(G-l)=L(G)-1
\end{array} \right.
\ee
where
$X(G, \theta|l)=P(G, \theta)|_{\alpha_l=0}$.
\label{lem1}
\end{lm}

\noindent
{\bf Proof}.

If $L(G-l)=L(G)$, then the line $l$ belongs to all trees of the graph $G$ and
thus $P(G,\theta)$ does not depend on $\a_l$. It implies that
$P(G,\theta)=P(G-l,\theta)$.

Consider now the case when $L(G-l)=L(G)-1$.
Let us  take the derivative of both sides of \eq{eq:sumtrees} with respect to
$\alpha_l$. It is clear that some terms will not give any contribution since
${\partial\ov\partial\alpha_l} \det\A [T^*| {T'}^*]=0$ if
$l\in T$ or $l\in T'$. Thus, we have
\be
{\partial\ov\partial\alpha_l} P(G,\theta)=\sum\limits_{
\stackrel{T,T'\in{\cal T}(G)}{\scriptstyle{l\notin T\cup T'}} }
(-)^{T+T'}\det\e[{T}]\det\e[{T'}]{\partial\ov\partial\alpha_l}
\det\A [T^*| {T'}^*]
\label{eq:der}
\ee
By Laplace expanding $\det\A [T^*| {T'}^*]$ about the row containing 
$\alpha_l$ we see that the only piece that survives is the
minor corresponding to $\alpha_l$, {\it i.e.}
\be
{\partial\ov\partial\alpha_l}\det\A [T^*| {T'}^*]=
\det{\hat\A}[T^* | {T'}^*]
\ee
where ${\hat \A}$ is obtained from $\A$ by removing the 
row and the column containing $\alpha_l$. 
Using the above relation the r.h.s. of \eq{eq:der} becomes
\bea
\sum\limits_{
\stackrel{T,T'\in{\cal T}(G)}{\scriptstyle{l\notin T\cup T'}} }\!\!\!\!\!\!\!&{}&\!\!\!\!\!\!(-)^{T+T'}
\det\e[{T}]\det\e[{T'}]
\det{\hat\A}[T^*| {T'}^*] \nonumber\\
&=&\sum\limits_{\scriptstyle{T,T'\in{\cal T}(G-l)}}(-)^{T+T'}
\det\e[{T}]
\det\e[{T'}]\det\A(G-l)[T^*| {T'}^*]\non
&=& P(G-l,\theta)
\eea
where for the first equality we used the obvious fact that
if a line does not belong to $T$ and $T^*$ of a graph $G$, then it does not 
belong to  $G$.
~~~~~~~~~~~~~~{\rm {\bf q.e.d.}}

For later purposes, let us introduce the following notation:
\be
X(G,\theta|M)=P(G,\theta)|_{\alpha_l=0,~\forall l\in M}
\label{eq:defX}
\ee
where $M$ is a set of lines of the graph $G$. This notation will be 
useful in Section \ref{sec7}.

Consider now the following question.
 Assume that we rescale
all of the Schwinger parameters  except the one associated to the 
line $l$ by the same factor $\rho$. What is  the scaling
of $P(G,\theta)$ as $\rho\rightarrow 0$?
Before answering this question,
let us consider a simpler problem of rescaling all of the Schwinger parameters
by  $\rho$.

It is easy to see from  \eq{eq:sumtrees} that the smallest 
number of $\alpha$'s in the monomials forming $P(G,\theta)$ 
equals the number of 
loops $L(G)$ minus the rank $r(G)$ of the intersection matrix $I_{mn}$.  
Thus, $P(G,\theta)$
is a polynomial of degree $r(G)$ in $\theta$:
\be
P(G,\theta)=\sum_{n=0}^{r(G)/2} \theta^{2n} P_{2n}(G)
\label{poltheta}
\ee
By the dimensional argument we find from \eq{poltheta}
the following scaling 
\be
P(\rho G, \theta)=\rho^{L(G)-r(G)}\,\,\left[ P_{r}(G)+{\cal  O}(\rho)\right]
\label{rhoGscale}
\ee
where the notation $\rho G$ stands for the rescaling of all Schwinger
parameters $\alpha$ by $\rho$. 
Note that the rank $r(G)$ of the intersection matrix of a graph $G$ is 
equal to twice the
genus of the graph $G$.\cite{we,mrs}. 
From  the definition of $c(G)$ it follows that
\be
r(G)= c_G(G)
\ee

We are interested  in the scaling of 
$P_{r}(G)$ only, because it gives the leading singular
term in the integrand of Feynman integral.

\begin{lm}
 Let $l$ be a line of a  graph $G$. In the expression for
$P_r(G)$ let us rescale all Schwinger parameters except the one 
associated to the line $l$:
$\a_j \rightarrow \rho \a_j$  for all $j \ne l$. Then the following
relations hold:

\noindent
(1) If $L(G-l)=L(G)$, then
\be
P_r(\alpha_l,\rho(G-l))=\rho^{L(G)-r(G)} P_r(G)
\label{relone}
\ee

\noindent
(2) If $L(G-l) \ne L(G)$ and  $r(G-l) < r(G)$, then \eq{relone} holds.

\noindent
(3) If $L(G-l) \ne L(G)$ and $r(G-l) = r(G)$, then
\be
P_r(\alpha_l, \rho(G-l))=\rho^{L(G-l)-r(G)}
\left[\alpha_l P_r(G-l)+{\cal O}(\rho)\right]
\label{reltwo}
\ee
\label{lem2}
\end{lm}

\noindent
{\bf Proof.} 

\noindent
(1) $L(G-l)=L(G)$

From lemma \ref{lem1} we have $P(G,\theta)=P(G-l,\theta)$. 
Thus \eq{relone} holds.

\noindent
(2)$L(G-l)\ne L(G)$ and  $r(G-l) < r(G)$  

From the lemma \ref{lem1} we have in this case
$$
P(G,\theta)= \a_l P(G-l,\theta) + X(G|l)
$$
By taking  $r(G)$-th derivative w.r.t. $\theta$  of both
sides of this equation 
 and using the fact that $P_{r(G)}(G-l)=0$ (  which follows
from  the
condition  $r(G-l) < r(G)$ ),
we find
\be
P_r(G)={1\ov r!}\partial^r_\theta X(G|l)
\ee
Thus $P_r(G)$ is 
independent of $\alpha_l$. Thus \eq{relone} holds in this case.

\noindent
(3) $L(G-l)\ne L(G)$ and $r(G-l) = r(G)$

From the lemma \ref{lem1} we have in this case
$$
P(G,\theta)= \a_l P(G-l,\theta) + X(G|l)
$$
By taking  $r(G)$-th derivative w.r.t. $\theta$ of both sides
of this equation we get
\be
P_r(G)=\alpha_l\,P_r(G-l)+
{1\ov r!}\partial^r_\theta X(G|l)
\ee
Rescaling all $\a$  except $\alpha_l$ yields
\be
P_r(\alpha_l,\rho (G-l))=\rho^{L(G-l)-r(G-l)}
\left[ \alpha_l P_r(G-l)+{\cal O}(\rho)\right]
\ee
since,   according to  lemma \ref{lem1},
$X(G|l)$ is independent of $\alpha_l$.~~~~~~{\bf q.e.d.}

\begin{lm}
Let $G$ be a  NQFT Feynman graph with
the  external momenta set to zero.
Consider a set of lines $l_1, l_2,\ldots, l_k \in G$. Let us define
$G_i,~i=1,\ldots,k$  as 
\be
G_k \equiv G,~~~~~~~G_{i-1}=G_i - l_i,~~i=1,\ldots,k
\ee 
Let $K_1$ and $K_2$ be the sets of lines of $G$ defined by
\eq{K1set} and \eq{K2set}.
If $l_1, l_2,\ldots, l_k \in K_1(G)\cup K_2(G)$,
then the following relations hold
\be
L(G_i)-c_G(G_i) = L(G)-c_G(G),~~~~~~i=0,\ldots,k
\label{pervich}
\ee
\label{lem3}
\end{lm}

\noindent
{\bf Remark.}
At  this   point  one might find it useful
to check the relations \eq{pervich} for a particular graph.
Consider the graph $G$ in figure \ref{commutant}. 
We have $K_2(G)=\{  3,4,5,6,7,8 \}$.
Let us take
$l_1=8$, $l_2=7$ and $l_3=6$. Then 
$$G_3=G,~~~G_2=\{ 1,2,3,4,5,7,8\},
$$
$$
G_1=\{ 1,2,3,4,5,8\},~~~G_0=\{ 1,2,3,4,5 \}
$$
Since  
$$
L(G_3)=5,~~~L(G_2)=4,~~~L(G_1)=3,~~~L(G_0)=2
$$
and
$$
c_G(G_3)=4,~~c_G(G_2)=3,~~c_G(G_1)=2,~~c_G(G_0)=1
$$
the relations \eq{pervich} are satisfied.

\noindent
{\bf Proof.}

The proof relies on two auxiliary statements:

\noindent
{\bf (1)} Let  ${\cal L}\subset G$ be any loop which contains a line
$l\in K_2(G)$. Then the associated cycle $\C(\L)\in H_1(\Sigma(G))$ is
non-trivial.\footnote{For the definition of $\C(\L)$ see the remark following
definition 1.}

\noindent
Suppose that $\C(\L)$ is trivial. The condition $l\in K_2(G)$ implies
that there exists a loop $\L' \ni l$ containing $l$  with the 
property  that  the  associated  cycle  $\C(\L')$  is  non-trivial
 and 
$\C(\L'')\ne \C(\L')$
for any loop $\L''\not\ni l$. Using the rule of addition of  cycles
it is easy to see that $\C(\L\cup\L' - l)=\C(\L')$.  This is
a contradiction, since $\L \cup \L' -l$ does not contain the line $l$.
Thus  $\C(\L)$ is non-trivial.

\noindent
{\bf (2)} If  $l_1,l_2\in K_2(G)$ and  $c_G(G-l_1-l_2)=c_G(G-l_2)$,  then
$L(G-l_1-l_2)=L(G-l_2)$.

\noindent
Suppose that $L(G-l_1-l_2)\ne L(G-l_2)$, i.e. $l_1$  belongs  to
a loop in $G-l_2$. Let  $\L$  be  such  a  loop.  The  condition
$l_1\in K_2(G)$
and the statement (1) implies that $\C(\L)$ is non-trivial.
The condition $c_G(G-l_1-l_2)=c_G(G-l_2)$ implies that  there is a loop
$\L'\subset G-l_2$ with the properties: (a) $l_1 \not\in \L'$, (b)
$\C(\L')$ is non-trivial and $\C(\L')=\C(\L)+\Delta \C$, where
$\Delta \C \in H_1(\Sigma(G))$ is associated to a loop in
$G-l_1-l_2$.

\noindent
The condition $c_G(G-l_2)< c_G(G)$ implies that for any loop $\L''\subset
G$, $\L'' \ni  l_2$, the following statement holds:
$\C(\L'')$ is not a linear combination of the homology cycles  formed
by  the loops  of  $G-l_2$.  In other words, $\C(\L'')$ is independent
of the  homology cycles  associated to loops  of  $G-l_2$.

\noindent
Let us compare the numbers of independent homology cycles
$c_G(G)$ and
$c_G(G-l_1)$. 
For this purpose
let us choose  an arbitrary loop  $\L''\subset G$, $\L''\ni l_2$.
In $G$ there are $\C(\L)$, $\C(\L'')$ and some other cycles
on which $\C(\L)$ and $\C(\L'')$ are independent of. 
Which cycles do we have in   $G-l_1$? 
There are two cases to consider:  (a) $l_1\not\in \L''$  and 
(b) $l_1\in \L''$.  In the case (a) we have $\C(\L')=\C(\L) +\Delta \C$ and
$\C(\L'')$. In the case (b)  we have $\C(\L')=\C(\L)+ \Delta \C$ and
$\C(\L)+\C(\L'')$. Thus in both cases the number of independent
cycles of $G-l_1$, $c_G(G-l_1)$, is equal to $c_G(G)$. This is
a contradiction to the assumption $l_1\in K_2(G)$. 
Thus $L(G-l_1-l_2)=L(G-l_2)$.

By  assumption we have $l_1,\ldots, l_k \in K_1(G_k) \cup  K_2(G_k)$.
Let us remove $l_k$  from $G_k$  and  see what happens  to a line
$l_i\in K_2(G_k)$ ($i< k$). There are only three possibilities:

\noindent
(a) $l_i\in K_1(G_{k-1})$

\noindent
(b) $l_i\in K_2(G_{k-1})$

\noindent
(c) $l_i\not\in  K_1(G_{k-1})$  and $l_i\not\in K_2(G_{k-1})$

Let $B_{k-1}$ be the set of lines $l_i\in K_2(G_k)$ ($i< k$) with
the property (c).  Thus we have
\be
l_1,\ldots,l_{k-1}  \in  K_1(G_{k-1})\cup  K_2(G_{k-1})\cup B_{k-1}
\ee
Similarly, the set $B_{i-1}$ ($1\le i < k$) is defined as follows.
Let us remove $l_i$ from $G_i$ and see what happens to a line
$l_j\in K_2(G_i)\cup B_i$ ($j<i$). There are only three possibilities:

\noindent
(a') $l_j\in K_1(G_{i-1})$

\noindent
(b') $l_j\in K_2(G_{i-1})$

\noindent
(c') $l_j\not\in  K_1(G_{i-1})$ and $l_j\not\in K_2(G_{i-1})$.

Let $B_{i-1}$ be the set of lines $l_j\in K_2(G_i)\cup B_i$ ($j<i$)
with the property (c'). Thus we have
\be
l_1,\ldots ,l_{i-1} \in K_1(G_{i-1})\cup K_2(G_{i-1}) \cup B_{i-1}
\ee
From the above definition of $B_i$ ($0\le i <k-1$) it follows
that $B_i \subset K_2(G)$.

The  lemma  follows  from  the  following  two  statements:

\noindent
{\bf (3)} If $L(G_{i-1})=L(G_i)$,  then  ${c}_G(G_{i-1})={c}_G(G_i)$.

\noindent
This statement is  evident. The condition 
$L(G_{i-1})=L(G_i)$ (or, equivalently, $l_i\in K_1(G_i)$)
implies that there is no loop
in $G_i$ to which 
$l_i$ belongs to. Therefore,  no cycle can be lost by removing $l_i$ from 
$G_i$.  Thus, ${c}_G(G_{i-1})={c}_G(G_i)$.

\noindent
{\bf (4)} If $L(G_{i-1})=L(G_i)-1$, then ${c}_G(G_{i-1})={c}_G(G_i)-1$.

\noindent
 The condition 
$L(G_{i-1})=L(G_i)-1$ 
implies that 
$l_i\in K_2(G_i)$ or $l_i\in B_i$. If~ $l_i\in K_2(G_i)$ then, by the
 definition of the set $K_2(G_i)$, we have
${c}_G(G_{i-1})<{c}_G(G_i)$. But by disconnecting
a single loop the cycle number can decrease by at most one.
Thus we have ${c}_G(G_{i-1})={c}_G(G_i)-1$. 
If $l_i\in B_i$ then 
 according to 
 statement (2) by removing $l_i$ the independent cycle associated to loops 
in $G_i$ containing $l_i$ will be lost. This implies again that ${c}_G(G_{i-1})={c}_G(G_i)-1$.
~~~~~~~~~~~~~{\bf q.e.d.}

\newtheorem{th}{Theorem}
\begin{th}
Let $G$ be a general NQFT Feynman graph.
Consider the change of variables \eq{permute} specified
by a permutation $\pi$. Let $\Pi$ be the set of all permutations of
$\{ 1, 2,\ldots, I\}$. Then for any $\pi \in \Pi$ the following 
relation holds
\be
P_{2g}(G)
=\LB \prod_{i=1}^I\beta_i^{2(L(\gamma_i)-c_G(\gamma_i))}\RB
[1+{\cal O}(\beta^2)]
\label{eq:ref}
\ee
where $g=r(G)/2$.
\label{ther1}
\end{th}

\noindent
{\bf Remark.} 
At  this  point 
one  might  find  it  useful  to  check \eq{eq:ref} for some particular
graphs, say for the figure \ref{commutant1212}(a). 
The homogeneous polynomial $P_4$ from \eq{homopoly} for the latter graph
is 
\be
P_4=(\a_1+\a_2+ \a_3+\a_4) (\a_5+\a_7) (\a_8+\a_9)+
  (\a_1+\a_2) (\a_3+\a_4)(\a_5+\a_7+\a_8+\a_9) 
\ee
Consider the subgraph formed by the lines $1,2,3,4$. Let us
rescale the corresponding Schwinger parameters by $t$:
$$\a_1, \a_2, \a_3, \a_4 \rightarrow t\a_1, t\a_2, t\a_3, t\a_4$$
It is easy to see that $P_4\sim t=t^{2-1}$  for $t\sim 0$. Similarly,
for any subgraph $\gamma$ we have $P_4\sim t^{L(\gamma)-c(\gamma)}$.

\noindent
{\bf Proof.}

We will prove the relation \eq{eq:ref} by induction.
It is not difficult to
check that \eq{eq:ref} holds for all one and two loop diagrams.

Let us assume that  \eq{eq:ref} holds for all diagrams
with number of loops less or equal to $L$
and prove the relation 
\eq{eq:ref}
for diagrams with $L+1$ loops.
Let $G$ be an arbitrary $L+1$ graph. We will prove that \eq{eq:ref}
holds for $G$.

Let $I$ be the total number of lines of the graph $G$. 
Consider the change
of variables \eq{permute} specified
by a permutation $\pi$.

If all the lines $1,\ldots,I$ of the graph $G$ are from 
the set $K_1(G)\cup K_2(G)$,
then  $L(\gamma_i)=c(\gamma_i)$ for all 
$i\in \{ 1,\ldots,I\}$, since any loop of  $G$ wraps
a non-trivial cycle. 
Using the dimensional argument, we see from
 \eq{poltheta} that  $P_{2g}(G)$ does not depend on 
 $\alpha$. More precisely, 
\be
P_{2g}(G)= {\rm det} ~I
\ee
where $I$ is a nondegenerate  intersection matrix of size 
$2g\times 2g$. Changing the basis to  the canonical one:
$\{ a_1,b_1,\ldots,a_g,b_g\}$,
it is easy to see that $P_{2g}(G)=1$.

Thus let us  consider the case when there is at least one line 
which does not  belong to the set $K_1(G)\cup K_2(G)$.
Let $\pi_n$ be the first line in the sequence $\pi_I,\pi_{I-1},\ldots,
\pi_{n+1},\pi_n,\ldots,$  which  does not belong to $K_1(G)\cup K_2(G)$, i.e.
\be
\pi_n \notin K_1(G) \cup K_2(G),~~~~~~~~\pi_i \in K_1(G)\cup K_2(G),~~n+1 \le i \le I
\label{K1cupK2}
\ee

From the lemma \ref{lem2}  and \eq{K1cupK2} it follows that $P_{2g}(G)$ does
not depend on $\a_{\pi_{n+1}},\ldots,\a_{\pi_I}$. Thus
$(\beta_{n+1}\beta_{n+2}\cdots \beta_I)$ trivially factors out of
$P_{2g}(G)$:
\be
P_{2g}(G)=(\beta_{n+1}\beta_{n+2}\cdots \beta_I)^{2(L(G)-c_G(G))}
P_{2g}(G | \beta_{n+1}=1,\ldots,\beta_I=1)
\ee
Since $P_{2g}(G)$ is a homogeneous polynomial of degree $L(G)-c_G(G)$
 in $\a_{\pi_1},\ldots,\a_{\pi_n}$, we have
\be
P_{2g}(G | \beta_{n+1}=1,\ldots,\beta_I=1)=\beta_n^{2(L(G)-c_G(G))}
P_{2g}(G | \beta_{n}=1,\ldots,\beta_I=1)
\ee

Using lemma \ref{lem3}  one can   show that 
\be
(\beta_{n+1}\beta_{n+2}\cdots \beta_I)^{2(L(G)-c_G(G))}=
\prod_{i=n+1}^I \beta_i^{2(L(\gamma_i)-c_G(\gamma_i))}
\ee

From the case 3 of lemma \ref{lem2}
 ( with $l=\pi_n$ and $\rho = \beta_{n-1}^2$ ) 
we find
\bea
&&P_{2g}(G | \beta_{n}=1,\ldots,\beta_I=1) \non
&&~~= \beta_{n-1}^{2(L(G-\pi_n)-
c_{G-\pi_{n}}(G-\pi_n))} [ P_{2g}(G-\pi_n | \beta_{n-1}=1,\ldots,\beta_I=1 ) +
{\cal O}(\beta_{n-1}^2) ]
\eea
The second term in the above equation is subleading in $\beta_{n-1}^2$.
Thus we consider the first term. It can be rewritten as
\be
\beta_{n-1}^{2(L(G-\pi_n)-
c_{G-\pi_{n}}(G-\pi_n))}  P_{2g}(G-\pi_n | \beta_{n-1}=1,\ldots,\beta_I=1 ) 
= P_{2g}(G-\pi_n | \beta_{n}=1,\ldots,\beta_I=1 ) 
\ee
since $P_{2g}(G-\pi_n)$ is a homogeneous polynomial of degree 
$L(G-\pi_n)-
c_{G-\pi_n}(G-\pi_n)$ in $\a$. Since $G-\pi_n$ is an $L$ loop graph,
 \eq{eq:ref} holds for it. 
If a line $l\in G$  is from $K_1(G)\cup K_2(G)$, then 
$l\in K_1(G-\pi_n)\cup K_2(G-\pi_n)$.
Thus it is easy to see that 
$P_{2g}(G-\pi_n)$ does not depend on $\a_{\pi_{n+1}},\ldots,\a_{\pi_I}$.
We thus have
\be
 P_{2g}(G-\pi_n | \beta_{n+1}=1,\ldots,\beta_I=1 ) =
\LB \prod_{i=1}^{n-1} \beta_i^{2(L_{G-\pi_n}(\gamma_i)-c_{G-\pi_n}
(\gamma_i))}\RB [1+{\cal O}(\beta)]
\ee
where $L_{G-\pi_n}(\gamma)$ and $c_{G-\pi_n}(\gamma)$
are the number of loops
and the cycle numbers of graph $\gamma$ with respect to the graph $G-\pi_n$.
It is clear that for any subgraph $\gamma\subset G$ which
does not contain line $\pi_n$, we have
\be
L_G(\gamma) = L_{G-\pi_n}(\gamma)
\ee
From the assumption  $g(G-\pi_n)=g(G)$ it follows that
for any subgraph $\gamma\subset G$ which
does not include line $\pi_n$ we have
\be
c_{G-\pi_n}(\gamma)=c_G(\gamma)
\ee
since by removing the line $\pi_n$ we do not disconnect any cycle
of $\Sigma_g(G)$. 
~~~~~~~~~~~{\bf q.e.d.}

Note that theorem 1 together with  
\eq{rhoGscale} implies that \eq{eqtwo} holds.

\begin{th}
 Let $G$ be a 1PI graph. 
Consider the change of variables \eq{permute} specified
by a permutation $\pi$. Let $\Pi$ be the set of all permutations of
$\{ 1, 2,\ldots, I\}$. Then for any $\pi \in \Pi$ the following 
relation holds
\be
Q(p,\a,\Theta)= \LB \prod_{i=1}^I \beta_i^{-2j(\gamma_i)} \RB 
                  [ f_G(\pi, p, \Theta)  + {\cal O}(\beta^2)]
\label{lemma4}
\ee
where $f_G$ is a non-negative function
\be
f_G(\pi, p, \Theta) \ge 0.
\label{fofGeq}
\ee
\label{ther2}
\end{th}

\noindent
{\bf Proof.}

Let $G_{ij}$ be the graph obtained from $G$ by connecting external
lines $i$ and $j$ (see figure \ref{defj}). As shown in appendix A 
(see also ref.\cite{mrs}) $Q$ is given by  
\be
Q(p,\a,\Theta)=  - \sum_{i\ne j} p_i^{\mu} \LB { P(G_{ij},\Theta | \alpha_{ij}=0 )
 \ov P(G, \Theta)}\RB^{\mu\nu} p_j^{\nu}
\label{Qmrs}
\ee
and it is non-negative
\be
Q(p,\a,\Theta) \ge 0
\label{Qge}
\ee
For a
given permutation $\pi$, let $\gamma_{k_0}$ 
in \eq{nested} be the smallest subgraph such that $j(\gamma_{k_0})=1$.
In other words, 
$$
j(\gamma_l) = 0,~~~~1 \le l \le k_0-1
$$
\be
j(\gamma_l) = 1,~~~~k_0 \le l \le I
\ee
Let $i, j$ be two external lines of  $G$ such that
\be
c_{G_{ij}}(\gamma_{k_0}) = c_G(\gamma_{k_0})+1
\label{cijeq}
\ee
The existence of at least a pair of such lines follows from the 
fact that $j(\gamma_{k_0})=1$. 
From theorem 1 we have\footnote{For the graph $G_{ij}$ one should
restrict to the subset of permutations $\{ \pi | \pi_{I+1}=l_{ij} \}$, 
where $l_{ij}$ 
is the new line formed by the joining of external lines $i$ 
and $j$. Note that, by assumption \eq{cijeq}, we have $g(G_{ij})> g(G)$.
Thus, according to lemma \ref{lem2}, $P_{2g(G_{ij})}(G_{ij})$ is 
independent of $\a_{ij}$. If $j(G)=0$, then  $g(G_{ij})= g(G)$ for any
$i\ne j$ and $P_{2g(G_{ij})}(G_{ij})$ depends on $\a_{ij}$. 
It is clear that if $g(G_{ij})= g(G)$, then for any subgraph $\gamma\subset
G$ we have $c_{G_{ij}}(\gamma)=c_{G}(\gamma)$. 
 Thus,   if $j(G)=0$,  then
for any $i,j$ the leading
term in $P(G_{ij}, \theta | \alpha_{ij} = 0 )$ is subleading compared
to the leading term in $P(G, \theta)$. 
\label{ftbetalijeq1}
}
\be
P(G_{ij}, \theta | \alpha_{ij} = 0 )= \theta^{c(G)+2} \LB \prod_{l=1}^I 
\beta_l^{2(L_{G_{ij}} (\gamma_l)- c_{G_{ij}}(\gamma_l))} \RB 
[1+\O
(\beta^2 )]
\ee
where subscript $G_{ij}$ in $L_{G_{ij}}$ and $c_{G_{ij}}$ 
means that the corresponding quantities are defined with respect to 
the graph $G_{ij}$.
Combining the above equation with \eq{eqtwo} and using  the
relations 
\be
L_{G_{ij}} (\gamma_l) = L_G(\gamma_l),~~~~ 1 \le l \le I
\ee
and
$$
c_{G_{ij}}(\gamma_l) = c_G (\gamma_l),~~~1 \le l \le k_0-1
$$
\be
c_{G_{ij}}(\gamma_l) = c_G (\gamma_l)+1,~~~k_0 \le l \le I
\ee
we find
\be
{ P(G_{ij}, \theta | \alpha_{ij}=0 ) \ov P(G,\theta) }= \theta^2
  \LB \prod_{l=1}^I \beta_l^{-2j(\gamma_l)} \RB [1+\O(\beta^2)]
\ee
The r.h.s. of the above equation does not depend on the particular
pair of lines $i, j$ that we chose as long as $c_{G_{ij}}(\gamma_{k_0})>c(\gamma_{k_0})$. 
It means that the leading contribution to
$Q$ of
 any pair of lines $i, j$ 
which satisfy \eq{cijeq}  is 
$$
-p_i^{\mu}(\Theta^2)^{\mu\nu}p_j^{\nu} \LB \prod_{l=1}^I \beta_l^{-2j(\gamma_l)} \RB 
$$
Let 
$Y$
be the set of all ordered pairs of external lines of graph $G$ which satisfy
condition \eq{cijeq}. 
It is easy to see that  \eq{lemma4}
holds, with the function $f_G$ given by
\be
f_G(\pi,p,\Theta)= -\sum_{(i,j)\in Y} p_i^{\mu}(\Theta^2)^{\mu\nu}p_j^{\nu}
\label{withfun}
\ee
The relation \eq{fofGeq} follows from \eq{Qge} and the following
statement.
Let $F(\beta)\ge 0$ be a non-negative function for $\beta \ge 0$.
Then the leading term in the expansion of $F(\beta)$ around
$\beta = 0$ is non-negative.
~~~~~~~~{\bf q.e.d}

\begin{lm} 
For any 1PI graph $G$, there exists a unique decomposition
\be
{\cal E}(G)=\bigcup_{s=1}^{n(\gamma)} {\cal E}_s(\gamma)
\label{decompos1}
\ee
for some $n(\gamma)\ge 1$.
\label{lem4}
\end{lm}

\noindent
{\bf Proof.}

If $j(\gamma)=0$, then the lemma holds trivially with
$n(\gamma)=1$. 
Thus consider the case $j(\gamma)=1$. 
From the definition 2 of index $j$ it
follows that there exists a pair of external lines $i,k$ such
that $c_{G_{ik}}(\gamma)>c_G(\gamma)$. Now, let us look for all external
lines $l\ne i$ with the property $c_{G_{il}}(\gamma)=c_G(\gamma)$.
Denote by ${\cal E}_1$ the latter set of lines together with
the line $i$. 

\begin{figure}[hbtp]
\begin{center}
\mbox{\epsfxsize=8truecm
\epsffile{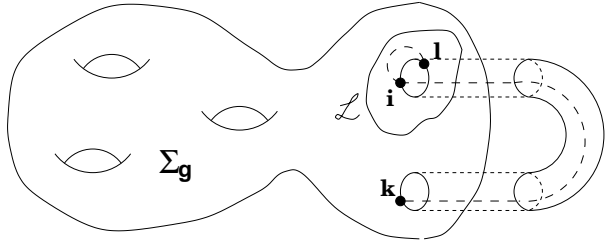}}
\end{center}
\caption{Graph $G_{ik}$.}
\label{par1ps}
\end{figure}

Let us  show that for any line $l$ from the set  ${\cal E}_1$ we 
have
$c_{G_{kl}}(\gamma)>c_G(\gamma)$. 
From the fact that $c_{G_{ik}}(\gamma) > c_G(\gamma)$ one can infer
that there exists a loop ${\cal L}$ around the new handle which was
created after joining external lines $i$ and $k$ as in figure \ref{par1ps}.
From the fact that  $c_{G_{il}}(\gamma)=c_G(\gamma)$ for all $l\in {\cal E}_1$
we can infer that ${\cal L}$ encloses all the lines from the set ${\cal E}_1$.
Thus we have $c_{G_{kl}}(\gamma)>c_G(\gamma)$ for all $l\in {\cal E}_1$.
~~~~{\bf q.e.d.}

\begin{th}
 Let $I_G$ be the Feynman integral  
associated with a 1PI graph $G$ in a massive
non-derivative scalar quantum field theory 
 over non-commutative $\re^d$.
There are three cases:

\noindent
(I) If $j(G)=0$, then $I_G$ is convergent if and only if
   $\omega(\gamma)-c_G(\gamma)d<0$ for all $\gamma\subseteq G$.

\noindent
(II) If $j(G)=1$ and the external momenta are non-exceptional, then
$I_G$ is convergent  if 
for any subgraph $\gamma\subseteq G$ 
at least one of the following conditions is satisfied:(1)
$\omega(\gamma)-c_G(\gamma)d<0$, (2) $j(\gamma)=1$.

\noindent
(III) If $j(G)=1$ and the external momenta are exceptional, then
$I_G$ is convergent if
$\omega(\gamma)-c_G(\gamma)d<0$ for all $\gamma\subseteq G$.
\label{ther3}
\end{th}

\noindent
{\bf Proof.}

\noindent
{\bf Case I.}
As explained earlier, \eq{eqfour} follows from theorem 1 and theorem 2.
Since $j(G)=0$, we have $j(\gamma)=0$ for all subgraphs $\gamma \subset G$.
Thus the exponential factor in \eq{eqfour} is independent of
$\beta$. It is  easy to see from the form of the prefactor in 
\eq{eqfour}  
that the theorem holds in this case.

\noindent
{\bf Case II.}
In this case the exponential factor in \eq{eqfour} depends on
$\beta$ and it is capable of suppressing arbitrary powers of
divergence coming from the prefactor in \eq{eqfour}. Thus if 
$f_G(\pi, p, \Theta) > 0$ for all $\pi$, 
then  $I_G$ converges. It remains to show
that $f_G(\pi, p, \Theta) = 0$ if and only if the external momenta $p$
are exceptional. 

$f_G$ is given by \eq{withfun}.
According to  lemma \ref{lem4}, for any subgraph $\gamma\subseteq G$ 
there exists
a unique decomposition \eq{decompos1}  of the set of external lines
of $G$.

Let us show that for the subgraph $\gamma_{k_0}$ from theorem 2,
the decomposition \eq{decompos1} has $n(\gamma_{k_0})=2$.
By assumption (see theorem 2),
for a given permutation $\pi$, $\gamma_{k_0}$ is the smallest subgraph
such that $j(\gamma_{k_0})=1$. Thus $n(\gamma_{k_0})>1$.
Consider the subgraph $\gamma_{{k_0}-1}$
which is obtained from $\gamma_{k_0}$ by deleting line $\pi_{k_0}$.
By assumption we have $j(\gamma_{{k_0}-1})=0$. Thus $n(\gamma_{{k_0}-1})=1$.
By disconnecting
a line, $n$ may decrease by at most one.\footnote{This can be shown
as follows. The lines from ${\cal E}_s(\gamma)$ are enlosed in the loops as
in figure \ref{par1ps}. By disconnecting a line the number of loops can decrease
by at most one. Suppose that ${\cal L}_1$ and ${\cal L}_2$
enclose  ${\cal E}_1(\gamma)$ and   ${\cal E}_2(\gamma)$, respectively.
Now remove a line $l$ from $\gamma$.  If  ${\cal L}_1$ and ${\cal L}_2$
share the line $l$, 
then  ${\cal E}_1(\gamma)$ and   ${\cal E}_2(\gamma)$ will combine
to form a set ${\cal E}_*(\gamma - l)$ from \eq{decompos1} for $\gamma-l$.} 
We thus conclude
that $n(\gamma_{k_0})=2$.

Let us define  momenta ${\cal P}_s(\gamma_{k_0}),~~s=1,2$, as
\be
{\cal P}_s(\gamma_{k_0}) = \sum_{i\in {\cal E}_s(\gamma_{k_0})} p_i
\ee
We thus have
\be
f_G(\pi, p, \Theta)= 2{\cal P}_1^{\mu}(\gamma_k) 
(\Theta^2)^{\mu\nu} {\cal P}_1^{\nu}(\gamma_k)
\label{lastequal}
\ee 
Since $\Theta$ is assumed to be non-degenerate,
 \eq{lastequal} imples that $f_G=0$ if and only if ${\cal P}_1(\gamma_{k_0})=0$,
i.e. the external momenta are exceptional.

\noindent
{\bf Case III.}
If $\omega(\gamma)-c(\gamma)d<0$ for all $\gamma\subseteq G$,
then it is easy to see from \eq{eqfour} that $I_G$ will be convergent
even if the exponent in \eq{eqfour} does not depend on $\beta$. 
~~~~{\bf q.e.d.}

\newsection{Convergence theorem for non-commutative theories--General case\label{sec7}}
In this section we prove the convergence theorem for the Feynman
integrals in 
arbitrary quantum  field theories over non-commutative 
$\re^d$ with the massive
propagators. 
It is assumed that the reader is familiar with
the constructions and  proofs given sections 5 and 6.
Thus  we will omit  the details of the proofs of
some of the  statements which are minor modifications of
those of section 6.

\subsection{Outline of the proof}
Consider  an arbitrary graph $G$ in a general theory and 
let $\{ q_l\}$ be
the set of internal momenta which appear  in the numerator  of the
Feynman integral. Let ${\cal I}_G$ be the integrand of the Feynman 
integral $I_G$ in the $\alpha$ representation.
Let us recall a trick introduced in ref.\cite{mrs} which enables
one to relate  ${\cal I}_G$ to  ${\cal I}_{G^+}$ for a graph $G^+$ in
a non-derivative scalar theory.

The new graph $G^+$ is defined as follows. For each $q^l$, let $v$ be
the vertex which $q$ leaves. Now include two new external momenta,
a momentum $r^{l}$ which leaves vertex $v$  immediately  counterclockwise
from $q$  and a momentum  $-r^{l}$ which leaves  $q$ immediately 
clockwise from $q$. We will denote by $l'$ and $l''$ the new external lines
carrying momenta $r^l$ and $-r^l$, respectively.\footnote{
From now on  the letter $r$ with an index $l$ will be reserved for the 
momentum carried  by the external lines $l'$ and $l''$ attached to the
line $l$ with ${\rm deg}(l)>0$.}   
  The only effect of these momenta is to multiply
the integrand  with an extra phase
\be
e^{i r^{l}_{\mu} \Theta^{\mu\nu} q^l_{\nu}}
\ee
Thus, if the function of 
momenta in the numerator of the Feynman integral  for $G$ is
some polynomial $P(q_{\mu},p_{\mu})$,  we  have
\be
{\cal I}_G(\alpha,\Theta)=P(-i (\Theta^{-1})_{\mu\nu}
\partial_{r^l_{\nu}}
, p_{\mu} )~ {\cal I}_{G^+}
(\alpha,\Theta)|_{r^{l}_{\mu}=0}
\ee

Let us denote by ${\cal E}'(G)$ the set of extra external lines 
added to $G$ to make $G^+$, i.e.
\be
{\cal E}'(G) = {\cal E}(G^+)\setminus {\cal E}(G)
\ee

The general form of
the leading term in the integrand of the Feynman integral
 in $\beta$ variables
 reads
$$
\LB \prod_{l=1}^I \beta_l^{2l-1}\RB
\LB \prod_{l=1}^I \beta_l^{2(L(\gamma_l)-c_G(\gamma_l))} \RB^{-d/2}
$$
\be
\times \left[ \prod_{l\in G}
\LB {\partial \ov \partial r_l }\RB^{{\rm deg}(l)}\right]
~\ex - [Q(p,\a,\Theta)+i {\tilde Q}(p,\a,\Theta)] \bigg|_{r=0}
\label{analyzthis}
\ee

In lemma \ref{QO1} it will be shown that ${\tilde Q}$ is $\O(1)$
in the $\beta$ expansion.  Thus, as far as the singular
terms are concerned, we can omit ${\tilde Q}(p,\a,\Theta)$ in the exponent
in \eq{analyzthis}.

In  lemma \ref{K2}   it  will be shown  that
if $l\in K_2(G)$, then the $r_l$-dependent terms in $Q(p,\a,\Theta)$
are nonsingular.  
Thus, as far as the singular terms are concerned,
\eq{analyzthis} can  equivalently be written in the form
$$
\LB \prod_{l=1}^I \beta_l^{2l-1}\RB
\LB \prod_{l=1}^I \beta_l^{2(L(\gamma_l)-c_{G}(\gamma_l))} \RB^{-d/2}
$$
\be
\times
\left[ \prod_{l\in G}
\LB {\partial \ov \partial r_l }\RB^{{\rm deg}(l)~{\rm ind}_{K_2}(l)}
\right]
~\ex 
\LB  
\sum_{i,j} p_i^{\mu} \LB { P(G_{ij},\Theta | \alpha_{ij}=0 )
\ov P(G, \Theta)}\RB^{\mu\nu} p_j^{\nu}
\RB\bigg|_{r=0}
\label{analyzthis2}
\ee
where ${\rm ind}_{K_2}$ is defined in \eq{indK2}.

In general there are three types of terms $p_i p_j$ in the exponent
in \eq{analyzthis2}: 

\noindent
(a)  $i, j\in {\cal E}(G)$

\noindent
(b)  $i\in {\cal E}'(G)$,  $j\in {\cal E}(G)$,

\noindent
(c)  $i, j\in {\cal E}'(G)$,

Let us consider these three cases.

\noindent
{\bf Type a.}
These terms are independent of the momenta $r_l$ and thus the
derivatives in \eq{analyzthis2} do not act on them. These  terms
are considered in theorem \ref{ther2}. From theorem \ref{ther2} we have 
\be
p_i (\Theta^2) p_j \prod_{l=k_0}^I \beta_l^{-2}
\label{typec}
\ee

\noindent
{\bf Type b.}
Let $\gamma_{k_0}$ be the subgraph considered in theorem 2.
Consider the terms proportional to  $r_i$ in
\eq{analyzthis2}. 
In theorem \ref{rpterm} it will be shown that 
the leading term in the expansion of 
\be
\sum_j r_i Q_{ij} p_j
\ee
in $\beta$,
after the summation over 
all $j\in {\cal E}(G)$ 
and using the momentum conservation for $p_j,~~j\in {\cal E}(G)$,
is
 of the form
\be
\LB  \prod_{l=k_1}^I \beta_l^{-2}\RB
R_1(\beta_1,\beta_2,\ldots, \beta_{k_1-1})
\label{singular2}
\ee
for some $k_1 \ge k_0$,
where $R_1$ is a polynomial in variables $\beta_1,\beta_2,\ldots, 
\beta_{k_1-1}$.

Comparison of \eq{singular2} and \eq{typec} shows that type {\bf b} terms
in the exponent in \eq{analyzthis2} are harmless because $k_1 \ge k_0$.
Namely,  the derivatives acting on the exponent in
\eq{analyzthis2} will bring down  type {\bf b} terms and may, 
in principle, cause the divergence. But there will
always be (as long as the external momenta of $G$ are non-exceptional)
the type {\bf a} terms in the exponent which will suppress these potential 
divergences  because $k_1 \ge k_0$.
 Thus, type {\bf b} terms are harmless.

\noindent
{\bf Type c.}
Let $\gamma_{k_{{\bar\imath}{\bar\jmath}}}$ be the smallest
subgraph in \eq{nested} which contains both lines
${\bar \imath}\notin K_2$ and  ${\bar \jmath}\notin K_2$
in loops. 
In  lemma \ref{rsquare}  it  will  be  shown  that  the  singular
term  in  the  exponent  of \eq{analyzthis2}  is  of  the  form
\be
\LB r_i (\Theta^2) r_j 
\prod_{l=k_2}^I \beta_l^{-2}\RB
R_2(\beta_1,\beta_2,\ldots, \beta_{k_2-1})
\label{singular}
\ee
for  some  $k_2  \ge k_{{\bar\imath}{\bar\jmath}}$,
where  $R_2$  is  a  polynomial in variables $\beta_1,\beta_2,\ldots,
\beta_{k_2-1}$.

From \eq{singular} it follows that the derivatives acting on the $r^2$
terms in the
exponent  in \eq{analyzthis2} cannot bring down the terms which are
more  singular  than
\be
\prod_{l=1}^I
{1\ov \beta_l^{\#_l}}
\ee
where $\#_l$  is defined
as the following sum over lines of the subgraph $\gamma_l$:
\be
\#_l=\sum_{i\in \gamma_l} 
{\rm deg}(i)~{\rm ind}_{K_2}(i)
\ee
Theorem \ref{CTHEOR} follows from
the latter definition, the definition 5 for $\omega$,
\eq{analyzthis2} and the fact that type {\bf b} terms are harmless.

\pagebreak

\subsection{Lemmas and theorems}
In this section we prove nine lemmas and two theorems which are 
inter-related as follows:

$$
~~~~~~~{\rm Lemma}~\ref{rsquare}~~~~~~~~~~~~~~~{\rm Lemma}~\ref{newPr}~~~~~~~~~~~~~~~~~
$$
$$
\downarrow~~~~~~~~~~~~\nwarrow~~~~~~~~~\swarrow~~~~~~~~~~~~~~~~~~ 
$$
$$
{\rm Lemma}~\ref{K2}~
 \rightarrow ~{\rm Theorem}~\ref{CTHEOR}~\leftarrow {\rm Theorem}~4~\leftarrow {\rm
Lemma}~\ref{Xlemma2}~\leftarrow {\rm Lemma}~\ref{Xlemma1}
$$
$$
\uparrow~~~~~~~~~~~~\swarrow~~~~~~~~~\nwarrow~~~~~~~~~~~~~~~~~~ 
$$
$$
~~~~~~~~~~~~~~~~~~~~~~~~~~~~{\rm Lemma}~\ref{QO1}~~~~~~~~~~~~{\rm Lemma}~\ref{Prequal}~+~{\rm Lemma}~
\ref{SS}~+~{\rm Lemma}~\ref{S'}
$$
Lemma \ref{newPr} is a slight generalization of 
theorem 1 from section 6. Theorem 1 is used in 
lemma \ref{rsquare} and theorem 5.

\begin{lm}
 Consider the expression
\be 
Q=-\sum_{i,j} p_i^{\mu} \LB { P(G_{ij},\Theta | \alpha_{ij}=0 )
\ov P(G, \Theta)}\RB^{\mu\nu} p_j^{\nu}
\label{expresion}
\ee
Let ${\bar \imath}$ be an internal line of graph $G$ with ${\rm deg}({\bar
\imath})>0$. There are three types of terms in \eq{expresion}:
terms proportional to $r_{\bar \imath}^2$, terms
proportional to $r_{\bar \imath}$ and terms independent of $r_{\bar \imath}$.
If ${\bar \imath}\in K_2(G)$, then
$r_{\bar \imath}$-dependent
terms are of order $\O(1)$ in the $\beta$-expansion.
\label{K2}
\end{lm}

\noindent
{\bf Proof.}

Consider first $r_{\bar \imath}^2$ term.
Let
$i_1\in {\cal E}'(G)$ and $i_2\in {\cal E}'(G)$ be
the external lines carrying momenta $r_{\bar \imath}$ and $-r_{\bar
\imath}$,
respectively. If ${\bar \imath}\in K_2(G)$, then 
there exists a cycle supported only by ${\bar \imath}$ (i.e. there exists 
an element $a_0\in H_1(\Sigma_g(G))$ which is associated only to a loop containing 
the line ${\bar \imath}$). 
This implies that 
the joining of lines $i_1$ and $i_2$ will not introduce any extra 
cycle (i.e. the set of independent
 nontrivial cycles
of the surface $\Sigma(G)$ associated with the graph $G$ is the same
as that of the surface $\Sigma(G_{i_1i_2})$ 
associated with the graph $G_{i_1i_2}$).
Therefore,
the coefficient of $r_{\bar \imath}^2$ in  \eq{expresion} 
is of order $\O(1)$ in the $\beta$-expansion (See footnote  
\ref{ftbetalijeq1} for the discussion of this point).

Consider now term linear in $r_{\bar \imath}$. This term 
comes from
the joining  of $i_1$ or $i_2$ with some other external
line $j\in {\cal E}(G^+)$. Since ${\bar \imath}\in K_2(G)$, 
$g(G_{i_1j})=g(G)$ if and only if  $g(G_{i_2j})=g(G)$.
If $g(G_{i_1j})=g(G)$, then the corresponding term in  \eq{expresion} 
is of order $\O(1)$ in $\beta$-expansion. 
Let us therefore consider the case $g(G_{i_1j})=g(G)+1$. The
assumption ${\bar \imath}\in K_2(G)$ implies that 
$g(G_{i_2j})=g(G)+1$.
The corresponding terms  in \eq{expresion} are proportional to
\be
{1\ov P(G)}(P(G_{i_1j})-P(G_{i_2j}))
\ee
$P(G)$ is given by \eq{eq:sumtrees}.
 Let us analyze the difference
$$
\LB 
\sum\limits_{T,T'\in{\cal T}(G_{i_1j})}(-)^{T+T'} \det\e[T]\,\,
\det\e[{T'}]\det\A (G_{i_1j})[T^*| {T'}^*]
\RB
$$
\be
-
\LB
\sum\limits_{T,T'\in{\cal T}(G_{i_2j})} (-)^{T+T'} \det\e[T]\,\,
\det\e[{T'}]\det\A (G_{i_2j})[T^*| {T'}^*]\RB
\label{difference}
\ee
Let $l_*$ be the internal line of the graph $G_{i_1j}$ ( $G_{i_2j}$ )
formed by joining the external lines
$i_1$ and $j$  ($i_2$ and $j$) of $G$.  
For the simplicity  of formulas, 
let us assume that  $l_*=1$ and  ${\bar \imath}=2$.
The matrices  $\A(G_{i_1j})$ and $\A(G_{i_2j})$ in \eq{difference}
differ only in the matrix elements $\A_{12}=-\A_{21}$. 
It is easy to see that if the joining of $i_1$  with $j$ creates the
intersection of lines 1 and 2 in  $G_{i_1j}$, 
, i.e. $I_{12}(G_{i_1j})=1$, then the lines 1 and 2 in  $G_{i_2j}$
do not intersect, i.e. $I_{12}(G_{i_2j})=0$. Similarly, we have: 
if $I_{12}(G_{i_2j})=1$,
then $I_{12}(G_{i_1j})=0$.
For simplicity we assume that
$I_{12}(G_{i_1j})=1$.

Consider now  \eq{difference}. Since the matrices $\A(G_{i_1j})$
and   $\A(G_{i_2j})$  differ only by the matrix elements $\A_{12}$ and
$\A_{21}$, all but the terms proportional to $I_{12}$ and $I_{12}^2$
cancel (because the lines $i_1$ and $i_2$ emerge from the same vertex
there is a term by term correspondence between the two sums in 
\eq{difference}). 
Using the Laplace expansion along
row 1 and column 1, it can be shown that
\bea
&&\det\A [T^*| {T'}^*](G_{i_1 j})-\det\A [T^*| {T'}^*](G_{i_2 j})\non
&&~~~=
\theta^2(I_{12}(G_{i_1j}))^2 \det 
\A [T^*,{\hat 1},{\hat 2}| {T'}^*,{\hat 1},{\hat 2}](G_{i_1 j})
 +\sum\limits_{n\ge 3} (-)^{n}\theta^2 I_{12}(G_{i_1j}) 
I_{1n}(G_{i_1j})\times \non 
&&~~~~~~\times [
\det \A [T^*,{\hat 1},{\hat 2}| {T'}^*,{\hat 1},{\hat n}](G_{i_1 j})
 +
\det \A [T^*,{\hat 1},{\hat n}| {T'}^*, {\hat 1},{\hat 2}](G_{i_1 j})]
\label{eq:difference2}
\eea
where  $\A [T^*, {\hat m}| {T'}^*, {\hat n}]
(G_{i_1 j})$ 
denotes
the matrix $\A [T^*| {T'}^*]$ with 
the rows and the columns corresponding to the lines 
$m$ and 
$n$, respectively, removed. 
The following argument shows that $(I_{12})^2$ term in \eq{eq:difference2}
is at most of the order $\theta^{2g(G)}$.  The matrix 
$\A [T^*| {T'}^*](G_{i_1 j})$ is
a sum of two matrices ( see \eq{ptwog} ). 
Any rank $2g(G_{i_1j})$ 
minor of the intersection matrix $I(G_{i_1j})$ contains rows and columns 
corresponding to lines 1 and 2. The reason is the following.
 The matrix $I(G_{i_1j})$ with
the rows 1,2 and columns 1,2 removed is the intersection matrix
$I(G-  {\bar \imath})$ for the graph $G-{\bar \imath}$. 
The rank of $I(G-  {\bar \imath})$ is $2g(G)-2$ since
 $g(G-{\bar \imath})=
g(G)-1.$\footnote{
The rosette (see ref.\cite{filk,we} for the definition) 
can be thought of as a set of curves on the Riemann 
surface $\Sigma_g(G)$, i.e. a set of 
maps $\phi_i:S^1\rightarrow \Sigma_g(G)$ with $i$ running from $1$
to the number of lines in the rosette. These maps can be decomposed 
in some orthonormal basis of the first homology group of $\Sigma_g(G)$:
$\phi_i=\sum\limits_{k=1}^{g} {\tilde a}_k \kappa_i^k +
\sum\limits_{k=1}^{g} {\tilde b}_k \rho_i^k$. 
The orthonormality is defined with respect to the scalar product given 
by the intersection form ($\langle {\tilde a}_i,\,
{\tilde b}_j\rangle=\delta_{ij}$,
$\langle {\tilde b}_i,\,{\tilde a}_j\rangle=-\delta_{ij}$, $\langle 
{\tilde a}_i,\,
{\tilde a}_j\rangle=0$, $\langle {\tilde b}_i,\,
{\tilde b}_j\rangle=0$).
Such a basis has, of course, $2g$ independent elements.
Because there are $2g$ independent elements in this vector space,
the intersection matrix $I_{ij}=\langle \phi_i,\,\phi_j\rangle$ has 
rank $2g$.

Let us assume that in some basis of $H_1(\Sigma_g(G))$ there is a {\it unique}
map $\phi_{i_0}$ which contains the basis element, say, 
${\tilde a}_{i_0}$. If we 
construct the intersection matrix $I'_{ij}=\langle 
\phi_i,\,\phi_j\rangle,\,i,j\ne i_0$
for the rosette without $\phi_{i_0}$ (i.e. for the graph $G-i_0$), it 
will {\it a priori} have rank $2g-1$ because there are $2g-1$
 independent 
 elements of the 
vector space buiding up the remaining maps $\phi_i$.
But $I'_{ij}$ is anti-symmetric. Thus its  rank is  $2(g-1)$.
\label{ftrk}}

The matrix 
$\A [T^*,{\hat 1},{\hat 2}| {T'}^*,{\hat 1},{\hat 2}](G_{i_1 j})$
in  \eq{eq:difference2}  does not contain rows and columns 
corresponding to lines 1 and 2. Thus the determinant of this matrix
is a polynomial of a degree not higher  than $\theta^{2g(G)-2}$.
Thus $(I_{12})^2$ term in \eq{eq:difference2}
is at most of the order $\theta^{2g(G)}$.
The same is true  for the second term in
\eq{eq:difference2}.

Thus we see that the largest power of $\theta$ in the difference
$P(G_{i_1j})-P(G_{i_2j})$ is $\theta^{2g(G)}$. Since 
$g(G_{i_1j})=g(G)+1$, we have
\be
P_{2g(G_{i_1j})}-P_{2g(G_{i_2j})}=0
\ee
Thus
$$
{1\ov P(G)} (P(G_{i_1j})-P(G_{i_2j}))
$$
is of order one in the $\beta$ expansion.~~~~~{\bf q.e.d.}

\noindent
{\bf Definition 8.}~{\it 
For a given permutation $\pi$ let us recursively define  the following
nested sets of lines of graph $G$
\be
\S_I\subseteq\S_{I-1}\subseteq\S_{I-2}\subseteq\dots \subseteq\S_2\subseteq\S_1
\ee
as follows.
\be
\S_I=\emptyset
\ee
and for $1\le i \le I-1$
\be
\S_{i}=\left\{
\begin{array}{cl}
\S_{i+1}\cup \{ \pi_{i+1}\}   & {\rm ~if~} r(G\setminus\S_{i+1}-\pi_{i+1})=r(G) \\
\S_{i+1}   &  {\rm ~if~} r(G\setminus\S_{i+1}-\pi_{i+1})<r(G)
\end{array}\right.
\label{eq:Sdef}
\ee
The complement  $\S_k^*$ of $\S_k$  is defined as follows.
\be
\S_I^* = \emptyset
\ee
for $k=I$,
and
\be
\S_k^*=\{\pi_{k+1},\pi_{k+2},\ldots,\pi_I\}\setminus\,\S_k
\ee
for $1\le k \le I-1$.
}

To illustrate the meaning of $\S_i$ and $\S_i^*$,   consider
the sequence of lines 
$$\pi_I,\pi_{I-1},\ldots ,\pi_{n_1},\ldots,\pi_{n_2},\ldots,\pi_{n_3},\ldots
\pi_1$$
with the following
properties:
\bea
&&\pi_I,\pi_{I-1},\ldots, \pi_{n_1+2} \in  K_2(G) ,  \non
&&\pi_{n_1+1}\notin K_2(G) , \non
&&\pi_{n_1},\pi_{n_1-1},\ldots, \pi_{n_2+2} \in K_2(G-\pi_{n_1+1}) ,\non
&&\pi_{n_2+1}\notin K_2(G-\pi_{n_1+1}),  \non
&&\pi_{n_2},\pi_{n_2-1},\ldots, \pi_{n_3+2} \in K_2(G\setminus \{\pi_{n_1+1},
\pi_{n_2+1}\}), \non
&&\pi_{n_3+1}\notin K_2(G \setminus \{\pi_{n_1+1},
\pi_{n_2+1}\}),  \non
&&~~~~~~~~ \vdots
\eea
Then the sets $\S_i$  read
\bea
&&\S_I = \S_{I-1} = \cdots = \S_{n_1+1} = \emptyset , \non
&&\S_{n_1}= \S_{n_1 -1} = \cdots = \S_{n_2+1} = \{ \pi_{n_1+1} \}, \non
&&\S_{n_2}= \S_{n_2-1 } = \cdots = \S_{n_3+1} = \{ \pi_{n_1+1}, 
\pi_{n_2+1}\},\non
&&~~~~\vdots
\label{sequenceS}
\eea
One can see from \eq{sequenceS} that $r(G\setminus \S_k )=r(G)$
for any $k$.   
The meaning of $\S_k^*$ is the following: if one removes
any  line 
$l\in \S_k^*$ from the graph $G\setminus\S_k$,
the genus of the resulting graph is lower than the genus 
of  $G$:  $r(G\setminus\S_k - l)< r(G)$.

The following lemma  is a slight generalization of theorem 1 from section 4.

\begin{lm}
\be
P(G,\theta)=\theta^r
\LB\prod\limits_{i=k+1}^I\beta_i^{2(L(G\setminus\S_i)-r(G))}\RB
\,P_r(G\setminus\S_k|
{\vec \beta}_{\S_k^*}=1)[1+\O(\beta)]
\label{eq:modexp}
\ee
where $r\equiv  r(G)$ and
 ${\vec \beta}_{\S_k^*}=1$ is a short-hand for $\beta_{l}=1 ,~\forall~l\in 
\S_k^*$.
\label{newPr}
\end{lm}

\noindent
{\bf Proof.}

The proof for \eq{eq:modexp} makes repeated use of  lemma \ref{lem2}. 
Let us explain in detail the step
from $G\setminus\S_k$ to  $G\setminus\S_{k-1}$. From lemma \ref{lem2} we have
\bea
P_r(G\setminus\S_k\!\!\!\!&|&\!\!\!\!{\vec \beta}_{\S_k^*}=1)=
\beta_{k}^{2(L(G\setminus\S_k)-r(G))}[1+\O(\beta)]\times\\
&\times&\left\{\begin{array}{cl}
P_r(G\setminus\S_k-\pi_k|{\vec\beta}_{\S_k^*}=1) & {\rm ~if~}
 r(G\setminus\S_k-\pi_k)=r(G)\\
P_r(G\setminus\S_k|{\vec\beta}_{\S_k^*}=1,\,\beta_{\pi_k}=1)& {\rm ~if~} 
r(G\setminus\S_k-\pi_k)<r(G)
\end{array}\right.\nonumber
\eea
But according to our definition of $\S_i$ and $\S_i^*$ we have:
\be
\S_{k-1}=\S_k\cup \{ \pi_k\} ,~~~\S_{k-1}^*=\S_k^*~~~{\rm if}~r(G\setminus\S_k-\pi_k)=r(G)
\ee
and 
\be
\S_{k-1}=\S_k ,~~~\S_{k-1}^*=\S_k^*\cup 
\{ \pi_k\}~~~{\rm if}~r(G\setminus\S_k-\pi_k)<r(G)
\ee
Thus we see that 
\be
P_r(G\setminus\S_k|{\vec\beta}_{\S_k^*}=1)=\beta_{k}^{2(L(G\setminus\S_k)
-r(G))}P_r(G\setminus\S_{k-1}|
{\vec\beta}_{\S_{k-1}^*}=1)
[1+\O(\beta)]
\ee
\noindent
{\bf q.e.d.}


Lemmas \ref{Xlemma1} and  \ref{Xlemma2} given below
are direct analogs of lemmas \ref{lem1} and  \ref{lem2}, respectively.

\begin{lm}
 Let $M$ be a set of lines of graph $G$. Let $k\notin M$
be a line of $G$. Then
\be
X(G,\theta|M) = \left\{
\begin{array}{cl}
X(G-k,\theta|M) & {\rm ~if~}L(G-k)=L(G) \\
 \alpha_k\,X(G-k,\theta|M) + X(G|M\cup \{ k\} ) &{\rm ~if~}L(G-k)=L(G)-1
\end{array} \right.
\label{eq:prop1X}
\ee
where $X(G|M\cup \{ k\})$ does not depend on $\alpha_k$.
\label{Xlemma1}
\end{lm}

\noindent
{\bf Proof.}

The proof is a repetition of the proof of lemma \ref{lem1} with the 
matrix ${\cal A}$ replaced by the matrix
${\cal A}|_{\a_l=0,~\forall l\in M}$.~~~~{\bf q.e.d.}

\begin{lm}
 Let $M$ be a set of lines of graph $G$. Let $k\notin M$
be a line of $G$. Consider the $\theta^{r(G)}$ term in the expansion
\be
 X(G,\theta | M)=\sum_{n=0}^{r(G)/2} \theta_i^{2n} X_{2n}(G|M)
\ee
In the expression for $X(G|M)$ let us rescale all Schwinger parameters
except the one associated to the line $k$. Then the following
relations hold:

\noindent
(1) If $L(G-k)=L(G)$, then
\be
X_r(\alpha_k,\rho(G-k)|M)=\rho^{L(G)-r(G)} X_r(G|M) 
\label{eq:reloneX}
\ee

\noindent
(2) If $L(G-k) \ne L(G)$ and  $r(G-k) < r(G)$, then \eq{eq:reloneX} holds.

\noindent
(3) If $L(G-k) \ne L(G)$ and $r(G-k) = r(G)$, then
\be
X_r(\alpha_k, \rho(G-k)|M)=\rho^{L(G-k)-r(G)}
\left[\alpha_k X_r(G-k|M)+\O(\rho)\right]
\label{eq:reltwoX}
\ee
\label{Xlemma2}
\end{lm}

\noindent
{\bf Proof.}

The proof is completely parallel to that of lemma \ref{lem2}.~~~{\bf q.e.d.}

\begin{lm}
Let $G_1$ and $G_2$ be two graphs.
Let $l_{1}\in G_1$ and $l_{2}\in G_2$
be lines, such 
that $G_1-l_{1}$ and $G_2-l_{2}$ 
differ only by the lines 
belonging to all trees of $G_1-l_{1}$ or $G_2-l_{2}$, and 
$r(G_i)=2+r(G_i-l_{i})$. Furthermore, assume that 
$I_{l_{1}n}(G_1)-I_{l_{2}n}(G_2)\not=0$  only if
$n\in K_2(G_1)\cap K_2(G_2)$.

\noindent
Then:

1) $P_r(G_1)=P_r(G_2)$ if there are no lines $n_i\in K_2(G_i)$
such that $r(G_i-l_{i})=r(G_i-l_{i}-n_i)$ and $L(G_i-l_{i})=L(G_i-l_{i}
-n_i)+1$.

2)  The relation $P_r(G_1)=P_r(G_2)$ still holds, even if 
there is a line $n_0\in K_2(G_1)\cap K_2(G_2)$ with the properties:
$r(G_i-l_{i})=r(G_i-l_{i}-n_0)$, $L(G_i-l_{i})=L(G_i-l_{i}-n_i)+1$ 
and $I_{l_{1}n_0}(G_1)=I_{l_{2}n_0}
(G_2)$. 
\label{Prequal}
\end{lm}

\noindent
{\bf Proof.}

Recall that 
\be
P(G_i,\theta)=\sum\limits_{T,T'\in{\cal T}(G_i)}(-)^{T+T'} 
\det \epsilon [T] \det\epsilon [{T'}]
\det \A[T^*| {T'}^*]
\ee

Let us  show that the contribution to $P_r(G_i)$ comes from trees such that
$\A[T^*|{T'}^*]$ contains the rows and columns corresponding 
to all the lines in $K_2(G_i)$. 
Clearly, the intersection matrix with the $n$-th row 
and column removed corresponds to the intersection matrix of the graph obtained
from the original one by removing the corresponding line $n$. From the fact that by 
removing any line from $K_2(G_i)$ the genus decreases, we infer that the rank of the 
intersection matrix decreases by removing both the row and column associated to any
line in $K_2(G_i)$ (see also footnote \ref{ftrk}). Thus, if both the row and the column 
associated to a
line in $K_2(G_i)$ are missing from $\A[{T^*| {T'}^*}]$, then the power
of $\theta$ will be less than $\theta^{r(G_i)}$ and the corresponding term will 
not contribute to $P_r(G_i)$.

Consider next the case when the row associated to a line $n$ in $K_2(G_i)$ is missing from
$\A[T^*|{T'}^*]$, but the column is present. In this situation we perform
the Laplace expansion following the  column corresponding to the line $n$. We get 
\be
\det \A[T^*|{T'}^*] =\theta\sum (-)^{a_m} I_{nm}\det 
\A[T^*,{\hat m}| {T'}^*,{\hat n}]
\ee
where $a_m$ is a number depending on $n, ~m,~T$ and $T'$.
From the assumption that the row associated to a line $n$ in $K_2(G_i)$ is missing from
$\A[T^*| {T'}^*]$ we see that both the row and the column associated 
to the line $n$ are missing from  $\A[T^*,{\hat m}| {T'}^*,{\hat n}]$.
Thus, the maximal power of $\theta$  that can come from such a term is $\theta^{r(G_i)-2}$.
Combining this with the overall $\theta$ we get a power less that $\theta^{r(G_i)}$
and therefore these terms will not contribute to $P_r(G_i)$.

Thus, we conclude that only the terms in which all rows and columns associated to lines 
in $K_2(G_i)$ are present in $\A[T^*| {T'}^*]$ can contribute to 
$P_r(G_i)$.

Let us now show that the lemma is valid.
From the assumption, elements that are different between the ${\cal A}$ matrices for the 
graphs
$G_i$ involve the intersection of the line $l_{i}$ and lines of the sets 
$K_2(G_i)$. 
Let us take a generic term, $\det{\cal A}[T^*| {T'}^*]$, 
in the expression for $P(G_i)$. Since we are looking at $P_r$, all rows and 
columns associated to lines in $K_2(G_i)$ must be present in ${\cal A}[T^*| {T'}^*]$, as 
shown above. This implies that neither $T$ nor $T'$ contain those lines. Such trees are 
trees of $G_i-K_2(G_i)$ and from the assumptions follows that there is a one to one 
correspondence between the set of these trees for $i=1$ and $i=2$.

For convenience, let us label by $1$ the line $l_{i}$ and
by $n$ running from $2$ to $n_i=card[K_2(G_i)]$ the other lines of 
$K_2(G_i)$. With this labeling, the difference between the matrices
$\A(G_i)$ involves the elements $\A_{1n}$ with 
$n$ running from $2$ to $n_i$.

We perform the Laplace expansion of our generic term, 
$\det{\cal A}[T^*|{T'}^*]$, following
the first row and the first column. What we get is the following:
\bea
\det{\cal A}[T^*| {T'}^*]&=&
\theta^2\sum\limits_{n, m=1}^{N_0} (-)^{a_1} I_{1m} I_{1n}
\det{\cal A}[T^*,{\hat 1},{\hat m}| {T'}^*,{\hat 1}, {\hat n}]\nonumber\\
+\theta^2\sum\limits_{n=1}^{N_0}\!\!\!\!\!&{}&\!\!\!\!\!\sum\limits_{m=N_0+1}^I (-)^{a_2}
I_{1n} I_{1m}[\det{\cal A}[T^*,{\hat 1},{\hat n}| {T'}^*,{\hat 1},{\hat m}]
+\det{\cal A}[T^*,{\hat 1},{\hat m}| {T'}^*,{\hat 1},{\hat n}]]\nonumber\\
&+&
\theta^2\sum\limits_{n, m=N_0+1}^I (-)^{a_3} I_{1m} I_{1n}
\det{\cal A}[T^*,{\hat 1},{\hat n}| {T'}^*,{\hat 1},{\hat m}]
\label{eq:junkexp}
\eea
where $N_0=card[K_2(G_1)\cap K_2(G_2)]$,
$a_1$, $a_2$, $a_3$ are integer numbers depending on $m$, $n$, $T^*$ and  ${T'}^*$.
Notice that the last term on the r.h.s. of \eq{eq:junkexp}
is the same 
 for both $i=1$ and $i=2$ since 
it does not involve the intersection of $l_{i}$ with lines in 
$K_2(G_1)\cap K_2(G_2)$ (by assumption we have  
$I_{l_{1}n}(G_1)-I_{l_{2}n}(G_2)\not=0$  only if
$n\in K_2(G_1)\cap K_2(G_2)$).

Now we distinguish the 2 cases stated in the beginning:

\noindent
1) any line $n\in K_2(G_i)$ is such that the genus of the graph obtained 
by removing it together with  the line $l_{i}$  equals the genus of the original 
graph $G_i$
minus 2. 

\noindent
2) there is a line $n_0\in K_2(G_1)\cap K_2(G_2)$ such that the genus of the graph 
obtained 
by removing it together with  the line $l_{i}$  equals the genus of the original 
graph $G_i$ minus 1.

Let us consider case 1. By assumption we have that
\be
\det{\cal A}[T^*,{\hat 1}, {\hat n}| {T'}^*,{\hat 1},{\hat m}]~~~~n,\,m=2,...,n_i
\ee
comes with at most $\theta^{2g(G_i)-4}$. Together with 
the $\theta^2$ coefficient, 
this makes less than the required $\theta^{2g(G_i)}$ for $P_r(G_{i})$. Thus, it does not 
contribute to $P_r(G_i)$.

Let us now look at the second term in \eq{eq:junkexp} for some
fixed $n$ between $1$ and $N_0$.
Such a term is:
\be
\theta^2 \sum\limits_{m=N_0+1}^I (-)^{a_2}
I_{1n} I_{1m}
\LB\det{\cal A}[T^*,{\hat 1},{\hat n}| {T'}^*,{\hat 1},{\hat m}]
+\det{\cal A}[T^*,{\hat 1},{\hat m}| {T'}^*,{\hat 1},{\hat n}]\RB
\label{eq:ssss}
\ee
Now we perform the Laplace expansion for each of the 2 pieces following
the row or column $n$, as appropriate. This operation gives:
\be
\theta^3 \sum\limits_{m=N_0+1}^I\sum\limits_{k=1}^I (-)^{a_2+a_4}
I_{1n} I_{1m}I_{n k}
\LB\det{\cal A}[T^*,{\hat 1},{\hat n},{\hat k}| {T'}^*,{\hat 1},{\hat m},{\hat n}]
+({\hat m}\longleftrightarrow{\hat k}) \RB
\label{eq:secondLaplaceexp}
\ee 
where $a_4$ depends on $k, m, n, T^*, {T'}^*$.
Each of these minors are otained from the original matrix by removing at least the first
and $n$-th row and column. But by assumption such a 
determinant brings at most $\theta^{2g(G_i)-4}$. Together with the prefactor
this makes less than $\theta^{2g(G_i)}$ required for $P_r(G_i)$ 
and thus it will not 
contribute. 

We are therefore left with:
\be
\det{\cal A}[T^*| {T'}^*]=
\theta^2\sum\limits_{n, m=N_0+1}^I (-)^{a_3} I_{1m} I_{1n}
\det{\cal A}[T^*,{\hat 1},{\hat m}| {T'}^*,{\hat 1},{\hat n}]
\ee
Since the sum does not run over the intersections between $l_{i}$ 
and $K_2(G_i)$, these terms are present for both $i=1$ and $i=2$..

Consider now case 2.
As in case 1, the last term in \eq{eq:junkexp} is the same for 
both $P_r(G_1)$, but we can get nontrivial contributions to $P_r$ from the first two terms. 
In particular, we have the terms:
\be
\det{\cal A}[T^*{\hat 1}, {\hat n_0}| {T'}^*,{\hat 1},{\hat n_0}]
\ee
\be
\det{\cal A}[T^*{\hat 1}, {\hat n_0}| {T'}^*,{\hat 1},{\hat k}],~~k=2,..., n_i,~~k\ne n_0
\ee
\be
\det{\cal A}[T^*{\hat 1}, {\hat n_0}| {T'}^*,{\hat 1},{\hat m}],~~m=n_i+1,...,I
\ee
But, because $n_0\in K_2(G_1)\cap K_2(G_2)$ and 
$I_{l_{1}n_0}(G_1)=I_{l_{2}n_0}(G_2)$ the first and the last term are present both 
in $P_r(G_1)$ and $P_r(G_2)$. For the second term, by performing the Laplace expansion 
as in \eq{eq:secondLaplaceexp} we find that it does not contribute 
to $P_r(G_i)$.

We conclude therefore that the second statement of the lemma holds as well.~~
{\bf q.e.d.}

\begin{lm}
Let $\gamma_{k_0}$ be the smallest subgraph
of $G$ that has $j(\gamma_{k_0})=1$ (let $i_0$ be the external line
such that $c_{G_{i_0 j}}(\gamma_{k_0})> c(\gamma_{k_0})$) and let $\gamma_{k_r}$ be 
the smallest subgraph
of $G$ that has $c_{G_{\pm rj}}(\gamma_{k_r})> c_G(\gamma_{k_r})$.
{
We assume that $\gamma \subset \gamma_{k_0}$.
}

Then, $G_{rj}\setminus\S_k(G_{rj}),~\forall~j=1,...,E$
satisfy the conditions of lemma \ref{Prequal} for $k<k_0$.
\label{SS}
\end{lm}

\noindent
{\bf Proof.}

Let $i_0$ be the external line
such that $c_{G_{i_0 j}}(\gamma_{k_0})> c(\gamma_{k_0})$.
The line $l_{j}$ in lemma \ref{Prequal} is the line joining (the line carrying momentum) 
$r$ and external line $j$. 
If $k<k_0$ then $j(G_{rj}\setminus\S_k(G_{rj})\setminus\S^*_k(G_{rj}))=0$. 
By construction, removing any one of the lines in 
${\S}^*_{k}(G_{rj})$ changes the genus of $G_{rj}\setminus\S_k(G_{rj})$
Therefore, all the lines in ${\S}^*_{k}(G_{rj})$ belong 
to $K_2(G_{rj}\setminus\S_k(G_{rj}))$.

\noindent
{\bf I.}~~$G_{rj}\setminus\S_k(G_{rj})-l_{j}$ are identical up to lines belonging to all 
trees of $G_{rj}\setminus\S_k(G_{rj})-l_{j}$.

Proof by induction:

1) By assumption, $G_{rj}-l_{j}=G$ are identical. Thus, the first step in the induction 
needs no further proof.

2) Prove $k\Rightarrow k-1$, i.e. assume that $G_{rj}\setminus\S_k(G_{rj})-
l_{j}$ are 
identical 
up to lines belonging to all trees of $G_{rj}\setminus\S_k(G_{rj})-l_{j}$
and prove the same for $G_{rj}\setminus\S_{k-1}(G_{rj})-l_{j}$. For the general 
implication, 
we distinguish two possible mutually exclusive situations:

{\bf a)} the line $\pi_k$ in the definition of $\S_{k-1}(G_{rj})$
belongs to all the trees of  $G_{rj}\setminus\S_k(G_{rj})-l_{j}$

{\bf b)} the line $\pi_k$ in the definition of $\S_{k-1}(G_{rj})$
belongs to a loop in  $G_{rj}\setminus\S_k(G_{rj})-l_{j}$

\noindent
These two cases are mutually exclusive because, by assumption, all the 
graphs $G_{rj}\setminus\S_k(G_{rj})-l_{j}$ 
are identical up to lines belonging to all trees of 
$G_{rj}\setminus\S_k(G_{rj})-l_{j}$ which implies 
that we cannot have case {\bf a} for some $j$-s and case {\bf b} for the rest. 

For each of these situations we distinguish two subcases. 

For the case {\bf a}, the line 
$\pi_k$ can belong, for some $j$-s, 
to all the trees of $G_{rj}\setminus\S_k(G_{rj})$, 
while for the rest of $j$-s the line $\pi_k$ belongs to
all the trees of $G_{rj}\setminus\S_k(G_{rj})- l_{j}$.
In the first subcase $\S_{k-1}(G_{rj})=\S_{k}(G_{rj})\cup\{\pi_k\}$ and 
$\S^*_{k-1}(G_{rj})=\S^*_{k}(G_{rj})$, while in the second subcase 
$\S_{k-1}(G_{rj})=\S_{k}(G_{rj})$ and 
$\S^*_{k-1}(G_{rj})=\S^*_{k}(G_{rj})\cup\{\pi_k\}$. 
It follows then 
that in the first subcase $\pi_{k}$ does not appear in 
$G_{rj}\setminus\S_{k-1}(G_{rj})$, while the 
second  subcase $\pi_{k}$ has the property that $l_{j}$ does not belong to a loop in 
$G_{rj}\setminus\S_{k-1}(G_{rj})-\pi_k$ which implies that $\pi_{k}$ does not 
belong to a loop in 
$G_{rj}\setminus\S_{k-1}(G_{rj})-l_{j}$. Thus, we 
get that $G_{rj}\setminus\S_{k-1}(G_{rj})-l_{j}$
differ only by lines belonging to all trees of 
$G_{rj}\setminus\S_{k-1}(G_{rj})-l_{j}$.

For the case {\bf b}, the line  $\pi_k$ can belong either to $\S^*_{k-1}(G_{rj})$
or to $\S_{k-1}(G_{rj})$. 

In the first subcase, since $\pi_{k}$ belongs to a loop 
in $G_{rj}\setminus\S_{k}(G_{rj})-l_{j}$ it follows that  
$G_{rj}\setminus\S_{k}(G_{rj})-\pi_{k}$ contains $l_{j}$ in a loop. The fact 
that $\pi_{k-1}\in \S^*_{k-1}(G_{rj})$ (i.e. by removing $\pi_{k}$ an 
independent cycle on 
$\Sigma(G_{rj})$ is not associated to any loop in 
$G_{rj}\setminus\S_{k}(G_{rj})-\pi_{k}$) 
implies that $\pi_{k}$ belongs to a loop in $G_{rj}\setminus\S_k(G_{rj})$. From the 
assumption that 
$G_{rj}\setminus\S_k(G_{rj})-l_{j}$ are identical 
up to lines belonging to all trees of $G_{rj}\setminus\S_k(G_{rj})-l_{j}$ 
follows that
the line  $\pi_{k}$ belongs to a loop in $G_{rj}\setminus\S_k(G_{rj})$ for any $j$. 
From 
the addition 
of cycles now follows that $l_{j}$ belongs to a loop in  
$G_{rj}\setminus\S_{k-1}(G_{rj})-\pi_{k}$  for 
any $j$.\footnote{Recall that $\pi_{k-1}\in \S^*_{k-1}(G_{rj})$
also implies that $\S_{k}(G_{rj})=\S_{k-1}(G_{rj})$} Thus, we get 
that $G_{rj}\setminus\S_{k-1}(G_{rj})-l_{j}$ are identical 
up to lines belonging to all trees of $G_{rj}\setminus\S_{k-1}(G_{rj})-l_{j}$ for 
any $j$.

Let us point out that the two subcases are again mutually exclusive because, 
as shown above, 
if the first subcase is realized,
then it is realized for all $j=1,...,E$.

In the second subcase, since $\pi_{k}\in \S_{k-1}(G_{rj})$, $\pi_k$ does not appear in
$G_{rj}\setminus\S_{k-1}(G_{rj})$. Furthermore, since the two subcases are mutually 
exclusive, this subcase is
either realized for all $j=1,...,E$ or is not realized at all. It follows therefore that 
in the case {\bf b}, $G_{rj}\setminus\S_{k-1}(G_{rj})-l_{j}$ differ only by lines 
belonging to all trees of
$G_{rj}\setminus\S_{k-1}(G_{rj})-l_{j}$, as well.

\noindent
{\bf II.}~~$I_{l_{1}n}(G_{rj_1}\setminus\S_k(G_{rj_1}))-I_{l_{2}n}(G_{rj_2}
\setminus\S_k(G_{rj_2}))\not=0$ only if
$n\in \cap_{j=1}^E K_2(G_{rj}\setminus\S_k(G_{rj}))$.

We can construct the intersection matrix such that the intersections
between $l_{j}$ and the lines of $\gamma_{k_0}$ are the same 
for any $j$.\footnote{
We take $I_{l_* n}(G_{rj})=I_{l_* n}(G_{ri_0})+I_{l_* n}(G_{i_0} j)$
}
From the fact that $j(G_{rj}\setminus\S_k(G_{rj})\setminus\S^*_k(G_{rj}))=0$ for 
any $k<k_0$ we see that 
the difference in the intersection matrices for different $j$-s 
involve the intersection of $l_{j}$ and other lines in $\S^*_k(G_{rj})$.
But, as we have shown before, all the lines in ${\S}^*_{k}(G_{rj})$ belong 
to $K_2(G_{rj}\setminus\S_k(G_{rj}))$. Furthermore, we have shown above that
$G_{rj}\setminus\S_k(G_{rj})-l_{j}$ are identical up to lines belonging to all 
trees of $G_{rj}\setminus\S_k(G_{rj})-l_{j}$. Because of equality of momenta
there is no intersection between two lines that belong to a single loop.
Thus we have that 
$I_{l_{1}n}(G_{rj_1}\setminus\S_k(G_{rj_1}))-
I_{l_{2}n}(G_{rj_2}\setminus\S_k(G_{rj_2}))\not=0$  only if
$n\in \cap_{j=1}^E K_2(G_{rj}\setminus\S_k(G_{rj}))$.

\noindent
{\bf III.}~~ If $\gamma_{k_r}\subset \gamma_k=G_{rj}\setminus\S_k(G_{rj})\setminus
\S^*_k(G_{rj})$
then any line in $n\in\S^*_k(G_{rj})$ is such that 
$r(G_{rj}\setminus\S_k(G_{rj})-l_{j})\ne
r(G_{rj}\setminus\S_k(G_{rj})-l_{j}-n)$ because the lines violating this 
condition belong to $\gamma_{k_r}$\footnote{in detail, this goes as follows: 
introducing $l_{j}$
adds 2 cycles to the graph $G$ (say $a$ and $b$). One of them (say $a$) is
disconnected 
by removing $l_{j}$.
From the fact that $\gamma_{k_r}\subset\gamma_k$ is such that 
$c_{G_{\pm rj}}(\gamma_{k_r})> c_G(\gamma_{k_r})$
follows that this 
subgraph wraps the 
other cycle ($b$). This implies that there is no line 
in $\S^*_{k}(G_{rj})$ supporting cycle $(b)$.}

\noindent
{\bf IV.}~~If $\gamma_k=G_{rj}\setminus\S_k(G_{rj})
\setminus\S^*_k(G_{rj})\subset\gamma$
then there is a line $n\in \S^*_k(G_{rj})$ such that 
$r(G_{rj}\setminus\S_k(G_{rj})-l_{j})=r(G_{rj}\setminus\S_k(G_{rj})-l_{j}-n)$.
But, as we said before, intersection matrix can be constructed 
such that the intersections between $l_{j}$ and the lines 
of $\gamma_{k_0}$ are the same for any $j$. Thus the second 
statement of  lemma \ref{Prequal} is satisfied.

This shows that the graphs $G_{rj}\setminus\S_k(G_{rj})$ for $k<k_0$ satisfy the 
assumptions of lemma \ref{Prequal}.~~{\bf q.e.d.}

\noindent
{\bf Definition 9.}{\it
Following the model of $\S$
described in Definition 8, we define the following nested sets:
\be
\Z_{k_0}\subset\Z_{k_0-1}\subset\dots\subset\Z_{2}\subset\Z_{1}
\ee
\be
\Z_{k_0}=\emptyset
\ee
\be
\Z_i=\left\{
\begin{array}{cl}
Z_{i+1}\cup \{\pi_{i+1}\}   &{\rm ~if}~\pi_{k_0}~{\rm does~not~belongs~to~a~loop~in}~G\setminus
(S_{i+1}\setminus
\Z_{i+1})-\pi_{i+1}-l\\
Z_{i+1}   &  {\rm ~if~}~\pi_{k_0}~{\rm belongs~to~a~loop~in}~G\setminus(S_{i+1}\setminus 
\Z_{i+1})-\pi_{i+1}-l
\end{array}\right.
\label{eq:Zdef}
\ee
where $l$ is some fixed line in $G$.

We define also
\be
\S'_i=\S_i\setminus Z_i~~~~~~~~~~{\S'}_i^*=\S_i^*\setminus \Z_i
\ee
where $\S_i$ and $\S^*_i$ were defined before.}

\noindent
{\bf Remark.}~
From  the  above  definition  it  follows  that 
$\S'_i\subseteq \S_i$ with equality for all $i\ge k_0$ and that $\Z_{k_0-1}=\{\pi_{k_0}\}$.

Now we would like to prove an analog of lemma \ref{SS} for the sets $\S'$. The 
proof is similar 
to that of lemma \ref{SS} so we will not go into much details. However, there is an important 
difference which will be pointed out.

\begin{lm}
Let $\gamma_{k_0}$ be the smallest subgraph
of $G$ that has $j(\gamma_{k_0})=1$ (let $i_0$ be the external line
such that $c_{G_{i_0 j}}(\gamma_{k_0})> c(\gamma_{k_0})$) and let $\gamma_{k_r}$ be the 
smallest subgraph
of $G$ that has $c_{G_{\pm rj}}(\gamma_{k_r})> c_G(\gamma_{k_r})$.
{
We assume that $\gamma \subset \gamma_{k_0}$.
}

Then, $G_{rj}\setminus\S'_k(G_{rj})-\pi_{k_0},~\forall~j=1,...,E$
satisfy the conditions of lemma \ref{Prequal} for $k<k_0$ where in the definition of the sets $\Z$
the fixed line $l$ is chosen to be $l_{j}$.
\label{S'}
\end{lm}

\noindent
{\bf Proof.}

\noindent
{\bf I.}~~$G_{rj}\setminus\S'_k(G_{rj})-\pi_{s}-l_{j}$ are identical up to lines 
belonging to all 
trees of $G_{rj}\setminus\S'_k(G_{rj})-\pi_{s}-l_{j}$ for any $\pi_s\in \Z_k$.

The proof, just like for lemma \ref{SS}, is by induction. 

The first step follows immediately from lemma \ref{SS}. More precisely, using the remark that 
$\Z_{k_0-1}(G_{rj})=\{\pi_{k_0}\}$ follows that 
$G_{rj}\setminus\S'_{k_0-1}(G_{rj})-\pi_{k_0}=G_{rj}\setminus(\S_{k_0-1}(G_{rj})-\pi_{k_0})-\pi_{k_0}=
G_{rj}\setminus\S_{k_0-1}(G_{rj})$ and the last graph satisfies the desired relation as 
implied by lemma \ref{SS}.

For the general implication we start by making two simple observations:

\noindent
{\bf 1.}~we notice that, since by definition the elements of the set $\Z_k$ have the 
property that by removing any one of them, $\pi_l\in \Z_k$, the line $\pi_{k_0}$ does not 
belong to a loop in $G_{rj}\setminus\S'_k(G_{rj})-\pi_l$ it follows that 
in $G_{rj}\setminus\S'_k(G_{rj})-\pi_{l}$ 
none of the elements of $\Z_k$ belongs to a loop. 

\noindent
{\bf 2.}~we notice that if $\pi_k\in \Z_{k-1}$ then $\pi_k$ necessarily belongs to a loop in 
$G_{rj}\setminus\S'_k(G_{rj})$. 

With these two observations we follow the analysis in lemma \ref{SS} and assume that 
$G_{rj}\setminus\S'_k(G_{rj})-l_{j}$ differ only by lines belonging to all trees of 
$G_{rj}\setminus\S'_k(G_{rj})-l_{j}$ and show the same property for 
$G_{rj}\setminus\S'_{k-1}(G_{rj})-l_{j}$.
We distinguish two main situations:

\noindent
{\bf A)} $\pi_k\notin \Z_{k-1}$

\noindent
{\bf B)} $\pi_k\in \Z_{k-1}$

In case {\bf A} the analysis is identical to the one in lemma \ref{SS} and we 
will not repeat it.

In case {\bf B} we again distinguish the two situations from lemma \ref{SS}:

{\bf a)} the line $\pi_k$ belongs to all the trees of  
$G_{rj}\setminus\S'_k(G_{rj})-l_{j}$

{\bf b)} the line $\pi_k$ belongs to a loop in  
$G_{rj}\setminus\S'_k(G_{rj})-l_{j}$

Unlike lemma \ref{SS}, case {\bf a} has no subcases due to the observations in the 
beginning. 
In particular, $\pi_k$ cannot belong to all trees of 
$G_{rj}\setminus\S'_k(G_{rj})$ 
because in $G_{rj}\setminus\S'_k(G_{rj})$ the line $\pi_{k_0}$ belongs to a 
loop while in $G_{rj}\setminus\S'_k(G_{rj})-\pi_k$
it does not.

In the case {\bf b} we have the same two subcases as in lemma \ref{SS}: the 
line  $\pi_k$ can 
belong either to $\S^*_{k-1}(G_{rj})$ or to $\S_{k-1}(G_{rj})$. The analysis 
is similar and we will not repeat it.

Collecting everything we see that we have proven the desired result. Another 
important outcome of this 
discussion is that if  $\pi_k\in \Z_{k-1}$ there are three mutually exclusive 
possibilities (case {\bf a} and 
the two subcases of case {\bf b}). This fact, that
if one of them is realized for some $j$ then it is realized for all $j=1, ..., E$,
will be important in theorem \ref{rpterm}.

\noindent
{\bf II.}~~$I_{l_{1}n}(G_{rj_1}\setminus\S'_k(G_{rj_1})-\pi_s)-
I_{l_{2}n}(G_{rj_2}\setminus\S'_k(G_{rj_2})-\pi_s)\not=0$ only if 
$n\in \cap_{j=1}^E K_2(G_{rj}\setminus\S'_k(G_{rj}))$ where $\pi_s\in \Z_k(G_{rj})$.

This relation follows from the observation in the beginning that in 
$G_{rj}\setminus\S'_k(G_{rj})-\pi_s$
none of the lines in $\Z_k$ belongs to a loop. Thus, as far as the 
intersection matrix is concerned,
this graph is equivalent to $G_{rj}\setminus\S_k(G_{rj})$. This latter 
graph has the desires property
as shown in lemma \ref{SS}.

Points {\bf III.} and {\bf IV.} are identical to those in lemma \ref{SS} and 
we will not repeat 
the arguments.~~~{\bf q.e.d.}

\begin{th} 
Let $\gamma_{k_0}$ be the subgraph considered in theorem 2.
The leading term in the expansion of 
\be
\sum_j r_i Q_{{i}j} p_j
\label{eq:sumexp}
\ee
in $\beta$,
after the summation over 
all $j\in {\cal E}(G)$ 
and using the momentum conservation for $p_j,~~j\in {\cal E}(G)$,
is
 of the form
\be
\LB  \prod_{l=k_1}^I \beta_l^{-2}\RB
R_1(\beta_1,\beta_2,\ldots, \beta_{k_1-1})
\label{rpt}
\ee
for some $k_1 \ge k_0$,
where $R_1$ is a polynomial in variables $\beta_1,\beta_2,\ldots, 
\beta_{k_1-1}$.
\label{rpterm}
\end{th}

\noindent
{\bf Proof.}

Let us begin by pointing out that if $j(G)=1$ and $r(G_{\pm r j})=r(G)$ for all 
$j=1, ..., E$, 
then the theorem is trivially satisfied because each term in the sum in 
\eq{eq:sumexp}
is of order $\beta^0$. We  thus  assume that $r(G_{\pm r j})>r(G)$.

As in theorem 2, let $\gamma_{k_0}$ be the smallest subgraph of $G$ from \eq{nested}
that has $j(\gamma_{k_0})=1$. 
This implies  that if we join any two external lines $i, j$ from 
${\cal E}(G)$, we have 
$c_{G_{ij}}(\gamma_{k_0}-\pi_{k_0})= c(\gamma_{k_0}-\pi_{k_0})$. 
Using the decomposition in lemma \ref{lem4}, let ${i_0}$ be a line in 
$\E_1(\gamma_{k_0})$ such that 
$c_{G_{i_0 j}}(\gamma_{k_0}) > c(\gamma_{k_0})$ for all $j\in \E_2(\gamma_{k_0})$.

Consider now a pair of lines $i_1,i_2\in {\cal E}'(G)$ attached
to a line ${\bar \imath}$ with $deg({\bar \imath})>0$.
Let $\gamma_{k_r}$ be the smallest subgraph from \eq{nested}
for which $c_{G_{i' j}}(\gamma_{k_r})> c(\gamma_s)$ or 
$c_{G_{i'' j}}(\gamma_{k_r})> c(\gamma_s)$.
Let us assume, for concreteness, that $c_{G_{i' j}}(\gamma_{k_r})> c(\gamma_s)$ and that 
the momentum carried by the line $i'$ is $r$. To emphasize this, let us denote the graph
$G_{i' j}$ by $G_{r j}$. There are two possibilities:

\noindent
(1) $\gamma_{k_0} \subseteq  \gamma_{k_r}$  ( $k_0 \le k_r$ )

\noindent
(2) $\gamma_{k_r} \subset \gamma_{k_0}$ ( $ k_r < k_0$ )

In the first case it is easy to see that 
$$
{P(G_{\pm r j})\ov P(G)}=\prod\limits_{n=k_r}^I{1\ov\beta_n^2} ~~~~~~~
{P(G_{i_0 j})\ov P(G)}=\prod\limits_{n=k_0}^I{1\ov\beta_n^2}~~i_0\ne j
$$
while from the assumption that ${\gamma}_{k_0}\subset \gamma_{k_r}$ we see 
that $k_0\le k_r$.
Therefore, the term $p_{i_0}p_{j}$, $i_0\ne j$ which exists in the exponent will regulate
whatever we bring down from $r_i p_j$.

Imagine now that the lines carrying momenta $\pm r$ lie in the interior of 
$\gamma_{k_0}$ (i.e.
$\gamma_{k_r} \subset \gamma_{k_0}$).
For each one of them, say the one carrying momentum $+r$, we have the 
following contribution
to the $r\cdot p$ term:
\be
Q(+r,p)={r\cdot\sum\limits_{i=1}^E P(G_{r j})~p_j\ov P(G)}~~.
\ee

Using lemma \ref{newPr} (\eq{eq:modexp}) 
we can identify the parts that cancel in the $\beta$ 
expansion of
the $r\cdot p$ terms. Let us write it for $k=k_0$ and insert it in the previous equation. 
We get:
\be
Q(+r,p)=r\cdot\left(\prod\limits_{i=k_0+1}^I{1\ov \beta_i^2}\right)\sum\limits_{j=1}^E
{P_r(G_{rj}\setminus\S_{k_0}(G_{rj})|{\vec \beta}_{\S^*_{k_0}(G_{rj})}=1)
          \ov   
P_r(G\setminus\S_{k_0}(G)|{\vec \beta}_{\S^*_{k_0}(G)}=1)} p_j [1+\O(\beta)]
\ee
As explained in footnote \ref{ftbetalijeq1}, the permutation used for the graphs $G_{rj}$
are constrained to have the line joining $r$ and $j$ as $\pi_{I+1}$ and $\beta_{I+1}$ is 
set to $1$. 
We can now write the numerator for each term in the sum using 
lemmas \ref{lem1} and
 \ref{lem2},  to get
\bea
P_r(G_{rj}\setminus\S_{k_0}(G_{rj})|{\vec \beta}_{\S^*_{k_0}(G_{rj})}=1)\!\!\!&=&\!\!\!
\beta_{k_0}^{2(L(G_{rj}\setminus\S_{k_0}(G_{rj}))-r(G))}\times\\
\Big(P_r(G_{rj}\setminus\S_{k_0}(G_{rj})-\pi_{k_0}|{\vec \beta}_{\S^*_{k_0}(G_{rj})}=1)
\!\!\!\!&+&\!\!\!\!
X_r(G_{rj}\setminus\S_{k_0}(G_{rj})|{\vec \beta}_{\S^*_{k_0}(G_{rj})}=1|\pi_{k_0})\Big)
\nonumber
\eea
We remind the reader that, as introduced in \eq{eq:defX}, the notation 
$X_r(G_{rj}\setminus\S_{k_0}(G_{rj})|{\vec \beta}_{\S^*_{k_0}(G_{rj})}=1|\pi_{k_0})$ 
is defined to be 
$P_r(G_{rj}\setminus\S_{k_0}(G_{rj})|
{\vec \beta}_{\S^*_{k_0}(G_{rj})}=1)$ with 
$\alpha_{\pi_{k_0}}=0$.
Using lemmas \ref{Prequal} and \ref{SS} we have that 
\be
P_r(G_{rj_1}\setminus\S_{k_0}(G_{rj_1})-\pi_{k_0}|{\vec \beta}_{\S^*_{k_0}(G_{rj_1})}=1)=
P_r(G_{rj_2}\setminus\S_{k_0}(G_{rj_2})-\pi_{k_0}|{\vec \beta}_{\S^*_{k_0}(G_{rj_2})}=1)
\label{eq:prequalpr}
\ee
for all $j_1,\,j_2=1,\dots,E$.
Therefore, they all cancel\footnote{Let us point out that this cancelation may occur
even before reaching the line $\pi_{k_0}$. Here we treat the most singular scenario.} by 
momentum conservation:
\be
\sum\limits_{j=1}^E p_j=0
\label{eq:momc}
\ee
while the surviving terms, using the quantities introduced in Definition 9 and the remark 
following it, are 
\be
Q(+r,p)=r\cdot\left(\prod\limits_{i=k_0}^I{1\ov \beta_i^2}\right)\sum\limits_{j=1}^E
{X_r(G_{rj}\setminus\S'_{k_0-1}(G_{rj})|{\vec \beta}_{{\S'}^*_{k_0-1}(G_{rj})}=1|\Z_{k_0-1})
          \ov   
P_r(G\setminus\S_{k_0-1}(G)|{\vec \beta}_{\S^*_{k_0-1}(G)}=1)} p_j [1+{\cal O}(\beta)]
\label{eq:rat}
\ee

Thus, we need an expansion of 
$X_r(G_{rj}\setminus\S'_{k_0-1}(G_{rj})|{\vec \beta}_{{\S'}^*_{k_0-1}(G_{rj})}=1|\Z_{k_0-1})$. 
It turns out that it is more useful to analyze the whole sum in 
the numerator of \eq{eq:rat}. As it has became
obvious from \eq{eq:prequalpr} and \eq{eq:momc}, for $k<k_0$ the 
leading term 
in the expansion of the numerator of \eq{eq:rat}
does not come from $P_r(G_{rj}\setminus\S_{k}(G_{rj}))$ because, according 
to lemmas \ref{Prequal} and \ref{SS}, such terms cancel when the sum over momenta is 
performed. 
Furthermore, lemmas \ref{Prequal} and \ref{S'} implies that the leading term 
comes from graphs that contain the line $\pi_{k_0}$ in a loop (if $\pi_{k_0}$ does not 
belong to  
a loop in $G_{rj}\setminus\S'_{s}(G_{rj})-\pi_s$, then $X_r$ can be interpreted as 
a $P_r$ and from
lemma \ref{Prequal} they cancel by momentum conservation.).

Let us now find the leading term 
in the expansion of the numerator of \eq{eq:rat}. The claim is that it is 
given by
\bea
\sum\limits_{j=1}^E
&{}&\!\!\!\!X_r(G_{rj}\setminus\S'_{k_0-1}(G_{rj})|{\vec \beta}_{{\S'}^*_{k_0-1}(G_{rj})}=1|
\Z_{k_0-1})p_j=
\prod\limits_{i=s+1}^{k_0-1}\beta_i^{2[L(G_{rj}\setminus\S'_{i}(G_{rj}))-r(G_{rj})]}\,
\times\nonumber\\
&\times&\!\!\!\!
\sum\limits_{j=1}^E 
X_r(G_{rj}\setminus\S'_{s}(G_{rj}))|{\vec \beta}_{{\S'}^*_{s}(G_{rj})}=1|
\Z_s)p_j [1+{\cal O}(\beta)]~~.
\eea
Here we can take $\beta_i^{2[L(G_{rj}\setminus\S'_{i}(G_{rj}))-r(G_{rj})]}$
out of the sum because both $r(G_{rj})=r(G)+2$ 
and $L(G_{rj}\setminus\S'_{i}(G_{rj}))$ are the same for all $j$ as will be shown.

The proof for this relation uses lemma \ref{Xlemma2}. Let us explain in detail the step
from $G_{rj}\setminus\S'_{s}(G_{rj})$ to $G_{rj}\setminus\S'_{s-1}(G_{rj})$. Using 
lemma \ref{Xlemma2} we write:
\bea
\label{eq:mess}
&{}&\!\!\!\!\sum\limits_{j=1}^E X_r(G_{rj}\setminus\S'_{s}(G_{rj}))|
{\vec \beta}_{{\S'}^*_{s}(G_{rj})}=1|
\Z_s)p_j=\beta_{s}^{2[L(G_{rj}\setminus\S'_{s}(G_{rj}))-r(G_{rj})]}\times \\
\!\!\!\!&{}&\!\!\!\!\!\!\!\!\sum\limits_{j=1}^E\left\{
\begin{array}{cl}
		  &   {\rm ~if~}    r(G\setminus\S_s-\pi_s)=r(G)  \\
 X_r(G_{rj}\setminus\S'_{s}(G_{rj})-\pi_s|{\vec \beta}_{{\S'}^*_{s}(G_{rj})}=1|
\Z_s)p_j   &{\rm and}~\pi_{k_0}~{\rm belongs~to~ a}\\
&{\rm loop~in}~ G_{rj}\setminus\S'_{s}(G_{rj})-\pi_s,\\
\vspace{-5pt}                  &    \vspace{-5pt}                                  \\
		  &  {\rm ~if~}   r(G\setminus\S_s-\pi_s)<r(G)  \\
X_r(G_{rj}\setminus\S'_{s}(G_{rj})|{\vec \beta}_{{\S'}^*_{s}(G_{rj})}=1,{\beta_{\pi_s}}=1|
\Z_s)p_j   &{\rm and}~\pi_{k_0}~{\rm belongs~to~ a}\\
  &{\rm loop~in}~ G_{rj}\setminus\S'_{s}(G_{rj})-\pi_s,\\
  \vspace{-5pt}                &          \vspace{-5pt}                         \\
                  &  {\rm ~if~}   r(G\setminus\S_s-\pi_s)<r(G) \\
X_r(G_{rj}\setminus\S'_{s}(G_{rj})|{\vec \beta}_{{\S'}^*_{s}(G_{rj})}=1|
\Z_s,\pi_s)p_j   &{\rm and}~\pi_{k_0}~{\rm does~not~belong~to}\\
&{\rm a~loop~in}~ G_{rj}\setminus\S'_{s}(G_{rj})-\pi_s,\\
    \vspace{-5pt}              &      \vspace{-5pt}                                  \\
 X_r(G_{rj}\setminus\S'_{s}(G_{rj})-\pi_s|{\vec \beta}_{{\S'}^*_{s}(G_{rj})}=1|
\Z_s)p_j+                   &   {\rm ~if~}   r(G\setminus\S_s-\pi_s)=r(G)     \\
X_r(G_{rj}\setminus\S'_{s}(G_{rj})|{\vec \beta}_{{\S'}^*_{s}(G_{rj})}=1|
\Z_s,\pi_s)p_j             &{\rm and~}\pi_{k_0}~{\rm does~ not~belongs~to}\\
                                  &{\rm  a~ loop~in}~ G_{rj}\setminus\S'_{s}(G_{rj})-\pi_s 
\end{array}\right.\nonumber
\eea
where we omitted $[1+{\cal O}(\beta)]$ on the r.h.s.

Using the definitions of $\S, \S', \Z$ and their starred partners,
we can assemble the arguments of $X$ in the four cases in 
\eq{eq:mess} as 
follows:

\noindent
{\bf case 1.}~if $r(G_{rj}\setminus\S'_s-\pi_s)=r(G_{rj})$ 
and $\pi_{k_0}$ belongs to a loop in $G_{rj}\setminus\S'_{s}(G_{rj})-\pi_s$
then $\S_{s-1}=\S_s\cup \{\pi_s\}$, $\S^*_{s-1}=\S^*_s$ and $\Z_{s-1}=\Z_{s}$. Thus
we have 
\be
\S'_{s-1}=\S_{s-1}\setminus\Z_{s-1}=(\S_s\cup \{\pi_s\})\setminus\Z_s=\S'_s\cup \{\pi_s\}
\ee
\be
{\S'}^*_{s-1}=\S^*_{s-1}\setminus\Z_{s-1}=\S^*_s\setminus\Z_s={\S'}^*_s
\ee

\noindent
{\bf case 2.}~if $r(G_{rj}\setminus\S'_s-\pi_s)<r(G_{rj})$ 
and $\pi_{k_0}$ belongs to a loop in $G_{rj}\setminus\S'_{s}(G_{rj})-\pi_s$
then $\S_{s-1}=\S_s$, $\S^*_{s-1}=\S^*_s\cup\{\pi_s\}$ and $\Z_{s-1}=\Z_{s}$.
Thus we have 
\be
\S'_{s-1}=\S_{s-1}\setminus\Z_{s-1}=\S_s\setminus\Z_s=\S'_s
\ee
\be
{\S'}^*_{s-1}=\S^*_{s-1}\setminus\Z_{s-1}=(\S^*_s\cup\{\pi_s\})\setminus\Z_s={\S'}^*_s\cup\{\pi_s\}
\ee

\noindent
{\bf case 3.}~if $r(G_{rj}\setminus\S'_s-\pi_s)<r(G_{rj})$ and $\pi_{k_0}$ does not 
belongs to a loop in $G_{rj}\setminus\S'_{s}(G_{rj})-\pi_s$
then $\S_{s-1}=\S_s$, $\S^*_{s-1}=\S^*_s\cup\{\pi_s\}$ and $\Z_{s-1}=\Z_{s}\cup\{\pi_s\}$.
Thus, we have 
\be
\S'_{s-1}=\S_{s-1}\setminus(\Z_{s}\cup\{\pi_s\})=\S_s\setminus\Z_s=\S'_s
\ee
\be
{\S'}^*_{s-1}=\S^*_{s-1}\setminus\Z_{s-1}=(\S^*_s\cup\{\pi_s\})\setminus(\Z_s\cup\{\pi_s\})=
{\S'}^*_s
\ee
This situation is covered by case {\bf a} and first subcase of case {\bf b} in the 
proof of 
lemma \ref{S'}.
As pointed out before, these cases are mutually exclusive, i.e. each one either 
occurs for all $j=1,...,E$ or it does not accur at all.

\noindent
{\bf case 4.}~We begin by studying the first term. If $\pi_{k_0}$ does not belong to a loop 
in $G_{rj}\setminus\S'_{s}(G_{rj})-\pi_s$ then, from the observation 
{\bf 1.} in the proof of lemma
\ref{S'}, we get that
\be
X_r(G_{rj}\setminus\S'_{s}(G_{rj})-\pi_s|{\vec \beta}_{{\S'}^*_s(G_{rj})}=1|
\Z_s)=
P(G_{rj}\setminus\S'_{s}(G_{rj})\setminus\Z_s-\pi_s|{\vec \beta}_{{\S'}^*_s(G_{rj})}=1)
\ee
This situation is covered by second subcase of
 case {\bf b} in the proof of lemma \ref{S'}.
Since this subcase is exclusive with respect to the other cases in that 
lemma it 
follows that, 
if it occurs, it will occur for all $j=1,...,E$. Therefore, using the 
lemmas \ref{S'} 
and \ref{Prequal}, we find that the first term will cancel upon summing over momenta. 

For the second term we have:
$\S_{s-1}=\S_s\cup \{\pi_s\}$, $\S^*_{s-1}=\S^*_s$ and $\Z_{s-1}=\Z_s\cup\{\pi_s\}$.
This implies that
\be
\S'_{s-1}=\S_{s-1}\setminus\Z_{s-1}=(\S_s\cup \{\pi_s\})\setminus(\Z_s\cup\{\pi_s\})=
\S_s\setminus\Z_s=\S'_s
\ee
\be
{\S'}^*_{s-1}=\S^*_{s-1}\setminus\Z_{s-1}=\S^*_s\setminus(\Z_s\cup\{\pi_s\})=
\S_s^*\setminus\Z_s={\S'}^*_s
\ee

In combining all the above cases we have to use again lemma \ref{S'}. In 
particular, we use
that the union of cases 1 and 2, case 3 and case 4 in
 \eq{eq:mess} are 
mutually exclussive 
together with the fact that $G_{rj}\setminus\S'_k(G_{rj})-\pi_{s}-l_{j}$ are 
identical 
up to 
lines belonging to all trees of $G_{rj}\setminus\S'_k(G_{rj})-\pi_{s}-l_{j}$ for all 
$j=1,...,E$
and any $\pi_s\in \Z_k(G_{rj})$. These two properties allow us to pull out of the sum
the factor $\beta_{s}^{2[L(G_{rj}\setminus\S'_{s}(G_{rj}))-r(G_{rj})]}$.\footnote{
The fact that $G_{rj}\setminus\S'_k(G_{rj})-\pi_{s}-l_{j}$ are identical up to 
lines belonging to all trees of $G_{rj}\setminus\S'_k(G_{rj})-\pi_{s}-l_{j}$ for all 
$j=1,...,E$
and any $\pi_s\in \Z_k(G_{rj})$ implies, in particular, that the number of loops 
of these
graphs is the same for all $j$. Gluing $l_{j}$ to 
$G_{rj}\setminus\S'_k(G_{rj})-\pi_{s}-l_{j}$
we can have two situations: an extra loop is produced (situation 
covered by the first subcase of
case ${\bf b}$ of lemma \ref{S'}) or it is not (situation covered by case ${\bf a}$ of 
lemma \ref{S'}). Since these cases are mutually exclusive, it 
follows that 
the numbers of loops of 
$G_{rj}\setminus\S'_k(G_{rj})$ are the same for all $j=1,...,E$.}
Therefore, 
\eq{eq:mess} becomes:
\bea
&{}&\!\!\!\!\sum\limits_{j=1}^E X_r(G_{rj}\setminus\S'_{s}(G_{rj}))|
{\vec \beta}_{{\S'}^*_{s}(G_{rj})}=1|
\Z_{s})p_j=\beta_{s}^{2(L(G_{rj}\setminus\S'_{s}(G_{rj}))-r(G_{rj}))}
\times\nonumber\\
&{}&\hskip2truecm\sum\limits_{j=1}^E
X_r(G_{rj}\setminus\S'_{s-1}(G_{rj}))|{\vec \beta}_{{\S'}^*_{s-1}(G_{rj})}=1|
\Z_{s-1})p_j
\eea

Replacing all this in the ratio \eq{eq:rat} we get:
\bea
\sum\limits_{j=1}^E
&{}&\!\!\!\!{X_r(G_{rj}\setminus\S'_{k_0-1}(G_{rj})|{\vec \beta}_{{\S'}^*_{k_0-1}(G_{rj})}=1|
\Z_{k_0-1})
          \ov   
P_r(G\setminus\S_{k_0-1}(G)|{\vec \beta}_{\S^*_{k_0-1}(G)}=1)} p_j=\label{eq:expratio}\\
&{}&\!\!\!\!\left(\prod\limits_{i=s+1}^{k_0-1}{\beta_i^{2(L(G_{rj}\setminus\S'_i(G_{rj}))
-L(G-S_i(G))-2)}}\right)\sum\limits_{j=1}^E
{X_r(G_{rj}\setminus\S'_s(G_{rj})|{\vec \beta}_{{\S'}^*_{s}(G_{rj})}=1|\Z_s)
          \ov   
P_r(G\setminus\S_{s}(G)|{\vec \beta}_{\S^*_{s}(G)}=1)} p_j\nonumber
\eea

As pointed out before, from the definition of $\S_i$ and $\S'_i$ we see 
that $\S'_i\subseteq \S_i$
and the number of loops of $G_{rj}\setminus\S'_i(G_{rj})$ is always larger by at least 2
than the number of loops of $G\setminus\S_i(G)$ for all $i<k_0$ (1 loop from 
the joining $(rj)$ and at least 
another loop from never letting the line $\pi_{k_0}$ belong to all trees of 
$G_{rj}\setminus\S'_i(G_{rj})$).

\eq{eq:expratio} and the comment following it are 
valid for any $s=k_r,...,k_0-1$ where $\pi_{k_r}$ 
has the property that by removing it the line on which the derivatives act is 
disconnected from all 
loops. On the other hand we know that if a subgraph of
$G$ does not contain this line $k_r$ in a loop, then its cycle number computed with 
respect to 
any joining of $\pm r$ with external lines is the same as computed with respect to 
the original graph $G$.
Therefore, we conclude that the $\beta$ expansion of the terms linear in $r$ cannot 
be more singular  
than:
\be
f(\beta_1,...,\beta_{k_0-1})\prod\limits_{i=k_0}^I{1\ov\beta_i^2}
\ee
where $\gamma_{k_0}$ is the smallest subgraph of $G$ in the permutation $\pi$ 
that has $j(\gamma_{k_0})=1$. At the same time, the leading term independent of $r$ in the 
same permutation
$\pi$ is  
\be
f(\theta, p)\prod\limits_{i=k_0}^I{1\ov\beta_i^2}~.
\ee
This shows that, for the non-exceptional momenta, the $r\cdot p$ terms are always 
regulated by the $p^2$ terms.~~~~{\bf q.e.d.}

\noindent
{\bf Remark.}~From the details of the proof of theorem \ref{rpterm} follows that 
it is not necessary that the line carrying external momentum $r$ be associated
to lines in the set $\E'(G)$. All we have used is that $\sum_{j\ne r} p_j = 0$.
Therefore, $r$ can denote any fixed line as long as the sum over 
the momenta
of the lines 
different from $r$ appearing in \eq{eq:sumexp} vanishes.

\begin{lm}
The leading term in the expansion of 
${\tilde Q}(p,\alpha,\Theta)$ from \eq{nfeyn} in the powers of $\beta$
is $\O(1)$. 
\label{QO1}
\end{lm}

\noindent
{\bf Proof.}

\be
 - [p_a (Q_{ab}+i{\tilde Q}_{ab}) p_b - {(p_a e_b (Q_{ab}+i{\tilde Q}_{ab})
+i\Theta\cdot p_a n_a)^2
\ov \alpha_* + e_ae_bQ_{ab}} ]
\label{eq:newexp2}
\ee
The imaginary part of the exponent after integration is given by:
\be
 - p_a {\tilde Q}_{ab} p_b + 2{p_a e_b Q_{ab}(p_c e_f{\tilde Q}_{cf}
+\Theta\cdot p_c n_c)
\ov \alpha_* + e_ae_bQ_{ab}}
\label{eq:imag}
\ee
In the above expresion we treat $p$ and $r_i$ in an unified fasion. The only restriction 
is that $e_a$ is never associated to lines carrying momentum $r_i$. Furthermore, because 
the momenta $r_i$ are conserved in pairs, they will not appear in the 
term $\Theta\cdot p_c n_c$. We now split $r$ from $p$ and write separately the $r_i.r_j$ 
and $r_i.p$:
\be
-r_i{\tilde Q}_{ij} r_j + 2 {r_i e_b Q_{ib} \,\, r_j e_f{\tilde Q}_{jf}
\ov \alpha_* + e_ae_bQ_{ab}}
\label{eq:rirj}
\ee
\be
 - 2 r_i {\tilde Q}_{ib} p_b + 2{r_i e_b Q_{ib}(p_c e_f{\tilde Q}_{cf}
+\Theta\cdot p_c n_c)
\ov \alpha_* + e_ae_bQ_{ab}}+ 2{p_a e_b Q_{ab}  \,\,r_i e_f{\tilde Q}_{if}
\ov \alpha_* + e_ae_bQ_{ab}}
\label{eq:ripa}
\ee

We will use induction to show that both the $r_i r_j$ and $r_i p_a$ terms are of order 1
on the $\beta$ expansion.

By explicit computation it can be shown that all the 1-loop graphs satisfy this
assumption. We proceed therefore by assuming that for all L-loop graphs the imaginary
part of the $r_i r_j$ and 
$r_i p_a$ terms in the exponent are of order 1 and apply the two Bogoliubov operations
on these graphs\footnote{The two Bogoliubov operations are:
(1) a new vertex is added together with one line joining this vertex
to one of the old vertices;
(2) a new internal line is added for a given number of vertices.
\label{bogop}}. We will show that the resulting exponents satisfy the assumption.

It is easy to see that by performing the first operation, adding a vertex 
together with a line joining it to some graph $G$, we get the same scaling for 
the imaginary part as for $G$ itself. (actually this case is covered by the assumption 
since the operation does not change the number of loops).

Before we proceed to the second operation, let us notice that it is enough to show that
$(r_i e_b Q_{ib})/(\alpha_* + e_ae_bQ_{ab})$ and $(p_a e_b Q_{ib})/(\alpha_* + 
e_ae_bQ_{ab})$ are of order 1. It is actually easy to show that. Consider the first 
expression:

\noindent  $(r_i e_b Q_{ib})/(\alpha_* + e_ae_bQ_{ab})$ 

We first notice that the leading term in the denominator comes from $e_ae_bQ_{ab}$. The 
reason is the following: look at the term with $\beta_{I+1}$. If the line $\alpha_*$ 
is in $K_2(G_{ab})$, then $e_ae_bQ_{ab}$ is proportional to $1/\beta_{I+1}^2$ thereas 
$\alpha_*\sim\beta_{I+1}^2$. If the line $\alpha_*$ 
is not in $K_2(G_{ab})$, then $e_ae_bQ_{ab}$ is proportional to $\beta_{I+1}^0$ thereas 
$\alpha_*\sim\beta_{I+1}^2$. So we need to study only
\be
{r_i e_b Q_{ib}\ov e_ae_bQ_{ab}} 
\label{eq:r1}
\ee
Now, $e_ae_bQ_{ab}$ can be interpreted as the $p^2$ part of the exponent of the 
graph obtained from the original one by setting to zero all external momenta except
those of the 2 lines $e_a$. From theorem \ref{rpterm}
follows that $r_i e_b Q_{ib}$ is never
leading over $e_ae_bQ_{ab}$.\footnote{This is just the application of 
theorem \ref{rpterm} to 
the particular choice of external momenta $p_a = p e_a$.}
 The equations expressing 
this are 
a direct application 
of \eq{rpt}:
\be
e_ae_bQ_{ab} = \LB \prod\limits_{i=k}^{I+1} {1\ov \beta_i^2}\RB
[1+{\cal O}(\beta)]
\ee
\be
r_ie_bQ_{ib} = f(\beta_1,...,\beta_{k'-1})\LB
\prod\limits_{i=k'}^{I+1} {1\ov \beta_i^2}\RB
[1+{\cal O}(\beta)]
\ee
with $k'\ge k$ and $f$ being some polynomial. Thus, the ratio \eq{eq:r1}
 is
\be
f(\beta_1,...,\beta_{k'-1})\LB
\prod\limits_{i=k}^{k'-1}\beta_i^2 \RB [1+{\cal O}(\beta)]
\ee
which in all situations is of order larger or equal to 1.

Combining this with the inductive assumption we see that this implies that
the expresion in \eq{eq:rirj} is of order $\O(1)$ or higher.

Consider now the term $(p_a e_b Q_{ab})/(\alpha_* + e_ae_bQ_{ab})$.
By the same argument as before it is enough to study the ratio
\be
{p_a e_b Q_{ab}\ov e_ae_bQ_{ab}} 
\label{eq:r2}
\ee

Using theorem \ref{rpterm} and the remark following it on the numerator of 
\eq{eq:r2} we see that each term is, after summation over $e_b$, not leading over 
$\sum_{a,b} e_a Q_{ab} e_b$.  In formulas this means:
\be
e_ae_bQ_{ab} = \LB \prod\limits_{i=k}^{I+1} {1\ov \beta_i^2}\RB
[1+{\cal O}(\beta)]
\ee
\be
p_ae_bQ_{ab} = g(\beta_1,...,\beta_{k'-1}, p_a)\LB
\prod\limits_{i=k'}^{I+1} {1\ov \beta_i^2}\RB
[1+{\cal O}(\beta)]
\ee
with $k'\ge k$ and $g$ some polynomial. Thus, the ratio \eq{eq:r2}
 is
\be
g(\beta_1,...,\beta_{k'-1}, p_a)\LB
\prod\limits_{i=k}^{k'-1}\beta_i^2 \RB [1+{\cal O}(\beta)]
\ee
which is of order larger or equal to 1.

Combining this with the inductive assumption we see that this implies that
the expresion in \eq{eq:ripa} is of order larger or equal to 1.

We have therefore shown that all the $L+1$ loop graphs have the part proportional 
to $r_i$
in the  imaginary part of the exponent of order $\beta^0$ and higher in 
the $\beta$ expansion.
~~~{\bf q.e.d.}

\begin{lm}
Let $\gamma_{k_{ij}}$ be the smallest
subgraph in \eq{nested} which contains both of the lines
$i\notin K_2$ and  $j\notin K_2$
in the loops (if $i=j$ then $\gamma_{k_{ij}}$ is the smallest subgraph 
in \eq{nested} which contains
$i$ in a loop). 
Then,  the leading $r^2$ term in 
the exponent of \eq{analyzthis2}  in the $\beta$ expansion is of the form
\be
\LB r_i (\Theta^2) r_j \prod_{l=k_2}^I \beta_l^{-2}\RB
R_2(\beta_1,\beta_2,\ldots, \beta_{k_2-1})
\label{singular3}
\ee
for some $k_2 \ge k_{ij}$,
where $R_2$ 
is a polynomial in variables $\beta_1,\beta_2,\ldots, \beta_{k_2-1}$.
\label{rsquare}
\end{lm}

\noindent
{\bf Proof.}

\begin{figure}[hbtp]
\begin{center}
\mbox{\epsfxsize=3truecm
\epsffile{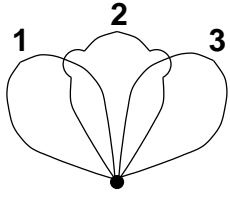}}
\end{center}
\caption{Graph for footnote \ref{ftsquare}}
\label{rsquareft}
\end{figure}

Consider first the case $i = j$. 
Let  $\imath', \imath'' \in {\cal E}'(G)$  be 
the external lines  attached to
the internal line $i\notin K_2$.
Applying theorem 1 to the numerator
and the denominator of the ratio
$P(G_{\imath' \imath''}| \a_{\imath'\imath''}=0)/P(G)$ as in theorem 2, we find
\be
{P(G_{\imath'\imath''  }| \a_{\imath'\imath'' }=0)\ov P(G)}=\theta^2 ~
{
\prod_{l=1}^I \beta_l^{2(L(\gamma_l)-c_{G_{ \imath'\imath'' }}(\gamma_l))}
\ov \prod_{l=1}^I \beta_l^{2(L(\gamma_l)-c_{G}(\gamma_l))}
}
[1+\O(\beta^2)]
\label{newratio}
\ee
It is easy to see that
none of the subgraphs of graph $G$ which do not contain line $i$ in a loop will wrap 
the new cycle of graph $G_{\imath'\imath'' }$
formed by the joining of lines $\imath', \imath'' $. That is we have
\be
c_{G_{\imath'\imath''}}(\gamma)=c_G(\gamma) 
\ee
for such  subgraphs. Thus, for some $k_2\ge k_{ij}$, we have 
\be
c_{G_{ \imath'\imath'' }}(\gamma_{l})=c_G (\gamma_{l}) +1
\ee
for all $l\ge k_2$  and 
\be
c_{G_{ \imath'\imath'' }}(\gamma_{l})=c_G (\gamma_{l}) 
\ee
for $1\le l \le k_2$.\footnote{$k_2$ need not be equal to $k_{ij}$ as the
following example shows. Consider the graph in figure \ref{rsquareft} and suppose
that $i=1$.  
Let us choose the identity permutation, i.e. $\a_1=\beta_1^2\beta_2^2\beta_3^2$,
$\a_2=\beta_2^2\beta_3^2$, $\a_3=\beta_3^2$.  The r.h.s. of \eq{newratio}
for this case reads: ${\beta_1^{2(1-1)} \beta_2^{2(2-2)} \beta_3^{2(3-3)}\ov
\beta_1^{2(1-1)}\beta_2^{2(2-2)}\beta_3^{2(3-2)}}={1\ov \beta_3^2}$.
In this case we have $k_{ij}=1$ and $k_2=3$.
\label{ftsquare}}
We thus conclude that \eq{singular3} holds
in the case $i = j$.

Consider now the case $i \ne j$. 
Let  $\imath',\imath''  \in {\cal E}'(G)$ and 
$\jmath',\jmath'' \in {\cal E}'(G)$ be 
the external lines  attached to
the internal lines $i\notin K_2$ and $j\notin K_2$,
respectively. Up to an overall factor, the
contribution to the exponent in \eq{analyzthis2} reads 
\be
r_i r_j e_a Q_{ab}e_b= r_i r_j [P(G_{\imath' \jmath'})- P(G_{\imath' \jmath''})- 
P(G_{\imath'' \jmath'})+ P(G_{\imath'' \jmath''})]/P(G)
\label{contribut}      
\ee
where $e_{\imath'}=1$, $e_{\imath''}=-1$ and the rest are zero.
Using theorem \ref{rpterm} and the remark following it we get that 
$Q_{\imath' \jmath'}-Q_{\imath' \jmath''}$ is not leading over $Q_{\jmath' \jmath''}$ and
similarly for the other three differences of $Q$-s appearing in 
\eq{contribut}.
Let $\gamma_{s_i}$ and $\gamma_{s_j}$ be the smallest subgraphs of $G$ that contain either
$i$ or $j$ in loops. There are three situations that can appear in a generic permutation:

- $\gamma_{s_i}\subset\gamma_{s_j}$ in which case $\gamma_{s_j}=\gamma_{k_{ij}}$ in the
text of the lemma

- $\gamma_{s_j}\subset\gamma_{s_i}$ in which case $\gamma_{s_i}=\gamma_{k_{ij}}$ in the
text of the lemma

- $\gamma_{s_i}=\gamma_{s_j}$ 

The first two situations are similar. For the first one, from theorem 
\ref{rpterm} we have that neither of  $Q_{\imath' \jmath'}-Q_{\imath' \jmath''}$
nor  $Q_{\imath'' \jmath'}-Q_{\imath'' \jmath''}$ is leading over 
$Q_{\jmath' \jmath''}$. Thus,
we get that the leading term in the expansion for small $\beta$ of 
\eq{contribut} is
\be
\LB r_i (\Theta^2) r_j \prod_{l=k_2}^I \beta_l^{-2}\RB
R_2(\beta_1,\beta_2,\ldots, \beta_{k_2-1})
\ee
with $k_2\ge s_j=k_{ij}$. 

For the second situation, from theorem 
\ref{rpterm} we have that neither of  $Q_{\imath' \jmath'}-Q_{\imath'' \jmath'}$
nor  $Q_{\imath' \jmath''}-Q_{\imath'' \jmath''}$ is leading over 
$Q_{\imath' \imath''}$. Thus,
we get that the leading term in the expansion for small $\beta$ of 
\eq{contribut}
is
\be
\LB r_i (\Theta^2) r_j \prod_{l=k_2}^I \beta_l^{-2}\RB
R_2(\beta_1,\beta_2,\ldots, \beta_{k_2-1})
\ee
with $k_2\ge s_i=k_{ij}$. 

The third situation is much simpler since it does not require the use of the details of 
theorem \ref{rpterm}. From the assumption that $\gamma_{s_i}=\gamma_{s_j}$ 
and theorem \ref{ther2} 
we see that
the leading term in the expansion of each of the four terms in \eq{contribut} 
is given by
\be
\LB r_i (\Theta^2) r_j \prod_{l=k_2}^I \beta_l^{-2}\RB
\ee
with $k_2\ge s_i=s_j=k_{ij}$. 

Combining the three cases, we conclude that \eq{singular3} holds
in the case $i \ne j$. ~~~~~{\bf q.e.d.}

\begin{th}[Convergence theorem]
  Let $I_G$ be the Feynman integral  
for a 1PI graph $G$ in a 
field theory over non-commutative $\re^d$ with the massive propagators.
There are three cases:

\noindent
(I) If $j(G)=0$, then $I_G$ is convergent if 
$\omega(\gamma)- c(\gamma)d<0$ for all $\gamma\subseteq G$.

\noindent
(II) If $j(G)=1$ and the external momenta are non-exceptional, then
$I_G$ is convergent  if 
for any subgraph $\gamma\subseteq G$ 
at least one of the following conditions is satisfied:(1)
$\omega(\gamma)-c(\gamma)d<0$, (2) $j(\gamma)=1$.

\noindent
(III) If $j(G)=1$ and the external momenta are exceptional, then
$I_G$ is convergent if
$\omega(\gamma)- c(\gamma)d<0$ for all $\gamma\subseteq G$.
\label{CTHEOR}
\end{th}

\noindent
{\bf Proof.} 

This theorem
follows from theorem 4 and
 lemmas \ref{K2}, \ref{QO1} and \ref{rsquare}  as explained in
 section 7.1.~~~~{\bf q.e.d.}

\newsection{Comments}

\noindent
$\bullet$  For  the  massive theories the integral \eq{nfeyn} is
convergent  at  the  upper limit ($\a \rightarrow \infty$) of the
integration. 
In  the  massless case, the IR power counting theorem of
ref.\cite{zimmer} does not hold in the non-commutative case. Consider
the graph in figure \ref{exampleps}. In the non-derivative massless 
scalar
NQFT in $d=4$ the corresponding Feynman integral reads
\be
 \int_0^{\infty} d\a_1 d\a_2 
{1\ov (\a_1\a_2 + 
\theta^2)^2} = \int_0^{\infty} {d\a_1 \ov \a_1}~ \int_0^{\infty}
d\a_2 {1\ov (\a_2 + 
\theta^2)^2}
\ee
This integral is divergent. In the massless commutative theory the graph
\ref{exampleps}  is IR convergent and UV divergent. In the massless
non-commutative theory it is  both IR  and UV divergent.

\noindent
$\bullet$  The divergences from $\tCom$
graphs can be removed by the mass renormalization as discussed in 
section 4. In the massless theories this procedure does not work.

\noindent
$\bullet$ 
The techniques developed in this paper can be extended to
 study the asymptotic dependence of 
Feynman diagrams upon external momenta. In the commutative
case this problem was  studied  in ref.\cite{weinberg}.

\noindent
$\bullet$
We formulated the convergence theorem for the nondegenerate matrix 
$\Theta$. It is not difficult to see that in the degenerate case
one has to modify the condition $\omega - c d<0$ and the definition
of exceptional external momenta.  The former condition is modified
to $\omega- c~ {\rm rank} \Theta<0$.  In the degenerate case, one
has to restrict to the momenta  along the nondegenerate directions in
\eq{deffour}.

\pagebreak

\noindent
{\bf Acknowledgements}  

We are especially grateful to W.~Siegel 
 for useful discussions.
 We would  also  like  to  thank  T.~Krajewski, D.~Kreimer,
E.~Miller,
M.~Ro\v{c}ek, G.~Sterman and P.~Zinn-Justin  for enlightening  discussions.
The  work  of I.C.  was  supported in  part by CRDF Award RP1-2108.

\bigskip

\setcounter{section}{0}
\setcounter{subsection}{0}

\appendix{}
Let  us   derive  relation \eq{Qmrs}. Consider
$$
Q=\sum_{i\ne j} p_i Q_{ij} p_j
$$
and 
$$
{\tilde Q}=\sum_{i\ne j} p_i {\tilde Q}_{ij} p_j
$$ 
from the exponent of \eq{nfeyn}. We can always write $Q$ and ${\tilde Q}$ 
in this way by using momentum conservation.

Consider two arbitrary external lines $i,j$ of a graph $G$
and form a new graph $G_{ij}$ as in figure \ref{defj}. 
In terms of Feynman integral
\eq{nfeyn} for the graph $G$, it corresponds to setting
 $p_i=-p_j=q$, including a Schwinger parameter $\alpha_{ij}$
corresponding to the new line formed from the joining of lines $i$ and $j$,
 and integrating over the momentum $q$. Since
${\tilde Q}_{ij}$ is anti-symmetric in the indices $i$ and $j$, it
does not contribute to the determinant coming from the integration 
over $q$. As a result, we find 
\be
P(G_{ij})= P(G) (\alpha_{ij} + Q_{ij} )
\ee
Thus 
$$Q_{ij}={P(G_{ij})|_{\alpha_{ij}=0}\ov P(G) }
$$

The proof of the inequality \eq{Qge} is  analogous to the one for
the commutative theories
given in ref.\cite{bogol}. Relation \eq{Qge} may be verified by
the method of induction, since its validity in the case of simplest
diagrams is evident. We shall carry out the induction by using the 
two Bogoliubov operations
introduced in footnote \ref{bogop}:
(1) a new vertex is added together with one line joining this vertex
to one of the old vertices;
(2) a new internal line is added for a given number of vertices.

It is easy to see that the inequality \eq{Qge}  holds
after the operation (1).
Thus consider operation (2). For the sake of definiteness we shall 
assume that the internal line with momentum $k$ is added between vertices
$1$ and $2$. The modified exponent reads
\be
- \alpha_*~k^2 - \sum_{a, b,a\ne b}(p_a + e_a k) 
(Q_{ab}+i{\tilde Q}_{ab}) (p_b+e_b k)- i\Theta_{\mu\nu} n^a p_a^{\mu} k^{\nu}
\ee
where the symbol $e_a$ is defined as follows
\be
e_1=1,~~e_2=-1,~~e_a =0 ~~(a>2)
\ee
and  $n_a$
  takes values $\pm 1$ or $0$
depending upon whether the line with momentum $p_a$ is intersected or not
by the added line $*$. As usual, the orientation of the intersection 
is given by the sign of $n_a$. Integrating over $k$, one finds
\be
 - [p_a (Q_{ab}+i{\tilde Q}_{ab}) p_b - {(p_a e_b (Q_{ab}+i{\tilde Q}_{ab})
+i\Theta\cdot p_a n_a)^2
\ov \alpha_* + e_ae_bQ_{ab}} ]
\label{eq:newexp1}
\ee
Now we need to show that the 
real part of minus this expression 
is larger or equal to zero, i.e.
\be
p_a Q_{ab} p_b - {(p_a e_b Q_{ab})^2\ov\alpha_* + e_a e_bQ_{ab}}  + 
{(\Theta\cdot p_a n_a+ p_a e_b {\tilde Q}_{ab} )^2
\ov \alpha_* + e_ae_bQ_{ab}} \ge 0
\label{eq:newexp}
\ee
The last term in the above expression 
is positive since it is the norm of a real vector.
By the inductive assumption we have
$$Q_{ab} q_a q_b \ge 0$$ for any $q_a$. As
in the commutative case ref.\cite{bogol}, it implies that
\be
(p_aQ_{ab}p_b)(e_aQ_{ab}e_b)-(e_aQ_{ab}p_b)^2\ge 0
\label{eq:identity}
\ee
Combining  \eq{eq:newexp} and \eq{eq:identity}
we arrive at the conclusion that the real part of the exponent
obtained by adding a line to a graph is 
non-negative.

\end{document}